\pdfoutput=1

\documentclass[11pt,twoside,a4paper,cmspaper,final,collab]{cms-tdr}
\def\svnVersion{210296}\def\cmsCernNo{2013-097}\def\cmsCernDate{\today}\def\cmsMessage{Published in the Journal of Instrumentation as \href{http://dx.doi.org/10.1088/1748-0221/8/09/P09009}{\doi{10.1088/1748-0221/8/09/P09009.}}}

\begin{document}\cmsNoteHeader{EGM-11-001}

\hyphenation{had-ron-i-za-tion}
\hyphenation{cal-or-i-me-ter}
\hyphenation{de-vices}
\RCS$Revision: 201567 $
\RCS$HeadURL: svn+ssh://svn.cern.ch/reps/tdr2/papers/EGM-11-001/trunk/EGM-11-001.tex $
\RCS$Id: EGM-11-001.tex 201567 2013-08-08 16:57:19Z alverson $
\providecommand{\re}{\ensuremath{\cmsSymbolFace{e}}}
\cmsNoteHeader{EGM-11-001}
\title{Energy calibration and resolution of the CMS electromagnetic calorimeter in $\Pp\Pp$ collisions at $\sqrt{s} = 7$\TeV}

\author{The CMS Collaboration}

\abstract
{
The energy calibration and resolution of the electromagnetic
calorimeter (ECAL) of the CMS detector have been determined using
proton-proton collision data from LHC operation in 2010 and 2011 at a
centre-of-mass energy of $\sqrt{s}=7$\TeV with integrated luminosities
of about 5\fbinv.
Crucial aspects of detector operation, such as the environmental
stability, alignment, and synchronization, are presented. The in-situ
calibration procedures are discussed in detail and include the
maintenance of the calibration in the challenging radiation
environment inside the CMS detector. The energy resolution for
electrons from $\cPZ$-boson decays is better than 2\% in the central
region of the ECAL barrel (for pseudorapidity $\abs{\eta}<0.8$) and is
2--5\% elsewhere. The derived energy resolution for photons from
125\GeV Higgs boson decays varies across the barrel from 1.1\% to
2.6\% and from 2.2\% to 5\% in the endcaps. The calibration of the
absolute energy is determined from $\cPZ \to \Pep\Pem$ decays to a
precision of 0.4\% in the barrel and 0.8\% in the endcaps.
}

\hypersetup{%
pdfauthor={CMS Collaboration},%
pdftitle={Energy calibration and resolution of the CMS electromagnetic calorimeter in pp collisions
 at sqrt(s) = 7 TeV},%
pdfsubject={CMS},%
pdfkeywords={CMS, electromagnetic calorimeter, electron, photon,
  energy calibration, energy resolution}}

\maketitle 

\setcounter{tocdepth}{2}
\setcounter{secnumdepth}{3}
\tableofcontents
\pagebreak

\section{Introduction}
\label{sec:intro}

The Compact Muon Solenoid (CMS) experiment~\cite{Chatrchyan:2008aa} is
designed to search for new physics at the \TeV energy scale, exploiting
the proton-proton and heavy-ion collisions produced by the Large
Hadron Collider (LHC)~\cite{Evans:2008zzb} at CERN. A key part of the
research programme is the investigation of electroweak symmetry breaking
through the direct search for the standard model (SM) Higgs boson. The
two-photon decay ($\PH\to\Pgg\Pgg$) is one of the most sensitive
channels in the search for a low-mass Higgs boson ($m_H<150\GeV$)
\cite{Seez90}, and was an essential contributor to the discovery
of the new boson in 2012~\cite{Chatrchyan:2012ufa, Chatrchyan:2013lba}.
Its distinctive
experimental signature is a narrow peak -- with a width dominated by
the instrumental resolution, the natural width of a low-mass Higgs
boson being less than 10\MeV -- in the invariant mass distribution of
two isolated photons with high momentum component transverse to the
beam axis, on top of a large irreducible background from direct
production of two photons. Events where at least one of the photon
candidates originates from misidentification of jet fragments
contribute to an additional, reducible background. The electromagnetic
calorimeter (ECAL) \cite{ECAL_tdr} of CMS has been specifically
designed to provide excellent invariant mass resolution, via precise
determination of energy and position, and fine transverse granularity
for photon identification purposes, to enhance the sensitivity to
the $\PH\to\Pgg\Pgg$ decay.
In this paper, we discuss the instrumental and operational aspects of
the CMS ECAL that are particularly relevant to the observation of the
$\PH\to\Pgg\Pgg$ decay. Emphasis is given to single-channel
response stability and uniformity within the ECAL, and to the
calibration of the energy of electrons and photons in CMS, as these
directly contribute to the overall energy resolution.

The central feature of the CMS detector is a superconducting solenoid
13\unit{m} long, with an internal diameter of 6\unit{m}. The solenoid generates a
3.8\unit{T} magnetic field along the axis of the LHC beams. Within the field
volume are a silicon pixel and strip tracker, a lead tungstate
scintillating crystal electromagnetic calorimeter and a
brass/scintillator hadron calorimeter. A lead/silicon strip preshower
detector is installed in front of the crystal calorimeter in the
forward sections. Muons are identified and measured in gas-ionization
detectors embedded in the outer steel magnetic flux return yoke. The
detector is subdivided into a cylindrical barrel part, and endcap
disks on each side of the interaction point. Forward calorimeters
complement the coverage provided by the barrel and endcap
detectors. CMS uses a two-level online trigger system to reduce the
event rate from about 20\unit{MHz} to about 300\unit{Hz}. The first level (L1)
uses custom electronics close to the detector to analyze coarse
information from the calorimeters and muon detectors to reduce the
rate to 100\unit{kHz} or less. The second level (known as the high-level
trigger) uses a computing farm to analyse the full information
from all subdetectors in order to make the final decision on whether
to record an event. A detailed description of the CMS detector can be
found in~\cite{Chatrchyan:2008aa}.

The CMS experiment uses a right-handed coordinate system, with the
origin at the nominal interaction point in the centre of CMS, the
$x$ axis pointing to the centre of the LHC ring, the $y$ axis pointing
vertically up (perpendicular to the LHC plane), and the $z$ axis along
the anticlockwise-beam direction. The pseudorapidity $\eta$ is defined
as $\eta=-\ln{[\tan(\theta/2)]}$, where $\theta$ is the
polar angle measured from the $z$ axis. The azimuthal angle, $\phi$,
is measured in the $x$-$y$ plane.

The installation of the ECAL crystal calorimeter inside the CMS
detector was completed in August 2008. The preshower detector was
installed in 2009. Early commissioning and initial calibrations were
performed with cosmic-ray muons \cite{Chatrchyan:2009hb} and using a
special data sample collected before collisions were achieved, where
bunches of 10$^9$ protons from the LHC were dumped in the collimators
150\unit{m} upstream of CMS. These results are summarized in
\cite{Chatrchyan:2009aj,Chatrchyan:2009ac,Chatrchyan:2009qm}.

The results presented in this paper make use of proton-proton
collision data from LHC operation in 2010 and 2011 at a centre-of-mass
energy $\sqrt{s} = 7\TeV$ with integrated luminosities of 36\pbinv
and 4.98\fbinv, respectively. The LHC bunch spacing was 50\unit{ns}
throughout this period. After a brief description of the CMS ECAL
(Section~\ref{sec:ecal}) we summarize its status during 2010
and 2011 (Section~\ref{sec:status}), paying particular attention to
the quantities influencing the energy
resolution. Section~\ref{sec:ReMoniCal} describes the monitoring and
calibration techniques employed, whilst Section~\ref{sec:ereso}
describes the energy resolution achieved. The energy resolution,
estimated from the analysis of $\cPZ$-boson decays into electrons, is
compared to Monte Carlo (MC) simulation. The energy resolution for
photons relevant to the $\PH\to\Pgg\Pgg$ analysis is discussed.

\section{The CMS electromagnetic calorimeter
\label{sec:ecal}}

 \begin{figure}[t]
  \begin{center}
    \includegraphics[width=0.7\linewidth]{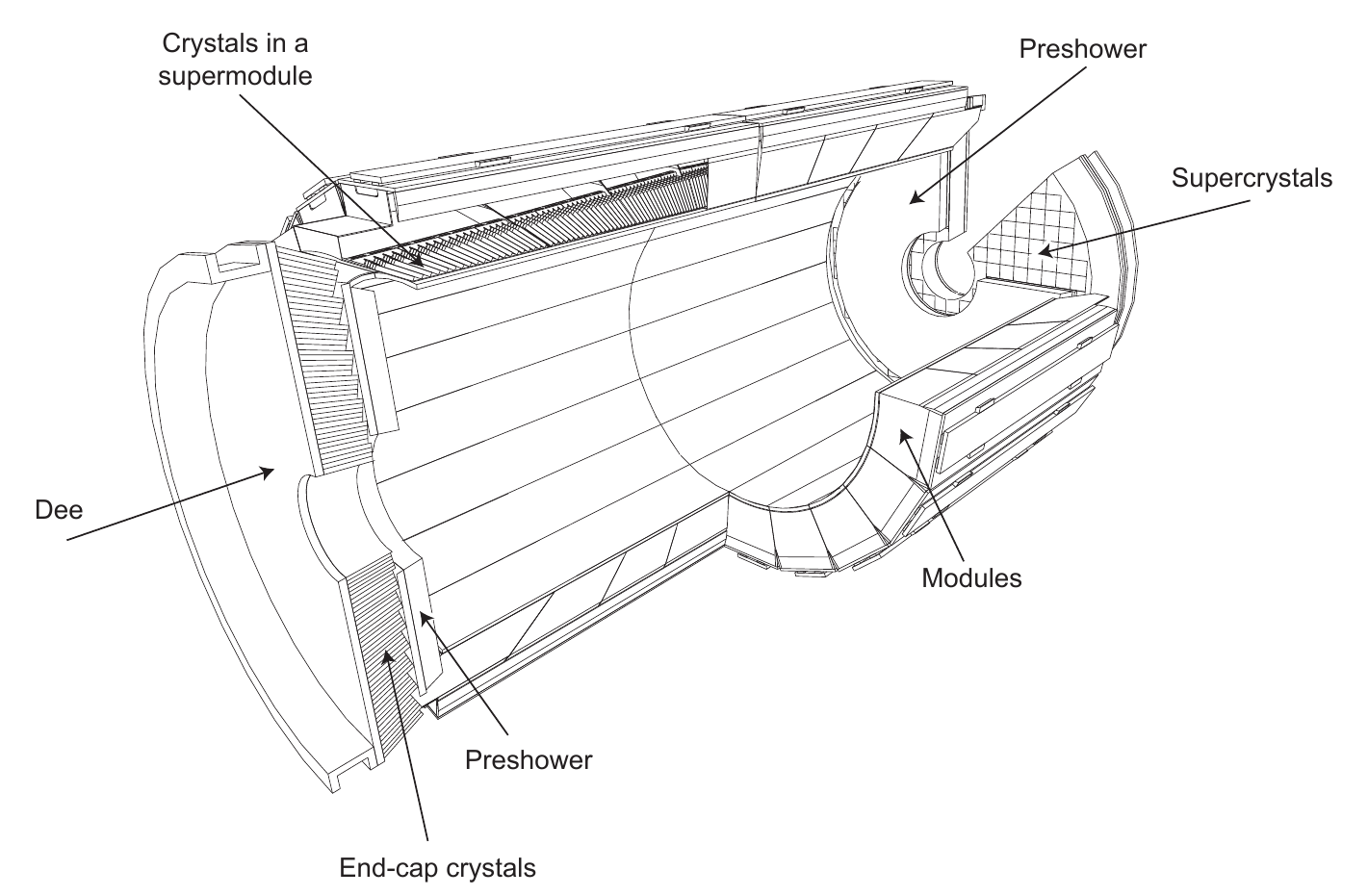}
  \end{center}
  \caption{Layout of the CMS ECAL, showing the barrel supermodules,
    the two endcaps and the preshower detectors. The ECAL barrel
    coverage is up to $\abs{\eta} = 1.48$; the endcaps extend the coverage
    to $\abs{\eta} = 3.0$; the preshower detector fiducial area is
    approximately  $1.65 < \abs{\eta} < 2.6$.
 \label{fig:ECAL} }
 \end{figure}

The CMS ECAL (Fig.~\ref{fig:ECAL})~\cite{Chatrchyan:2008aa,ECAL_tdr}
is a homogeneous and hermetic calorimeter containing 61200 lead
tungstate (PbWO$_4$) scintillating crystals mounted in the barrel
(EB), closed at each end by endcaps (EE) each containing 7324
crystals. A preshower detector (ES), based on lead absorbers equipped
with silicon strip sensors, is placed in front of the endcap crystals,
to enhance photon identification capabilities.
Avalanche photodiodes (APDs) \cite{Baccaro:1999ns,Antunovic:2005} and
vacuum phototriodes (VPTs) \cite{Bell:2004eu} are used as
photodetectors in the EB and EE respectively. The high-density
($8.28\,\mathrm{g/cm}^3$), short radiation length ($X_0=0.89$\unit{cm}), and small
Moli\`{e}re radius ($R_{\mathrm{M}}=2.2\unit{cm}$) of PbWO$_4$ allow the
construction of a compact calorimeter with fine granularity. The
PbWO$_4$ properties were improved during a long R\&D project in
collaboration with the producers in Russia (BTCP in Bogoroditsk) and
China (SIC in Shanghai), leading to the mass production of optically
clear, fast, and radiation-tolerant
crystals~\cite{Auffray:2002sm,Mao:2004}.

The PbWO$_4$ crystals emit blue-green scintillation light with a broad
maximum at wavelengths 420--430\unit{nm}. The quantum efficiency and
surface coverage of the photodetectors are such that a particle
depositing 1\MeV of energy in a crystal produces an average signal of
about 4.5 photoelectrons both in EB and EE. The stability of the
temperature and of the photodetector gain are critical for an accurate
determination of the energy deposited in the crystals, as described in
Section~\ref{sec:status}.
The crystals have to withstand the damage to the crystal lattice
caused by radiation expected throughout the duration of LHC
operation. The expected integrated ionizing dose in the ECAL is up to
4\unit{kGy} in the barrel and 200\unit{kGy} at $\abs{\eta}=3$ after 10 years of LHC
operation corresponding to an integrated luminosity of
500\fbinv~\cite{ECAL_tdr}. The expected hadron fluence varies
between about 10$^{13}\unit{cm}^{-2}$ in the barrel and
10$^{14}\unit{cm}^{-2}$ at $\abs{\eta}=3$. The main observable effect of
the radiation is a wavelength-dependent loss of crystal transparency
but without changes to the scintillation
mechanism~\cite{Adzic:2009aa}. A second effect of the
radiation is that the VPT response decreases with accumulated
photocathode charge to a plateau~\cite{ECAL_VPT_rad}. Radiation does
not affect the gain of the APDs but in large doses induces dark
currents which cause small reductions in the bias voltage at the APDs
if not compensated for. In order to measure and correct for response
change during LHC operation, the ECAL is equipped with a light
monitoring system~\cite{Anfreville:2007zz,Zhang:2005ip}.

The EB crystals have a truncated pyramidal shape and are mounted in a
quasi-projective geometry, to minimize inter-crystal gaps aligned to
particle trajectories. The geometric construction of the EE is based
on a right-sided crystal with two tapering sides. The EB uses 23\unit{cm}
long crystals with front face cross sections of around
$2.2\unit{cm}{\times}2.2\unit{cm}$, whilst the EE comprises 22\unit{cm} long crystals
with front face cross sections of $2.86\unit{cm}{\times}2.86\unit{cm}$.
In the EB the crystals are organized in 36 supermodules, 18 on each
side of the beam interaction point, and provide 360-fold granularity
in $\phi$ and 85-fold granularity in each eta direction up to $\abs{\eta}
= 1.48$. Each supermodule is made up of four modules along $\eta$.
The EE extends the coverage to $\abs{\eta}$ = 3.0, with the crystals
arranged in an $x$-$y$ grid to form an approximately circular shape.
The ES fiducial area is approximately $1.65 < \abs{\eta} < 2.6$. The ES
contains two active planes of silicon strip sensors and associated
mechanics, cooling and front-end electronics. The sensors have an
active area of $61\unit{mm}{\times}61\unit{mm}$, divided into 32 strips. The planes
closer to the interaction point have their strips aligned vertically
while the farther plane strips are horizontal, to provide accurate
position measurement and fine granularity in both coordinates.
Electron and photon separation is possible up to $\abs{\eta}=2.5$, the
limit of the region covered by the silicon tracker.

The ECAL barrel energy ($E$) resolution for electrons in beam tests
has been measured to be~\cite{Adzic:2007mi}:
\begin{equation}
\frac{\sigma_{E}}{E}=\frac{2.8\%}{\sqrt{E(\GeVns{})}}\oplus\frac{12\%}{E(\GeVns{})}\oplus 0.3\% ,\label{eq:energyres}
\end{equation}
where the three contributions correspond to the stochastic, noise, and
constant terms. This result was obtained reconstructing
the showers in a matrix of 3$\times$3 crystals where the electron
impact point on the calorimeter was tightly localized in a region of
4\unit{mm}$\times$4\unit{mm} to give maximum containment of the shower
energy within the 3$\times$3 crystal matrix. The stochastic term
includes contributions from the shower containment, the number of
photoelectrons and the fluctuations in the gain process. The noise
term of 12\% at 1\GeV corresponds to a single-channel noise of about
40\MeV, giving 120\MeV in a matrix of 3$\times$3 crystals. The
constant term, which dominates the energy resolution for high-energy
electron and photon showers, depends on non-uniformity of the
longitudinal light collection, energy leakage from the back of the
calorimeter, single-channel response uniformity and stability. The
beam test setup was without magnetic field, no inert material in front
of the calorimeter, and accurate equalization and stability of the
single-channel response (better than 0.3\%)~\cite{Adzic:2008zza}.
The specification for the ECAL barrel crystals was chosen to ensure
that the non-uniformity of the longitudinal light collection and the
energy leakage from the back of the calorimeter contributed less than
0.3\% to the constant term~\cite{Auffray:2002sg,ECAL_tdr}. The beam
test resolution studies show that this target was met.

During CMS operation, the contributions to the resolution due to
detector instabilities and to the channel-to-channel response spread
must be kept to within 0.4\%, in order to retain the excellent
intrinsic resolution of the ECAL. The `intercalibration constants',
used to equalize the channel-to-channel response, must be measured
with appropriate calibration procedures for single-channel response
and stability. Moreover, the intense field of the CMS magnet and inert
material upstream of the ECAL affect the stochastic term of the
resolution, for electrons and photons that interact before reaching
the calorimeter. Energy deposits from multiple interactions per LHC
bunch crossing (pileup) and APD dark current changes induced by
radiation damage contribute to the noise term, but these were
negligible in 2010 and 2011.

In studying the energy resolution of the ECAL inside CMS, discussed in
Section~\ref{sec:ereso}, the in-situ data have been compared to the
predictions of a full MC simulation of the CMS detector based on
\GEANTfour~\cite{Agostinelli:2002hh,Allison:2006ve}.
The simulation of the ECAL standalone response has been tuned to match
test beam results, upon a detailed simulation of the readout stage,
with inclusion of fluctuations in the number of photoelectrons and in
the gain process as well as a detailed description of the
single-channel noise. The simulation also includes a spread of the
single-channel response corresponding to the estimated
intercalibration precision for the 2010-2011 data, an additional
constant term of 0.3\% to account for longitudinal non-uniformity of
light collection, and the few non-responding channels identified in
data. Response variations with time are not simulated; response
corrections are applied to data at the single-crystal level.

\section{ECAL operation and stability}
\label{sec:status}
The ECAL has been efficiently operating since installation. The
percentages of responding channels in EB, EE and ES at the end
of 2011 were 99.1\%, 98.6\%, and 96.1\% respectively. The electronic
noise was stable during 2010 and 2011. At the start of ECAL operation
it was equivalent to an energy deposit in the crystals of about 42\MeV
per channel in the EB, and a transverse energy ($\ET$, defined as the
energy component transverse to the beam axis) deposit of about 50\MeV
per channel in the EE. A small fraction of channels, 0.1\% in the EB
and 0.4\% in the EE, have been classified as problematic, due to high
levels of electronic noise. These channels were suppressed in the
trigger and in the offline reconstruction.

Triggers for electron/photon candidates were provided by the
two-level trigger system of CMS. At L1, electromagnetic candidates are
formed from the sum of the transverse energy in two adjacent trigger
towers (\ie, arrays of 5$\times$5 crystals in EB). Coarse information
on the lateral extent of the energy deposit inside each trigger tower
is exploited to suppress spurious triggers, such as those arising from
direct ionization in the APD sensitive
region~\cite{ECAL_CMSNOTE_2010-12, Petyt:2012sf}. This feature has
allowed the single-photon L1 trigger to be operated unprescaled at a
low threshold of $\ET =15$\GeV in 2011. From data analysis, this
trigger has been verified to be fully efficient ($>$99\%) for $\ET >
20$\GeV, causing no inefficiencies to, \eg, the $\PH\to\Pgg\Pgg$
analysis, for which events are retained if the leading photon has
transverse energy $\ET > 35$\GeV.

The operating temperature of ECAL of 18$\,^\circ$C is maintained
by a dedicated cooling system~\cite{ECAL_cooling}. The temperature
dependence of the crystal light yield ($-2\%/^\circ$C) and of the APD
gain ($-2\%/^\circ$C) demand a precise temperature stabilization of
better than 0.05$\,^\circ$C in the EB. In the endcaps, the dependence of
the VPT response on the temperature is negligible, and a stabilization
of better than 0.1$\,^\circ$C for the crystals is sufficient. These
specifications limit the contribution to the constant term of the
energy resolution to be less than 0.2\%. The measured temperature
stability throughout 2010 and 2011 is better than 0.03$\,^\circ$C in
EB and 0.08$\,^\circ$C in EE.

The APD working point, of nominal gain 50, has been chosen to provide
a good signal-to-noise ratio with an acceptable sensitivity of the
gain to the bias voltage of 3.1\%/V. This is achieved with a high
voltage (HV) of around 380~V~\cite{Bartoloni:2007hx}. The contribution
of the gain variation to the constant term is required to be less than
0.2\%, implying an HV stability of around 65\unit{mV}. The measured
fluctuation during 2011 was around 33\unit{mV}. The VPTs operate in a
region where the response variation with HV is less than 0.1\%/V. The
stability of the EE HV supplies is better than 0.1\unit{V} over 100 days so
the contribution to the constant term from this source is negligible.

\begin{figure}[htbp]
 \begin{center}
   \includegraphics[width=0.8\linewidth]{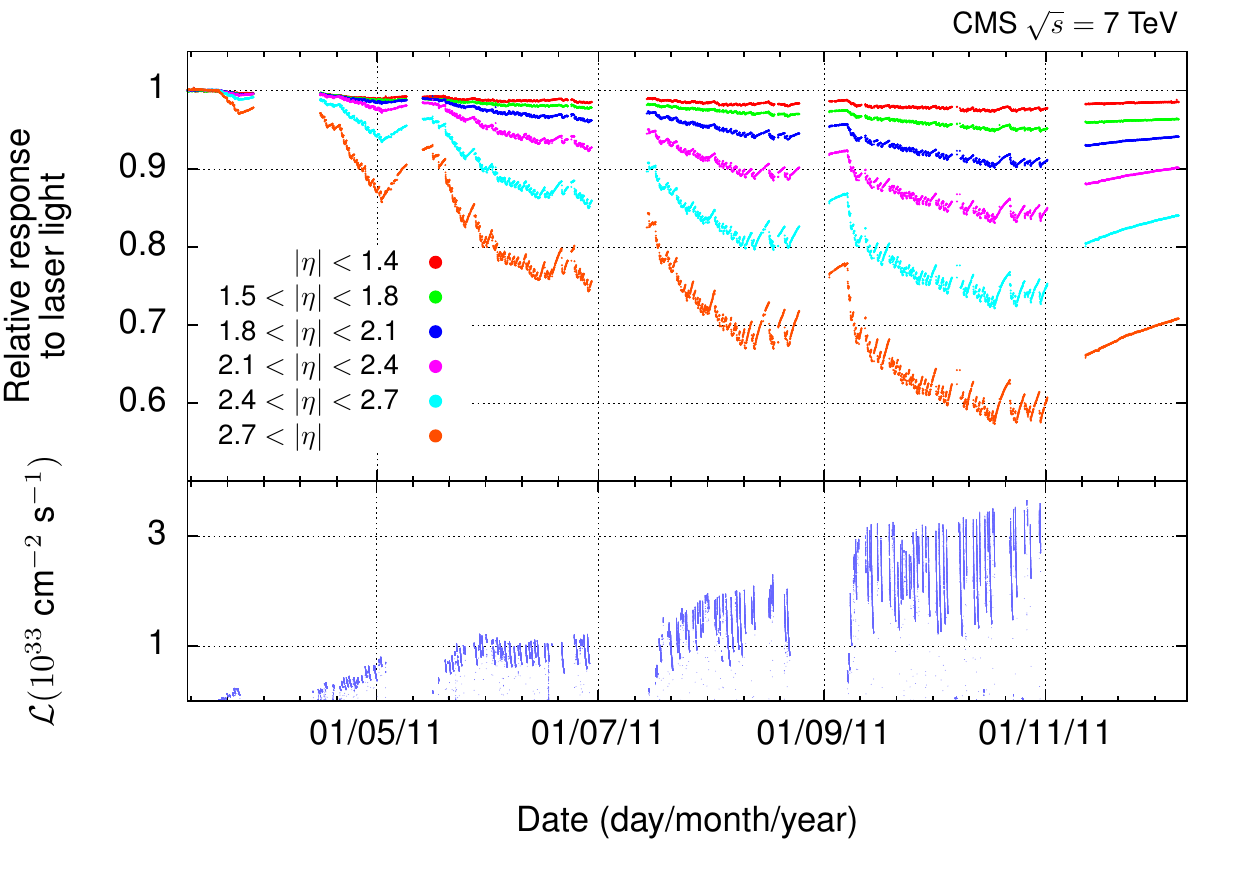}
 \end{center}
 \caption{Relative response to laser light during 2011, normalized to
   data at the start of 2011. An average is shown for each
   pseudorapidity range. The bottom plot shows the corresponding
   instantaneous luminosity. After the last LHC technical stop, a
   recovery of crystal transparency is observed during the low
   luminosity heavy-ion data-taking at the end of 2011.
\label{fig:tloss} }
\end{figure}

The ECAL response varies under irradiation due to the formation of
colour centres that reduce the transparency of the lead tungstate. The
crystal transparency recovers
through spontaneous annealing~\cite{Adzic:2009aa}. A
monitoring system, based on the injection of laser light at 440\unit{nm},
close to the emission peak of scintillation light from PbWO$_4$, into
each crystal, is used to track and correct for response changes during
LHC operation \cite{Anfreville:2007zz, Zhang:2005ip}. Additional laser, and
LED in the EE, light sources provide ancillary information on the
system stability.
The evolution of the ECAL response to the laser light in 2011 is shown
in Fig.~\ref{fig:tloss}, as a function of time. An average value is
shown for each of six pseudorapidity ranges. The data are normalized
to the measurements at the start of 2011. The corresponding
instantaneous luminosity is also shown. The response drops during
periods of LHC operation, but for a given dose-rate the compensating
self-annealing of the crystals reduces the rate of loss of response.
These observations correspond to changes in crystal transparency
\cite{Adzic:2009aa}, coupled with a more gradual loss in VPT
response in EE due to the radiation environment at the
LHC~\cite{ECAL_VPT_rad}. The average drop in response to laser
light, by the end of 2011, was 2--3\% in EB rising to 40\% in the
range $2.7 \le \abs{\eta} \le 3.0$ in EE.

The last data-taking period covered in Fig.~\ref{fig:tloss}, in
November 2011, was for low luminosity heavy-ion data-taking, when the
crystal transparencies partially recovered due to self-annealing.
During this period the precision of the monitoring system was
measured. The laser cycle provides a measurement from each channel
every 20 to 30 minutes. By taking three consecutive measurements, the
middle point can be compared to the interpolated value from the other
two. The RMS for the difference is on average $3 \times 10^{-4}$ for
each channel. This is well within the required precision of
0.2\%. The system stability was measured prior to proton-proton
collisions, for periods of 30 days, with 99.8\% of the monitored
channels in EB and 98.3\% in EE exhibiting stability within
requirements, of better than 0.2\%~\cite{Chatrchyan:2009qm}. Using
quasi-online processing of the monitoring data, single-channel
response corrections are delivered in less than 48\unit{h} for prompt
reconstruction of the CMS data. The complete set of corrections used
for final calibration of the ECAL is discussed in this paper.

\section{Reconstruction and energy calibration}
\label{sec:ReMoniCal}
The front-end electronics of the EB, EE, and ES use 12-bit
analogue-to-digital converters (ADC) to sample the analogue signals
from the detectors (APDs, VPTs, and silicon sensors) at 40\unit{MHz}. In EB
and EE ten consecutive samples are stored for each trigger received,
whilst in the ES only three samples are stored. The delays of the
EB/EE readout pipelines, common for 5$\times$5 channels, are adjusted
in steps of 1.04\unit{ns} such that the signal pulse is expected to start
from the fourth sample and the baseline pedestal value can be
estimated from the first three samples \cite{ECAL_CMSNOTE_2010-12}. In
the ES the pedestal is in the first
sample and the signal is in the two following samples. In both cases
the amplitude of the signal is reconstructed in the same way using a
linear combination of the samples: $A$ = $\sum_j w_j \cdot s_j$, where
$s_j$ is the sample value in ADC counts and $w_j$ is a weight,
optimized for noise reduction using the average pulse shapes measured
in beam tests in the respective detectors~\cite{Bruneliere:2006ra}.

The fast time constants of PbWO$_4$ scintillation and the response
of the readout electronics provide excellent time resolution
capabilities \cite{Chatrchyan:2009aj}. The signal arrival time is measured
from the relative phase of the signal samples to the expected shape of
an in-time signal, with an algorithm using ratios of consecutive
samples. Residual channel-to-channel time offsets are corrected with
appropriate constants derived from in-situ data~\cite{Chatrchyan:2009aj,
  ECAL_CMSNOTE_2010-12}. The timing resolution is measured from data
using electrons from $\cPZ$-boson decays ($\cPZ\to \Pep\Pem$). By comparing
the time difference between the channels with highest amplitude in
each of the two electron showers, we deduce the
single-channel timing resolution to be 190\unit{ps} and 280\unit{ps} in EB and EE
respectively, for the energy range of electrons from the $\cPZ$-boson
decays.
The timing information, combined with topological information of the
energy deposits, is exploited at reconstruction level to reject
signals inconsistent with the emission of scintillation light by
particles produced in $\Pp\Pp$ collision events. These spurious signals
include those arising from direct ionization in the APD sensitive
region that survive the rejection at trigger level. The residual
contamination of these spurious deposits has a negligible impact on
the current analysis~\cite{ECAL_CMSNOTE_2010-12, Petyt:2012sf}.

The ECAL crystals are approximately one Moli\`ere radius in lateral
dimension; thus high energy electromagnetic showers spread laterally over
several crystals. Furthermore, in CMS, the presence of material in
front of the electromagnetic calorimeter (corresponding to 1--2$\,X_0$
depending on the $\eta$ region) causes conversion of photons and
bremsstrahlung from electrons. The strong magnetic field of the
experiment tends to spread this radiated energy along $\phi$.
Clustering algorithms are used to sum together energy
deposits in adjacent crystals belonging to the same electromagnetic
shower. The clustering algorithm
proceeds first with the formation of ``basic clusters'', corresponding
to local maxima of energy deposits. The basic clusters are then merged
together to form a ``supercluster'', which is extended in
$\phi$, to recover the radiated energy. Because of the differences
between the geometric arrangement of the crystals in the barrel and
endcap regions, a different clustering algorithm is used in each
region. The clustering algorithm used in EB, called the `hybrid'
algorithm, is described in~\cite{CMS_TDR_v1}. In EE and ES, the
algorithm merges together fixed-size 5$\times$5 crystal basic clusters
and associates each with corresponding ES energy deposits.

The energy in a supercluster can be expressed as:
\begin{equation}
E_{\Pe,\Pgg} = F_{\Pe,\Pgg} \cdot \bigl[ G\cdot
\sum_{i } S_i(t) \cdot C_i \cdot A_i + E_{\mathrm{ES}} \bigr],
\label{eq:one}
\end{equation}
where the sum is over the crystals $i$ belonging to the
supercluster. The energy deposited in each crystal is given by the
pulse amplitude ($A_i$), in ADC counts, multiplied by ADC-to-GeV
conversion factors ($G$), measured separately for EB and EE, by the
intercalibration coefficients ($C_i$) of the corresponding channel,
and by $S_i(t)$, a correction term due to radiation-induced channel
response changes as a function of time $t$.
The preshower energy ($E_{\mathrm{ES}}$) computation and calibration
procedure are described in Section~\ref{sec:es}.
The term $F_{\Pe,\Pgg}$ represents the energy correction, applied to
the superclusters to take into account the $\eta$- and
$\phi$-dependent geometry and material effects as well as the fact
that electrons and photons shower slightly differently.
This factorization of the various contributions to the electromagnetic
energy determination enables stability and intercalibration to be studied
separately from material and geometry effects.

For the purpose of studying the ECAL calibration and performance, the
energy of both electrons and photons is estimated from the
supercluster energy. For electrons, this is different from the default
energy reconstruction in CMS, which uses the combination of the
supercluster energy and the momentum of the track matched to the
supercluster~\cite{Baffioni:2006cd}. This combination is mainly relevant
for transverse energies below 25\GeV.

Electron identification relies upon matching the measurements in the
ECAL and the Tracker to better than 0.02\unit{rad} in $\phi$ and 4$\times
10^{-3}$~units in $\eta$~\cite{EGM-10-004}. The accurate position
measurement of photons impacting on the calorimeter is used in
determining their direction with respect to the collision vertex,
which is located and, in case of multiple vertices, identified with
analysis-dependent algorithms exploiting track information
(e.g.~\cite{Chatrchyan:2012ufa, Chatrchyan:2013lba, HGG-PAPER}).
The accuracy of the measurement of the opening angle between the two
decay photons from the Higgs boson contributes to its reconstructed
invariant mass resolution. The ECAL alignment and position
resolution measurement is performed with isolated electrons from
$\PW$-boson decays using both the ECAL and tracker information. The
achieved position resolution in EB (EE) is 3 (5)\unit{mrad} in $\phi$ and
1$\times 10^{-3}$ (2$\times 10^{-3}$)~units in $\eta$, and matches the
position resolution of a MC simulation with perfectly aligned
geometry.
Efficient clustering and total energy measurement in the
endcaps requires the alignment between EE and ES to be known to better
than the ES strip pitch ($\approx$2\unit{mm}). The measured alignment
uncertainty is better than 0.15\unit{mm}.

\subsection{Corrections for changes in response, \texorpdfstring{$S_i(t)$}{Si}}
\label{sec:lasercorrection}

The ECAL light monitoring (LM) system~\cite{Anfreville:2007zz,Zhang:2005ip}
is used to determine corrections, denoted by $S_i(t)$ in
Eq.~(\ref{eq:one}), to response changes in the ECAL.
The laser light is injected through optical fibres in each EB and
EE crystal through the front and rear face respectively. The
spectral composition and the path for the collection of laser light at
the photodetector are different from those for scintillation light. A
conversion factor is required to relate the changes in the ECAL
response to laser light to the changes in the scintillation
signal. The relationship is described by a power law \cite{ECAL_tdr}:
\begin{equation}
\label{eq:alpha}
\frac{S(t)}{S_0} = \biggl(\frac{R(t)}{R_0}\biggr) ^\alpha,
\end{equation}
where $S(t)$ is the channel response to scintillation light at a
particular time $t$, $S_0$ is the initial response, and $R(t)$ and $R_0$ are
the corresponding response to laser light. The exponent
$\alpha$ is independent of the loss for small transparency losses.

The value of $\alpha$ has been measured in a beam test for a limited
set of crystals under irradiation. Average values of 1.52 and 1.0 were
found for crystals from the two producers, BTCP and SIC,
respectively~\cite{ECAL_alpha_1,Adzic:2006za,ECAL_alpha_3}. The values
are in qualitative agreement with a ray-tracing simulation
program~\cite{Gentit200235} and are due to the different initial
transparency of the two sets of crystals. The spread in $\alpha$ was
found to be 10\%~\cite{ECAL_alpha_3}, which arises from residual
differences in transparency and different surface treatments of the
crystals. Given the response loss to laser light, shown in
Fig.~\ref{fig:tloss}, the spread in $\alpha$ limits the precision of
the response correction by the end of 2011 running for a single
channel to 0.3\% in EB, and between 0.5\% and a few percent at
high pseudorapidity in EE.

\subsubsection{Validation of the response corrections using collision data}
\label{sec:wenu_laser}
The response corrections were tuned and validated using the energy
of electrons from $\PW$-boson decays, the reconstructed
mass from $\Pgh$ decays to two photons, and the energy resolution
measured with $\cPZ \to \Pep\Pem$ events. The tuning involves the
optimization of the value of $\alpha$, for BTCP and SIC crystals in EB
and EE separately, to obtain the best in-situ resolution of the
invariant mass of the $\cPZ$-boson.

\begin{figure}[bht]
\begin{center}
 \includegraphics[width=0.40\linewidth,height=5.6cm]{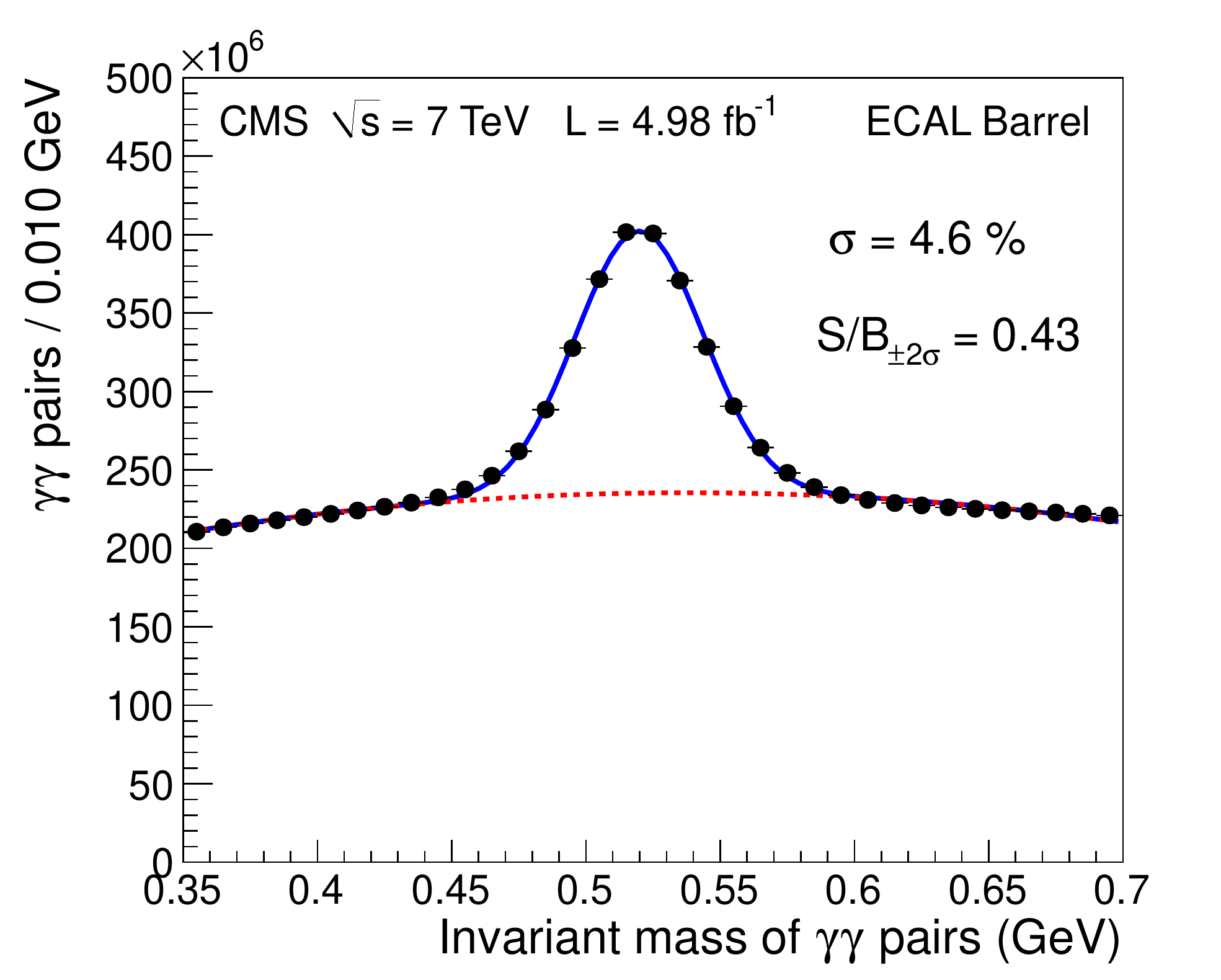}
 \includegraphics[width=0.59\linewidth,height=5.6cm]{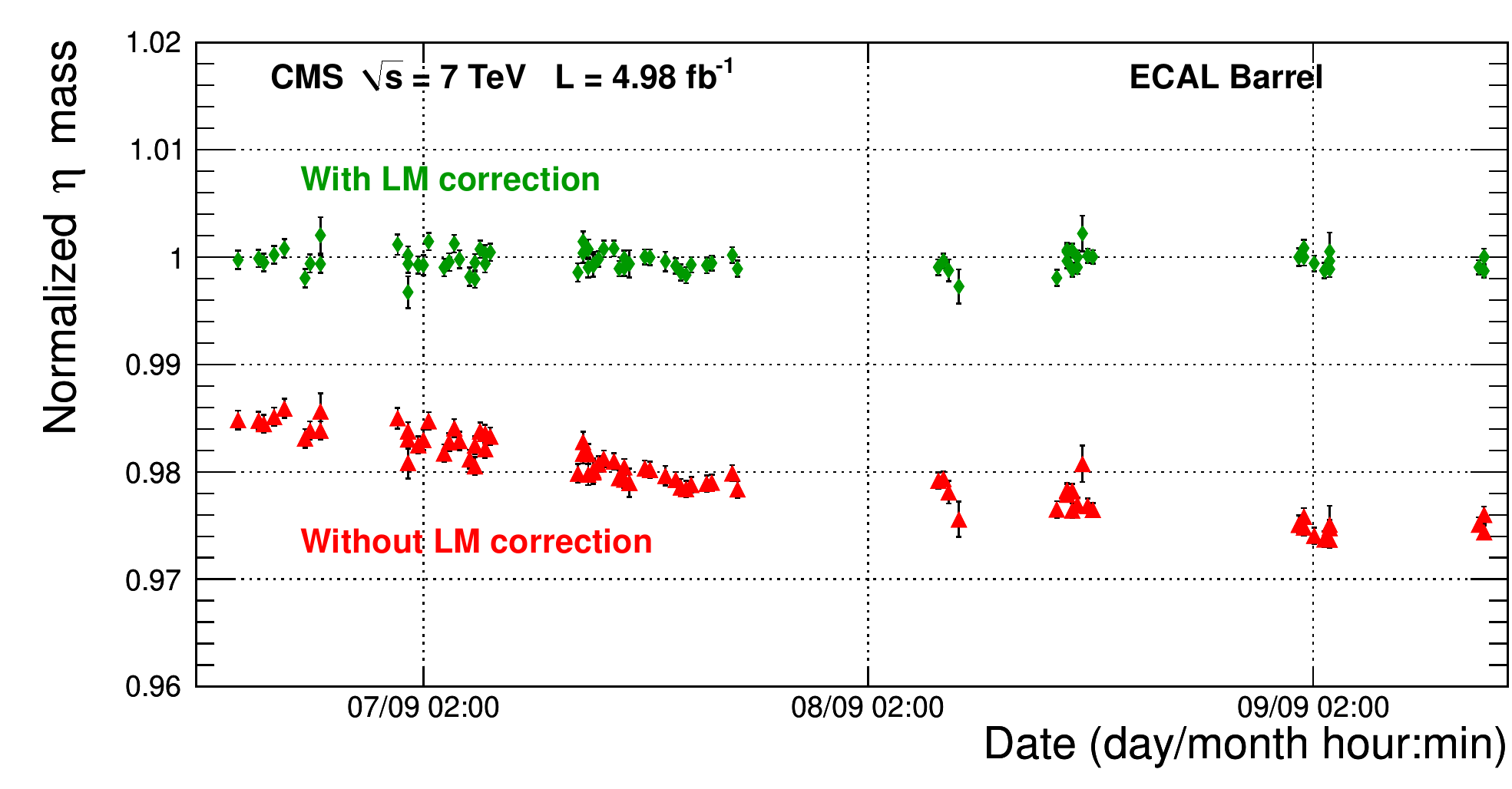}
\end{center}
\caption{\label{fig:etaEB}
Left: An example of the $\Pgh$-meson peak reconstructed from the invariant
mass of photon pairs in EB, with the result of the fit with a Gaussian
distribution (continuous line) and a polynomial function (dotted line);
Right: Stability of the $\Pgh\to\Pgg\Pgg$ mass measurement in EB as a
function of time, over a period of 60 hours, for data recorded in
September 2011. The plot shows the data with (green points) and
without (red points) LM corrections applied.
}
\end{figure}

The $\Pgh$ meson data are used to provide fast feedback, to validate
the LM corrections for prompt data reconstruction. The events are
selected online by a dedicated calibration trigger and recorded with
reduced event content.
A fit is carried out on the invariant mass distribution of the photon
pairs in the mass range of the $\Pgh$ meson. The fit comprises a
polynomial function to describe the background and a Gaussian
distribution to describe the resonance peak. Figure~\ref{fig:etaEB}
shows an example of the $\Pgh$-meson peak with the fit superimposed,
and the relative value of the fitted $\Pgh$ mass versus time in EB for
a period of 60 hours. The plot shows the data before (red points) and
after (green points) the LM corrections applied. A number of
measurements are possible for each LHC fill, owing to the high
rate for recording $\Pgh$ events. This permits short-term changes in
the ECAL response to be verified before prompt data reconstruction
takes place.

\begin{figure}[htb]
\begin{center}
\begin{tabular}{cc}
 \hspace{-0.5cm}
 \includegraphics[width=0.40\linewidth]{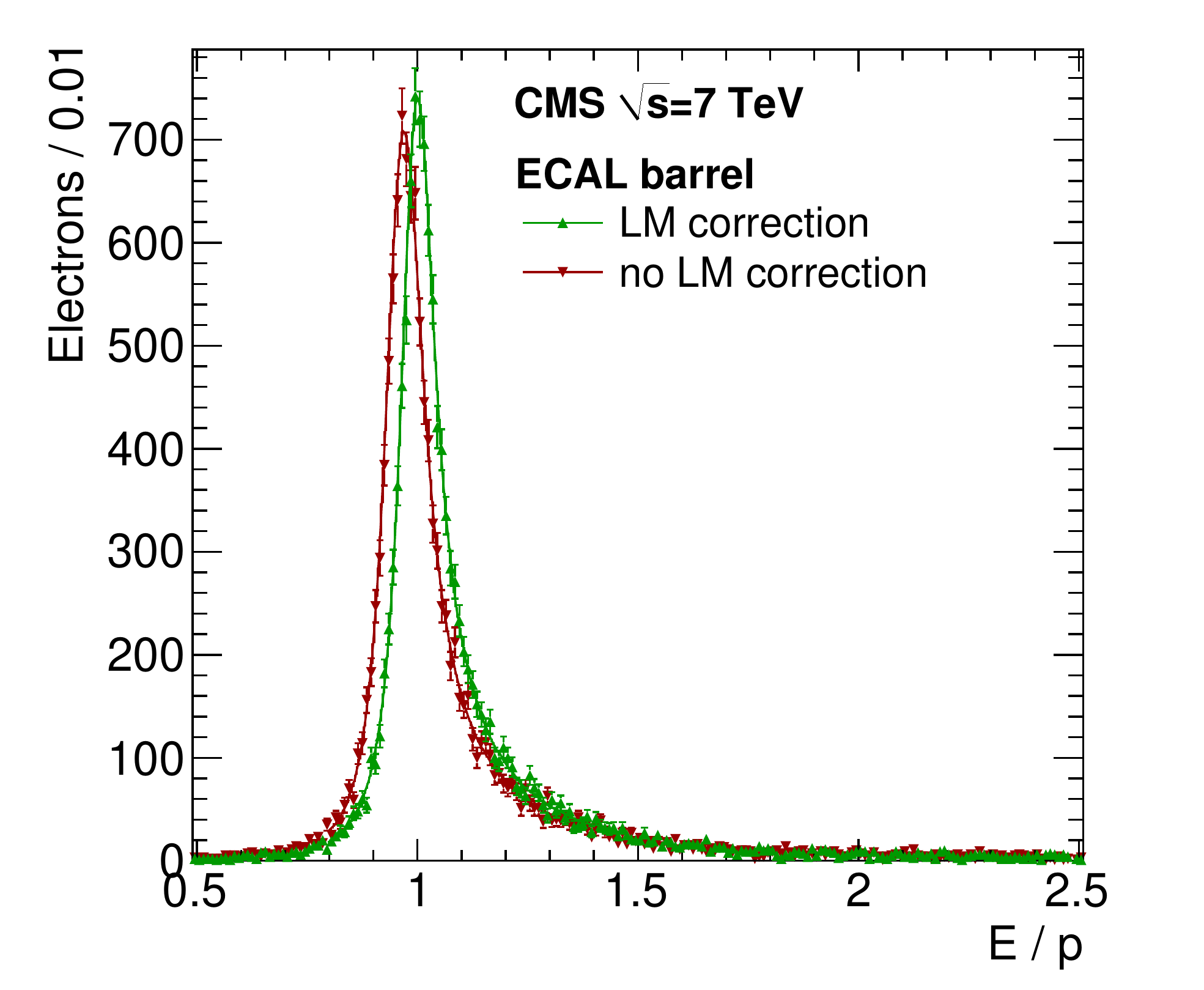} &
 \hspace{-1cm}
 \includegraphics[width=0.69\linewidth]{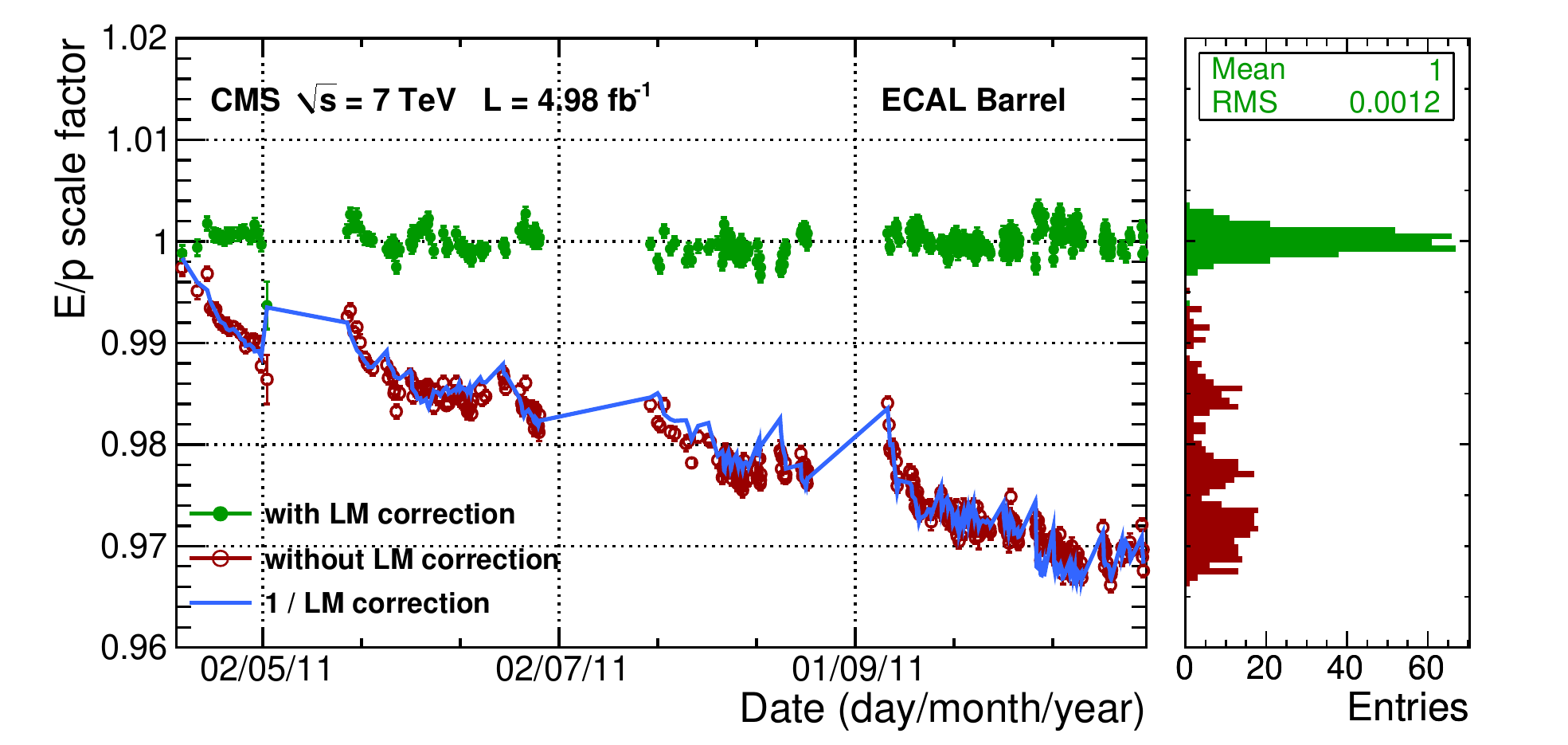} \\
 \hspace{-0.5cm}
 \includegraphics[width=0.40\linewidth]{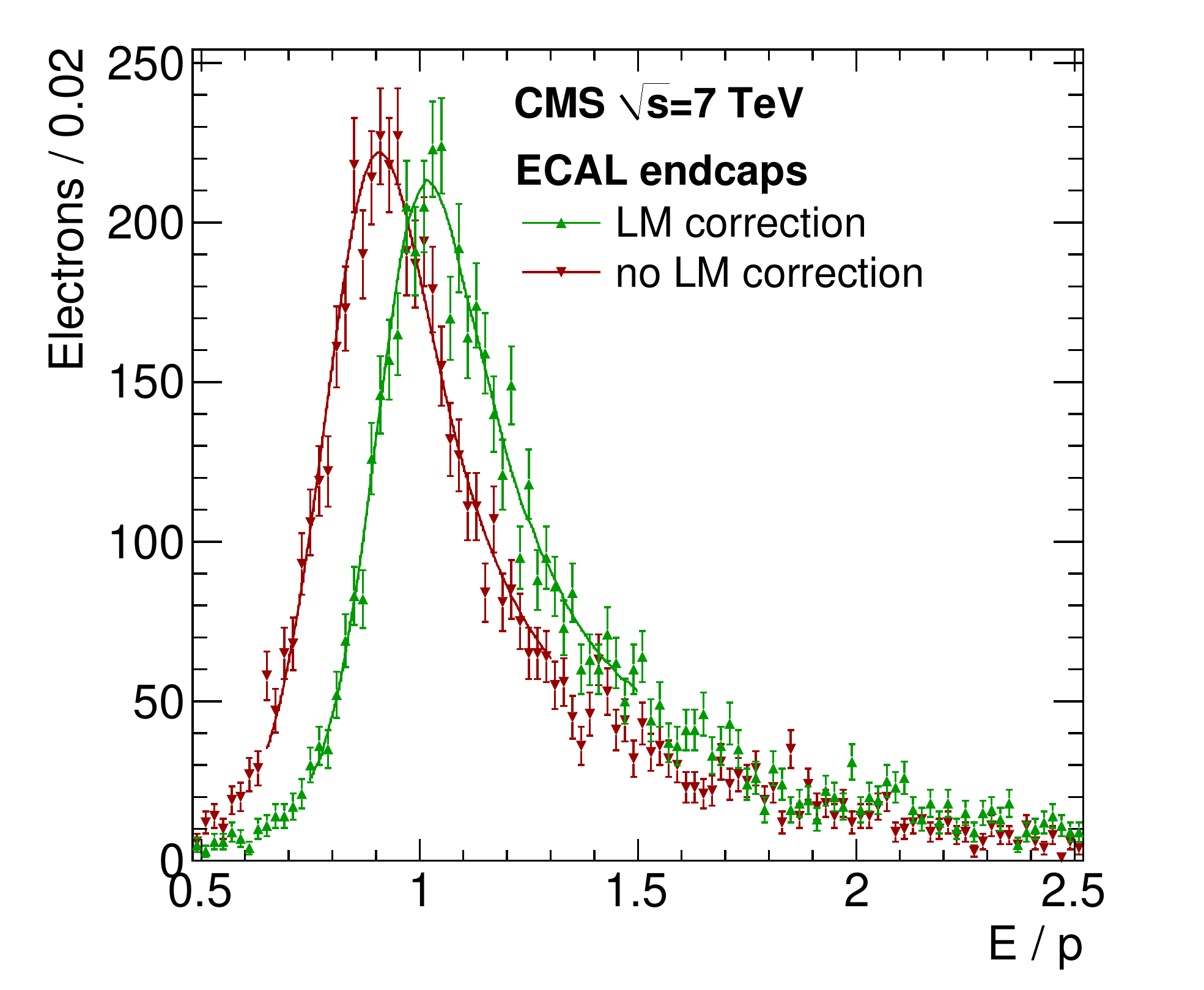} &
 \hspace{-1cm}
 \includegraphics[width=0.69\linewidth]{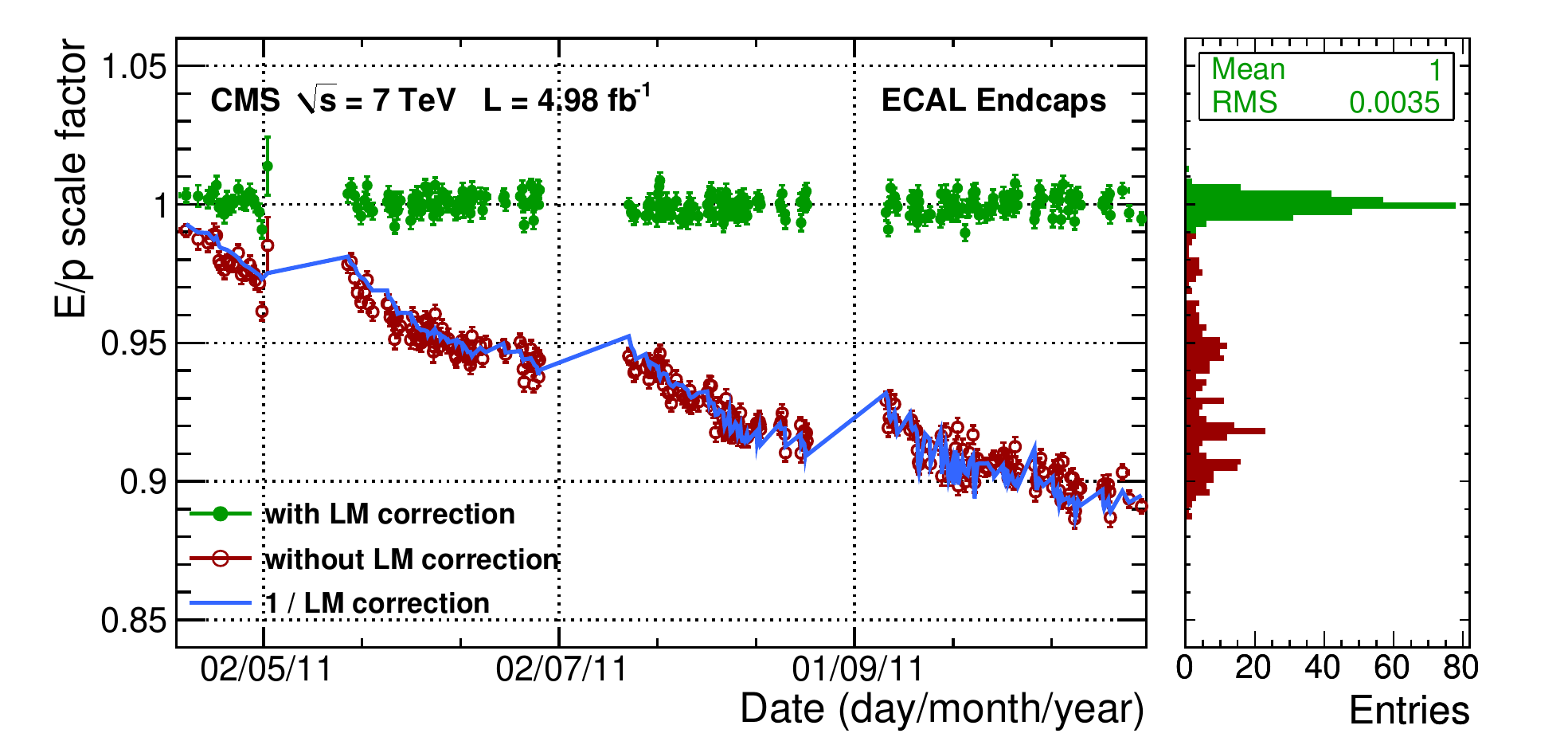}
\end{tabular}
\end{center}
\caption{\label{fig:wenuEB}
Relative energy response variation for EB (top) and EE (bottom)
determined from the $E/p$ analysis of electrons in $\PW$-boson decays.
Left: examples of fits to the $E/p$ distributions before (red) and
after (green) LM corrections. Middle: Response stability during the
2011 $\Pp\Pp$ data-taking period before (red open circles) and after
(green points) response corrections; the blue line shows the inverse
of the average LM corrections. Right: Distribution of the projected
relative energy scales.}
\end{figure}

Isolated electrons from $\PW$-boson decays are used to provide an energy
scale to validate response corrections over periods of days to
weeks. The event selection is described
in~\cite{EGM-10-004,Khachatryan:2010xn}.
The ratio of the electron energy, $E$, measured in the ECAL, to the
electron momentum, $p$, measured in the tracker, is computed in each
event, and a reference $E/p$ distribution is obtained from the entire
data set after applying LM corrections. The width of the $E/p$
reference distribution is dominated by the energy and momentum
resolution and is not biased by residual imperfections in the LM
corrections. This reference distribution is then scaled to fit
$E/p$ distributions obtained by dividing the same data in groups of
12000 consecutive events. The scale factors provide a measure of the
relative response and are shown in Fig.~\ref{fig:wenuEB} for 2011, as
a function of time. The data are shown before (red points) and after
(green points) LM corrections to the ECAL channel response are
applied. The magnitude of the average correction for each point is
indicated by the continuous blue line. A stable response to
electromagnetic showers is achieved throughout 2011 with an RMS of
0.12\% in EB and 0.35\% in EE. This method does not require a
knowledge of the absolute calibration of both the energy and the
momentum.

The response corrections for EE were calculated using an `effective'
$\alpha$ value of 1.16 for all BTCP crystals. This value of $\alpha$
was shown to give the most stable and optimal mass resolution as a
function of time by minimizing the resolution of the invariant mass
for $\cPZ \to \Pep\Pem$ decays, and evaluating the stability of the $E/p$
evolution with time for different values of $\alpha$. The value of the
effective $\alpha$ is smaller than the value measured in beam tests,
of 1.52. This is attributed to the larger crystal transparency losses
in EE and the VPT response losses. Large transparency losses reduce
the difference between the path lengths for injected light and
scintillation light. For the same path length $\alpha$ is expected to
be 1. VPT response losses give rise to a proportional loss of the ECAL
response, and correspond to $\alpha=1$.

\begin{figure}[htb]
\begin{center}
 \includegraphics[width=0.45\linewidth]{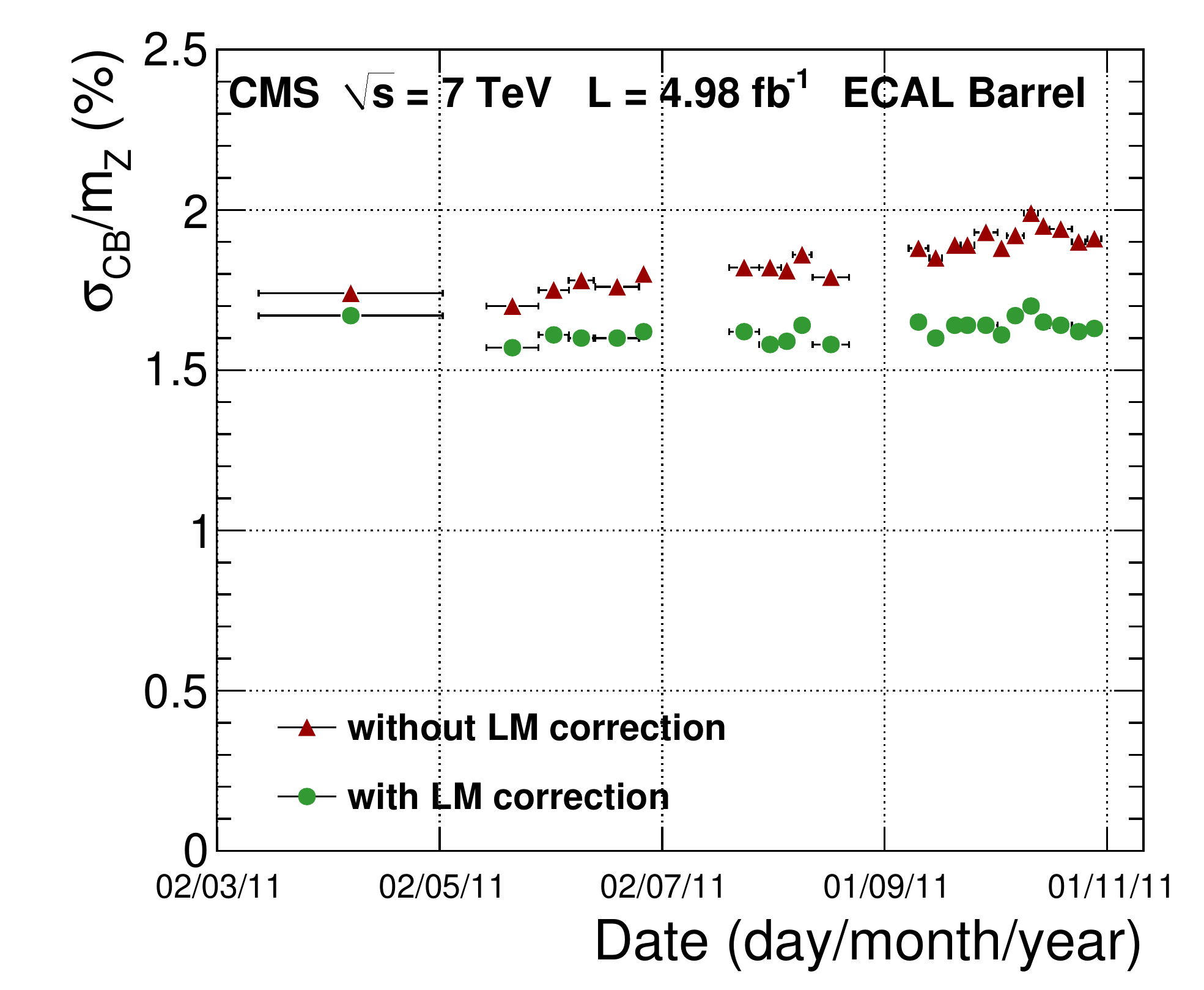}
 \includegraphics[width=0.45\linewidth]{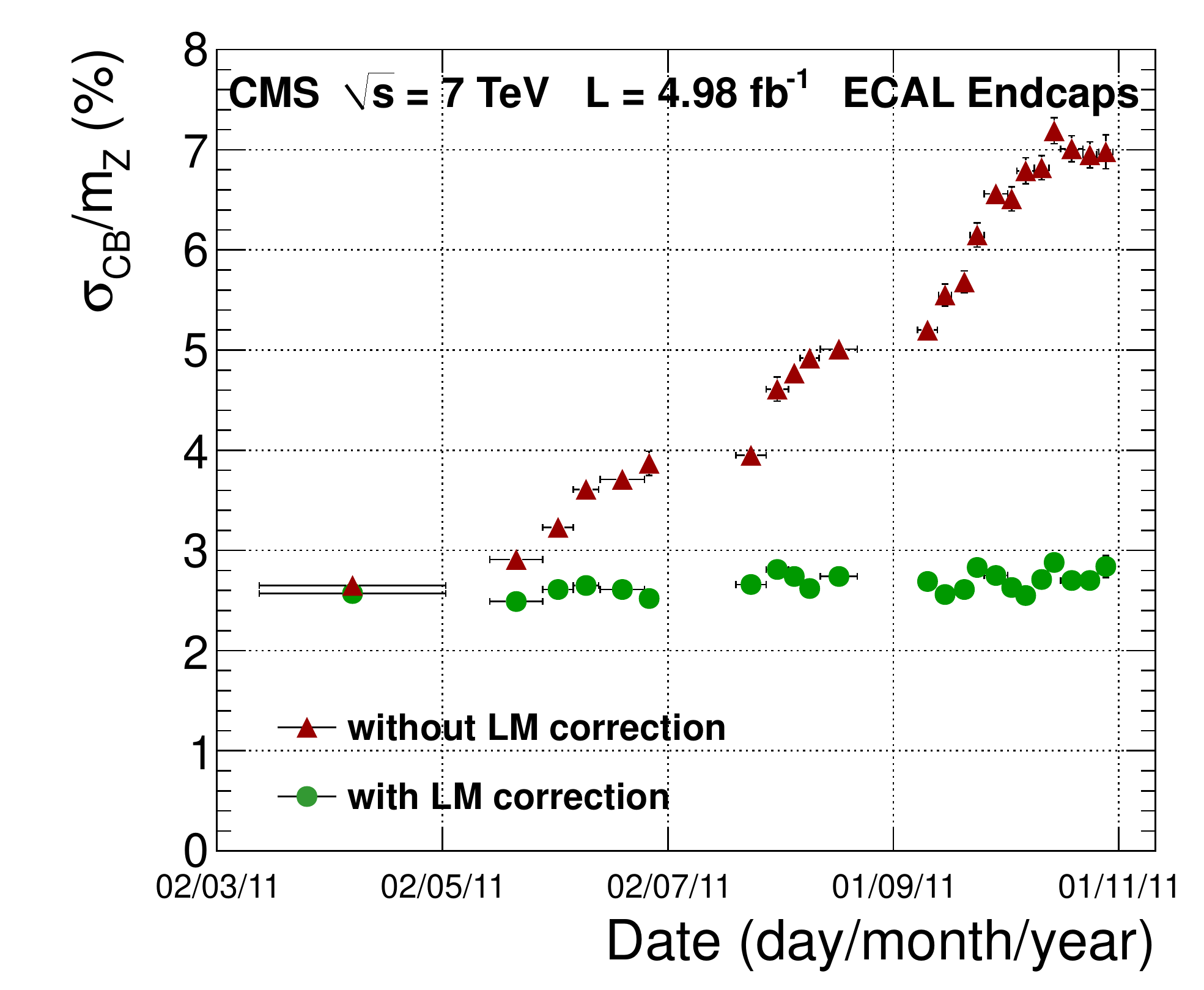}
\end{center}
\caption{\label{fig:Z_historyplot}
Mass resolution for the reconstructed $\cPZ$-boson peak, from $\cPZ \to
\Pep\Pem$ decays, as a function of time for EB (left) and EE (right)
before (red dots) and after (green dots) LM corrections are
applied.}
\end{figure}

The validation of the response corrections was also carried out by
monitoring the ECAL energy resolution during 2011 using events with
a $\cPZ$-boson decaying into two electrons. The selection of these events
is described in~\cite{EGM-10-004,Khachatryan:2010xn}. The invariant
mass was calculated from the energy deposits of the two electrons and
the angle between them using track and vertex information. The mass
resolution is dominated by the energy resolution of the electron
reconstruction.
Figure~\ref{fig:Z_historyplot} shows the contribution to the
instrumental mass resolution for the $\cPZ$-boson peak,
$\sigma_{CB}/M_{\cPZ}$, as a function of time for events with both
electrons in EB (left) or both in EE (right). The fits to the
$\cPZ$-boson peak, based on the Crystal Ball parameterization~\cite{CB}
of the resolution function, and the fit parameters are described in
Section~\ref{sec:zeecalib}. The mass resolution, after the application
of the response corrections, is stable within an RMS spread of 0.1\%
and 0.2\% for events with both electrons in EB or EE,
respectively. The observed spread of the points is consistent with the
uncertainty on the resolution from the fit.

\subsubsection{Response correction summary}

Excellent energy response and resolution stability have been achieved
for 2011 after the application of LM corrections. In EE an effective
value of $\alpha$ has been derived to stabilize and optimize the
invariant mass resolution with  $\cPZ \to \Pep\Pem$decays. The various
cross-checks, using reconstructed masses from particle decays, have
confirmed the validity of the LM corrections.

The contributions to the constant term of the energy resolution due to
the monitoring corrections at the single-crystal level comprise:
\begin{itemize}
\item
The precision of an individual LM correction measurement, which is better than
0.1\%, and the long-term instability of a single channel, which is
$<0$.2\% (Section~\ref{sec:status}).
\item
The 10\% spread in $\alpha$, from channel-to-channel, translates to a
contribution to the resolution of 0.3\% for EB by the end of 2011.
\item
In EE, the introduction of an effective $\alpha$ compensates for the
average VPT response loss, which is not separated from the
contribution due to crystal transparency change. Both the
channel-to-channel variation of the VPT loss and the channel-to-channel
difference in the value of $\alpha$ contribute to the single-channel
uncertainty on the value of the effective $\alpha$, which is estimated
to be approximately 10\%. Given the impact of the high LHC radiation
levels on the EE response, this uncertainty translates into a
contribution to the energy resolution of about 1.5\% on average, and
ranging from about 0.5\% at $\abs{\eta} \approx 1.6$ to about 2.5\% at
$\abs{\eta}\approx 2.5$ by the end of 2011.
\end{itemize}
In addition to the effects listed above, the residual instabilities of
0.12\% in EB and 0.35\% in EE in the mean-energy response observed
during 2011 (see Fig.~\ref{fig:wenuEB}) also contribute to the
constant term of the energy resolution.

\subsection{Single-channel intercalibration, \texorpdfstring{$C_i$}{Ci}}
\label{sec:inter}

The ECAL channels are calibrated by using relative and absolute
calibrations. Relative calibrations, $C_i$, between one channel and
another, are referred to as intercalibrations and are described in
this section. Absolute calibrations are obtained by referring the
intercalibrations to a mass scale by using $\cPZ$-boson decays, as
described Section~\ref{sec:scalecalib}. The intercalibration constants
in EB and EE are divided by their average value, to provide a set of
numbers with a mean value of unity. A number of methods are used for
intercalibration and are then combined to provide a weighted mean
intercalibration constant for each channel.

An initial set of calibrations, known as the `pre-calibration', were
obtained from laboratory measurements, beam tests, and from exposure to
cosmic rays. The laboratory measurements included the crystal light
yield and photodetector gain. Nine out of 36 EB supermodules and
about 500~EE crystals were intercalibrated with high-energy electrons
in beam tests. All channels in the EB supermodules were calibrated with
cosmic-ray muons ~\cite{Adzic:2008zza}. After installation at
the LHC, the ``beam splash'' events were used to improve further the
EB and EE calibrations~\cite{Chatrchyan:2009qm}. The intercalibration
constants from each method were cross-checked for consistency
and combined to provide a weighted average for the channel. The
precision of the intercalibration for each channel at the start of
7\TeV operation in 2010 is estimated to be:
\begin{itemize}
\item
EB: about 0.5\% for the nine supermodules calibrated in beam tests and
1.4\% to 1.8\%, depending on pseudorapidity, for the other 27
supermodules;
\item
EE: below 1\% for the $\approx $500 crystals calibrated in beam tests and
about 5\% for all other channels;
\item
ES: about 2.5\% in all silicon modules from the calibration with
cosmic rays prior to installation.
\end{itemize}

Intercalibration with collision data involves the following
methods~\cite{EGM-10-003}:
\begin{itemize}
\item The $\phi$-symmetry method is based on the expectation
  that, for a large sample of minimum bias events, the total deposited
  transverse energy should be the same in all crystals at the
  same pseudorapidity. In CMS this corresponds to crystals located in
  a particular $\eta$ ring. The method provides a fast
  intercalibration of crystals located within the same ring.
\item The $\Pgpz$ and $\Pgh$ calibrations use the invariant mass of
  photon pairs from these mesons to intercalibrate the channel
  response.
\item Intercalibrations with isolated electrons from $\PW$- and $\cPZ$-boson
  decays are based on the comparison of the energy measured in ECAL to
  the track momentum measured in the silicon tracker.
\end{itemize}
All these methods are used to intercalibrate channels at the same
pseudorapidity. Isolated electrons are also exploited to derive the
relative response of the various $\eta$ rings.

The precision of the intercalibrations quoted in the following
sections has been studied for each method with the aid of MC
simulations, and validated using the pre-calibration data and by a
channel-by-channel comparison of the intercalibrations derived with
each method.

\subsubsection{Intercalibration using the $\phi$-symmetry method}
\label{sec:physim}

The intercalibration in $\phi$ is taken from the ratio of the total
transverse energy deposited in one crystal to the mean of the total
transverse energy collected by all crystals at the same value
of~$\eta$~\cite{EGM-10-003}. Events used for this calibration are
acquired with a special minimum bias trigger. All single-crystal
energy deposits above 150\MeV in EB, and above 650\MeV in EE are
recorded, while the rest of the event is dropped to limit the trigger
bandwidth required.

The data analysis is restricted to deposits with transverse energies
between a lower and an upper threshold. The lower threshold is applied
to remove the noise contribution and is derived by studying the noise
spectrum in randomly triggered events. It is set to about six times
the channel RMS noise (\eg, 250\MeV for channels in EB). The upper
threshold is applied to minimize the fluctuations induced by rare
deposits of very high $\ET$ and is set to 1\GeV above the lower
threshold, in both EB and EE.
Because the transverse energy scalar sum is obtained from a truncated
$\ET$ distribution, a given fractional change in the $\ET$ sum does
not correspond to the same fractional change in the value of the
intercalibration constant. This is accounted for with an empirical
correction~\cite{FutyanSeez2002}. Corrections are also applied to
compensate for known azimuthal inhomogeneities of the CMS detector,
related to the intermodule gaps in the ECAL and to the tracker support
system.

\begin{figure}[hbtp]
\begin{center}
\includegraphics[width=0.49\linewidth]{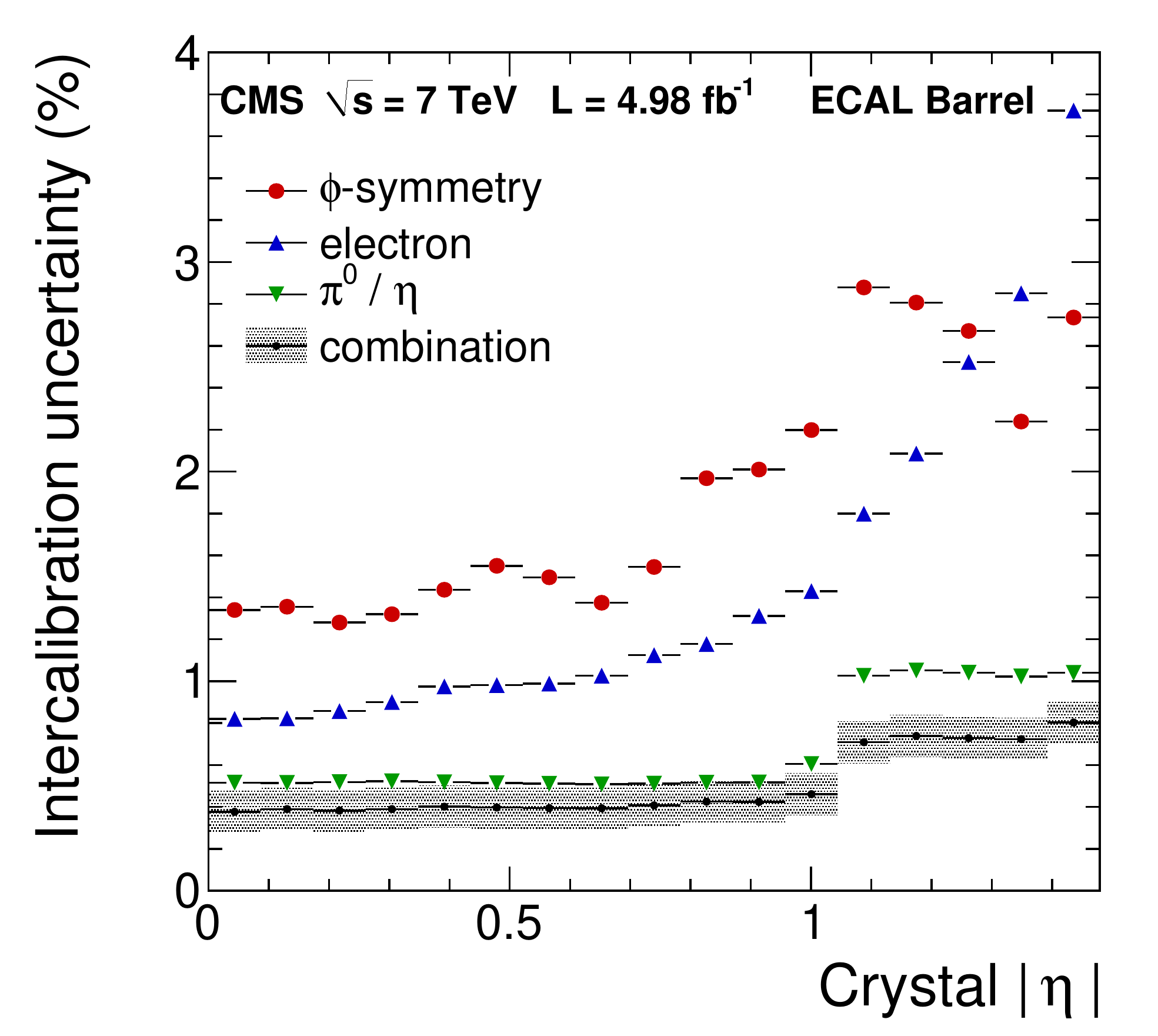}
\includegraphics[width=0.49\linewidth]{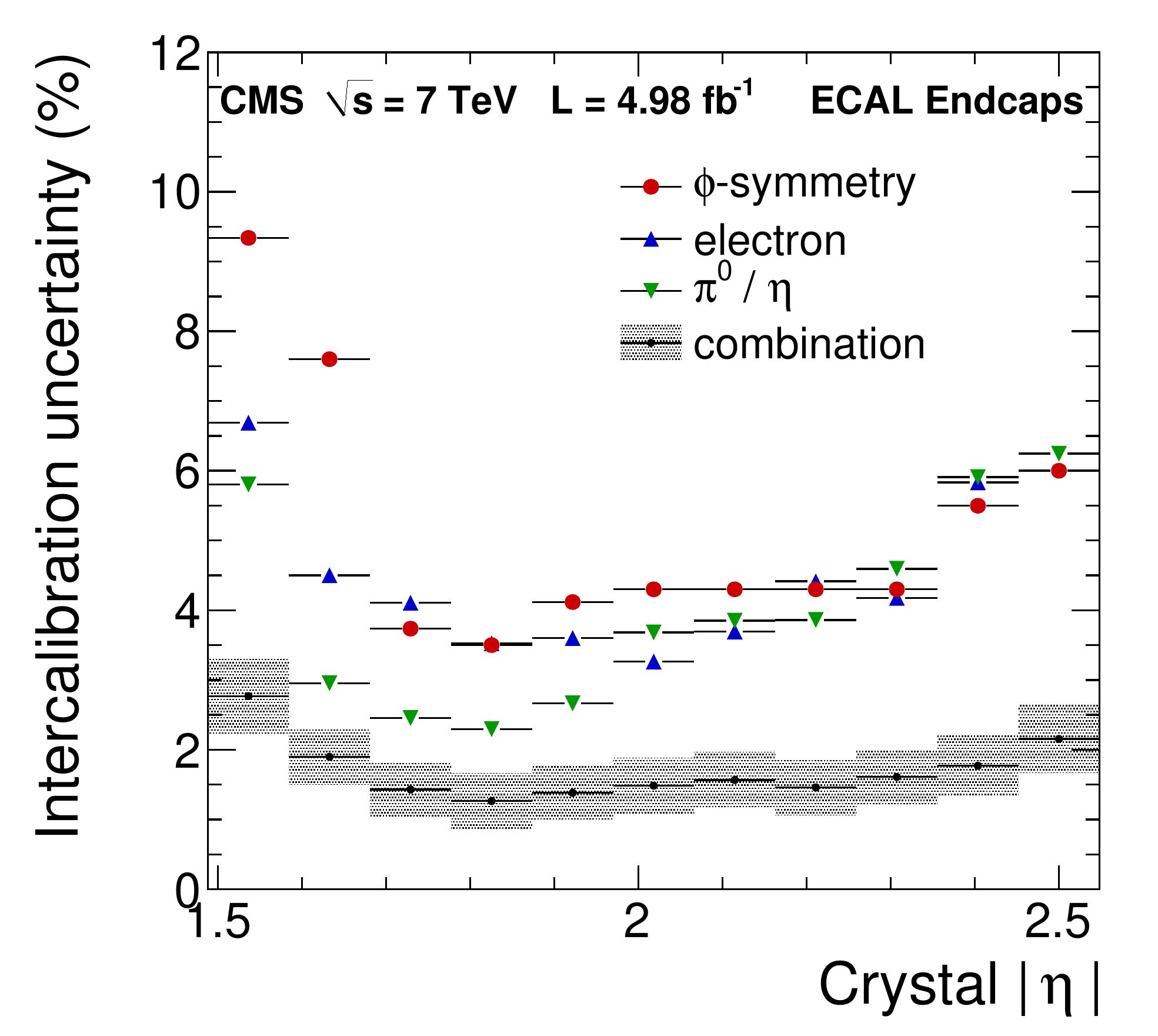}
\caption{The intercalibration precision obtained in 2011 using $\phi$
  symmetry, the $E/p$ ratio with electrons, $\Pgpz/\Pgh$ decays, and
  the resultant precision, with its uncertainty, for the combination
  of the methods, in EB (left) and EE (right).
\label{fig:comb_2011}}
\end{center}
\end{figure}

Figure~\ref{fig:comb_2011} shows the estimated precision (red circles)
for the $\phi$-symmetry intercalibration as a function of $\abs{\eta}$
for EB and EE in 2011. For a typical sample of about $10^8$ events,
the precision of the method is limited by a systematic uncertainty of
1.5\% in the central part of EB, growing to above 3\% at larger
$\abs{\eta}$, due to residual effects of the azimuthal inhomogeneities of
the material in front of ECAL. These are larger in the region where
the material budget is largest (see Fig.~\ref{fig:clustercorr}).
By using the ratio of $\phi$-symmetry intercalibrations over periods
of about one week, the systematic uncertainties from the
inhomogeneities largely cancel, and a relative precision between
successive periods of 0.3\% is achieved. This method is used to
monitor the stability of the intercalibration constants or to improve
the intercalibration constants obtained with other analyses.

\subsubsection{Intercalibration using $\Pgpz \to \Pgg\Pgg$ and $\Pgh \to \Pgg\Pgg$ decays}
\label{sec:pi0}

The decays of $\Pgpz$ and $\Pgh$ mesons to two photons are exploited
to intercalibrate the ECAL crystals using the peak of the $\Pgg\Pgg$
invariant mass distribution~\cite{EGM-10-003}. A special data stream
is used to profit from the copious production of $\Pgpz$ and $\Pgh$
mesons at the LHC. Candidate diphoton decays are directly selected
online from events passing the single-$\Pe/\Pgg$ and single-jet L1
triggers. After selection, only limited data, in the vicinity of the
photon candidates, are kept in order to collect $\Pgpz$ and $\Pgh$
meson candidates at a rate of the order of 10\unit{kHz} with minimal impact
on the CMS readout bandwidth and storage space.

The individual photon energy is obtained from the sum of energy in a
3$\times$3 matrix of crystals centred on the crystal with the highest
energy deposit (seed). The seed is required to have an energy greater
than 0.5\GeV. The single-crystal energy deposits are small and
corrections are applied to these deposits to account for the effects
of the noise suppression algorithm used in the readout~\cite{CMS_TDR_v1}.

For the $\Pgpz$ sample, the photons are required to have transverse
energies above 0.8\GeV in EB and 0.5\GeV in EE. The transverse energy
of the $\Pgpz$ candidate is required to be above 1.6\GeV in EB and
2.0\GeV in EE. For the $\Pgh$ sample, the photons are required to have
transverse energies above 0.8\GeV in EB and 1.0\GeV in EE. The
transverse energy of the $\Pgh$ candidate is required to be above
3.0\GeV in both EB and EE. Moreover, to suppress photons converted in
the material in front of the ECAL, the transverse shape of the energy
deposition is required to be consistent with that of an
electromagnetic shower produced by a photon and be isolated from other
ECAL energy deposits~\cite{EGM-10-003}. This calibration method is
only indirectly affected by tracker material, through an efficiency
loss and a worsening of the signal to background ratio in the detector
regions where the material thickness is large.

An iterative procedure is used to determine the intercalibration
constants. For each crystal, the invariant mass distribution is
obtained from all $\Pgpz/\Pgh$ candidates for which one of the photons
is centred on the crystal. The distribution is fitted with a Gaussian
function, for the signal, and a fourth-order polynomial for the
background. The intercalibration constants are updated iteratively to
correct the fitted mass value in each channel. The quality of the
calibration depends on the number of selected candidates per crystal
and on the signal-to-background ratio. The results from each resonance
are combined to form an average weighted by precision.

The precision of the intercalibration constants in 2010 was estimated
by comparing the $\Pgpz/\Pgh$ intercalibrations to those derived from
the pre-calibration and it was found to be at the systematic limit of
the methods employed.
Figure~\ref{fig:comb_2011} shows the estimated precision of
intercalibration constants in EB (left) and EE (right), in 2011, as a
function of pseudorapidity using the $\Pgpz/\Pgh$ method. The large
2011 data set provides intercalibration constants with a precision of
0.5--1\% each month in the EB, with a pattern along $\eta$ related to the
distribution of the tracker material. A precision of 2--4\% is
achieved every 2--3 months in the EE.

\subsubsection{Intercalibration using electrons from $\PW$- and $\cPZ$-boson decays}
\label{sec:wenu}

The ratio of the supercluster energy, $E$, of an electron measured by
ECAL to the momentum, $p$, measured by the tracker, is used to provide
$E/p$ intercalibrations in $\phi$ and along $\eta$. Isolated electrons
were selected from $\PW$-boson and $\cPZ$-boson decays, as described in
Section~\ref{sec:wenu_laser}. The data comprise 7.5 million isolated
electrons collected during 2011, corresponding approximately to each
crystal being struck by 100 electrons. The estimated background due to
misidentified jets is below 1\%.

The intercalibration constants in $\phi$ were calculated using an
iterative procedure that derives constants for all the channels
\cite{Chaturvedi:2000wj,Agostino:2006bb}. Once convergence was reached, the
constants were normalized to have a mean value equal to unity in each
$\phi$ ring at each position in $\eta$.
In each $\phi$ ring, corrections were applied to take into account the
effect of the supermodule boundaries in $\phi$ and $\phi$-dependent
variations of
the tracker momentum response. Variations in the amount of material in
different regions of $\phi$ and $\eta$ affect the electron momentum
measurement due to bremsstrahlung losses. The relative momentum
response was calibrated using electrons from $\cPZ$-boson decays using
2011 data. The invariant mass was reconstructed in 360 $\phi$-bins by
using the tracker momentum for the electron entering a specific
$\phi$-bin, and the ECAL energy for the other electron.
The square of the invariant-mass peak position in each $\phi$-bin is
proportional to the local momentum scale for the corresponding region of
the detector, because the mean contribution from the other electron
and from the angle is independent of the $\phi$-bin.
Correction factors are between $\pm$1.5\% in EB and $\pm$3\% in
EE. The resulting uncertainty on the relative momentum scale is 0.48\%
for EB and 1.4\% for EE. These uncertainties, added in quadrature to
the statistical uncertainty of the method, contribute less than about
10\% of the total uncertainty on the intercalibration constants
achieved with this method in 2011.

Figure~\ref{fig:comb_2011} shows the precision of the intercalibration
for EB and EE. The precision in the central barrel, for $\abs{\eta}<1.0$,
is 0.8--1.4\% and reaches 4\% at $\abs{\eta}=1.48$. The precision in
EE is better than 4\%, apart from the outer regions, which are
calibrated to $\approx 6$\%. The variation of the precision is due to
changes with $\eta$ of the $E/p$ resolution, and of the tracker
material budget, which impacts on the mean number of crystals per
supercluster. In contrast to the other methods, this intercalibration
method was still limited by the statistical precision in 2011.

\begin{figure}[hbtp]
\begin{center}
\includegraphics[width=0.88\linewidth]{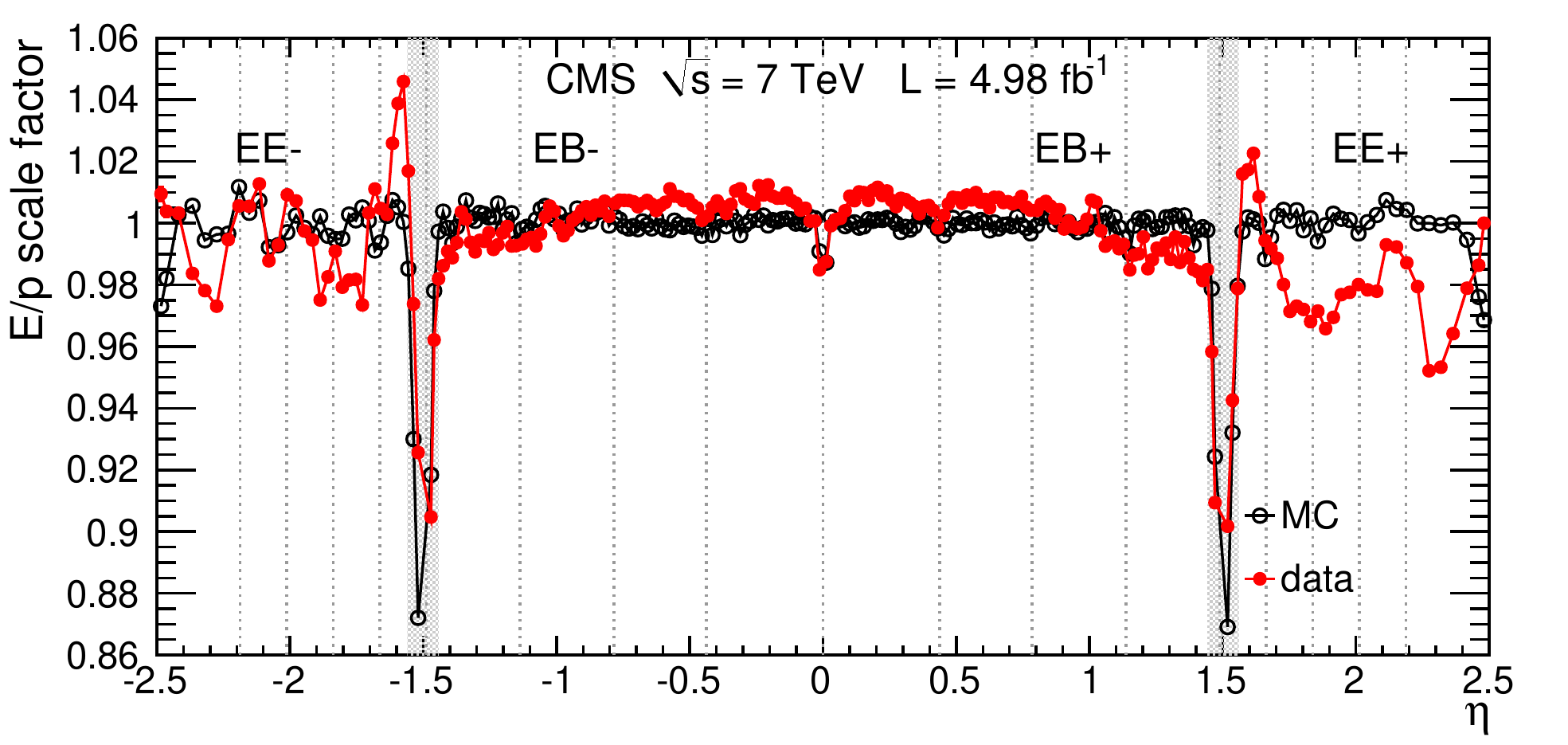}
\caption{Energy scale factors from the $E/p$ analysis of isolated
  electrons as a function of $\eta$ for data and MC simulation. The
  shaded regions, corresponding to the EB/EE interface regions, are
  usually excluded from the acceptance for physics analyses. The
  relative differences between MC simulation and data, as a function
  of $\eta$, are used to derive a set of relative energy scale
  calibrations to be applied to data.
\label{fig:E_over_p_feta}}
\end{center}
\end{figure}

Electrons from $\PW$- and $\cPZ$-boson decays are also used for the
relative calibration between the rings along $\eta$. An $E/p$
reference distribution obtained from the MC simulation is scaled to
fit the $E/p$ distributions in data from crystals in the same $\phi$
ring. Since the shape of the $E/p$ distribution varies along $\eta$,
MC reference distributions are made for four $\abs{\eta}$ regions in EB,
corresponding to EB modules, and for five $\abs{\eta}$ regions in
EE. For each $\phi$ ring a specific calibration of the local momentum
scale for electrons was derived from $\cPZ\to \Pep\Pem$ events, with the
method described above. Corrections to the supercluster energy,
described in Section~\ref{sec:energycorrections}, were also applied.
The scale factors extracted from the fit for each ring of crystals
along $\eta$ are shown in Fig.~\ref{fig:E_over_p_feta} for MC
simulation and data, as a function of electron pseudorapidity. The
shaded regions between EB and EE are usually excluded from the
acceptance of electrons and photons for physics analyses. The $E/p$
scale factors provide a measure of the relative response to electrons
along $\eta$. In MC simulation, they are consistent with unity, which
shows the self-consistency of the method for MC events. Results from
data have been used to scale the intercalibration constants in each
ring, although the observed $\eta$ dependence of the response in EB
and EE might indicate the need for further tuning of the energy
corrections in data. Deviations from unity for data in EE can be also
partly ascribed to the lower precision of pre-calibrations in the
endcaps.

\subsubsection{Combination of the intercalibration constants}

The precision of the combined intercalibration set is shown in
Fig.~\ref{fig:comb_2011}. The combination was obtained from a mean of
the intercalibration constants in fixed $\phi$ rings from the
$\Pgpz / \Pgh$, the $E/p$, and the $\phi$-symmetry methods, weighted
on the respective precisions. The intercalibration set established in
2010 was also included in the combination. The combined
intercalibration precision is 0.4\% for central EB crystals ($\abs{\eta}
<1$), and is 0.7--0.8\% for the rest of the EB up to $\abs{\eta} =
1.48$. In EE the precision is 1.5\% for $1.6<\abs{\eta}<2.3$ and better
than 2\% up to the limit of the electron and photon acceptance at
$\abs{\eta}=2.5$. The variation of the precision with pseudorapidity
arises partly from the size of the data sample, and partly from the
amount of material in front of the ECAL.

The precision of each intercalibration set used in the combination
has been derived by means of MC simulation studies. They were
validated at low instantaneous luminosity, prior to transparency
changes in ECAL, by measuring the spread of the in-situ constants with
respect to those derived at beam tests. In addition, the precision was
estimated from the cross-comparison of the results of the different
intercalibration techniques. In each $\phi$ ring, the variance of the
difference between the intercalibration constants for every pair of
intercalibration sets (\ie, $C_i(j)-C_i(k)$, where $i$ is a channel
index and $j$ and $k$ indicate the intercalibration method) was
derived. This variance was assumed to be the sum in quadrature of the
uncertainty of the constants in each set. Consequently, the precision
of each intercalibration set was extracted by solving three simultaneous
equations for the three variances. The values obtained with this
method were found to be consistent with the expected precisions based
on the simulation studies. The difference between the two estimates
has been used to derive the uncertainty on the precision of the
combined intercalibration set, shown by the grey band in
Fig.~\ref{fig:comb_2011}.

\subsubsection{Summary of the intercalibration precision}
The supercluster energy is determined from the energy deposited over
several crystals. As a consequence, the
contribution to the constant term of the energy resolution due to the
response spread of the individual channels is smaller than the spread
itself. Simulation studies show that the scale factor between the
uncertainty in the intercalibration and the constant term is
approximately 0.7, corresponding  to the average level of energy
containment in the central crystal of the supercluster. From the
results shown in Fig.~\ref{fig:comb_2011}, the contribution to the
constant term, due to the intercalibration precision, is about 0.3\%
in the central part of EB ($\abs{\eta} <1.0$) and 0.5\% for $1.0 < \abs{\eta}
< 1.48$. In EE the contribution is about 1.0\% for $1.6< \abs{\eta} <2.3$
and better than 1.5\% up to the limit of electron and photon
acceptance at $\abs{\eta} = 2.5$.

\begin{figure}[hbtp]
\begin{center}
\includegraphics[width=0.48\linewidth]{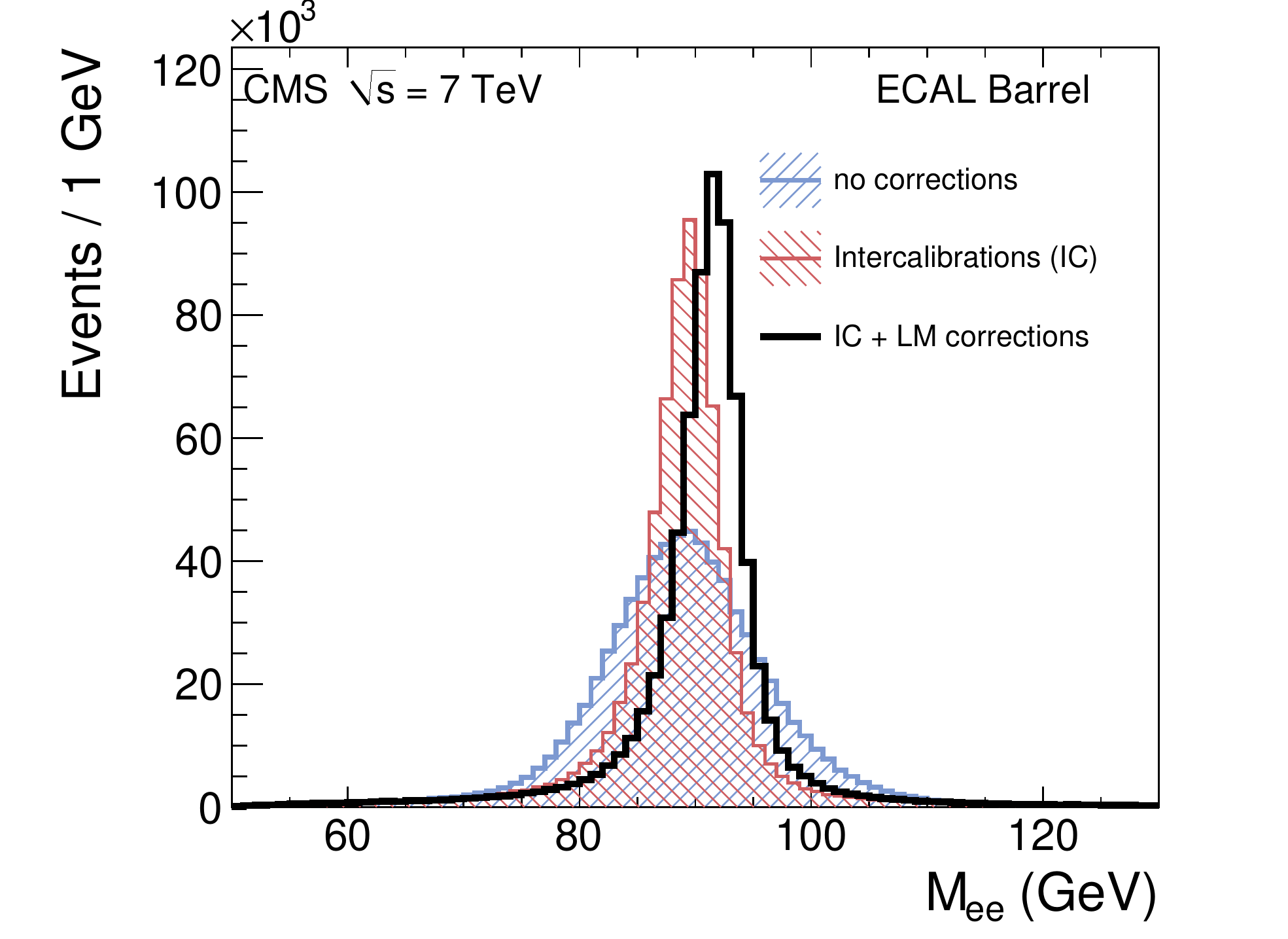}
\includegraphics[width=0.48\linewidth]{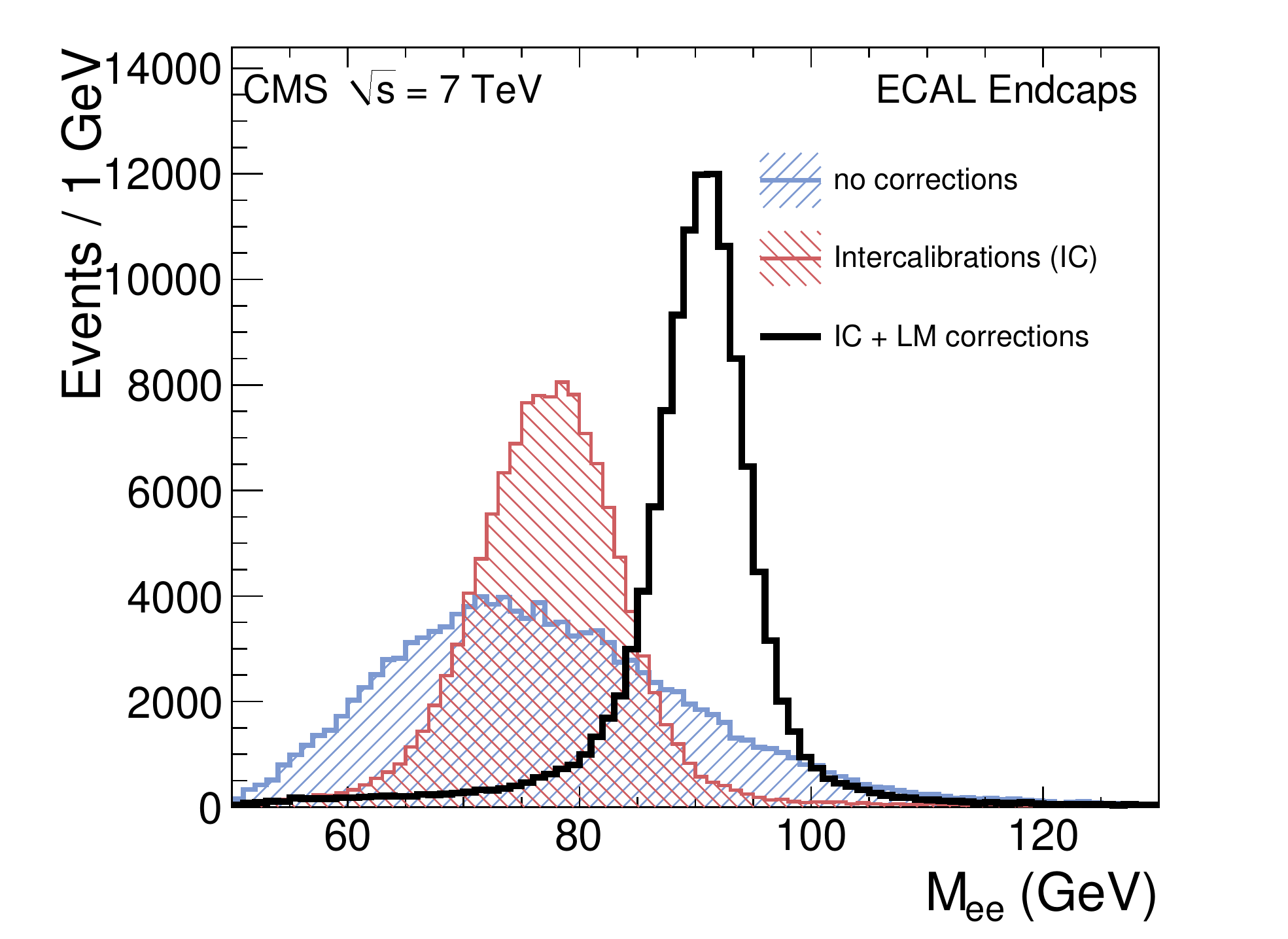}
\caption{Reconstructed invariant mass from $\cPZ\to \Pep\Pem$ decays, for
  single-channel corrections set to unity (blue), for final
  intercalibration (red), and for both final intercalibration and LM
  corrections (black), in the EB (left) and the EE (right).
\label{fig:zee_propaganda}}
\end{center}
\end{figure}

To illustrate the relative importance of the individual calibrations
and corrections, Fig.~\ref{fig:zee_propaganda} shows the dielectron
invariant mass distributions for various reconstruction scenarios: for
single-channel corrections set to unity (blue), for the final
intercalibrations (red), and for the final intercalibrations plus the
monitoring corrections (black) in the EB (left) and the EE (right).
In all the cases, supercluster-level corrections, $F_{\Pe/\Pgg}$ in
Eq.~(\ref{eq:one}) (see Section~\ref{sec:energycorrections}), were
included in the energy computation.

\subsection{Calibration of the preshower}
\label{sec:es}

The precision required for the calibration of the preshower is largely
determined by the fraction of energy deposited in the ES with respect
to that in the EE. Approximately 6--8\% of the shower energy (decreasing
with $\Pe/\Pgg$ energy) is deposited in the ES. As a consequence,
to limit the contribution to the combined EE+ES energy resolution to
0.3--0.4\%, therequired channel-to-channel calibration precision is
5\%.

Prior to installation, all the ES sensors were calibrated with
cosmic rays to an accuracy of 2.5\%. In situ, the ES sensors are
calibrated using charged pions and muons with momentum greater
than 1\GeV. These particles are close to being minimum ionizing
particles (MIPs), with an average momentum of about 6\GeV, and have a
signal-to-noise ratio greater than 10. The pulse height distribution
for each channel is fitted to a Landau distribution convolved with a
Gaussian function. The fitted peak position is taken as the
calibration. There is a good correlation between the cosmic ray and
in-situ calibrations. The precision of in-situ calibrations is ~2.2\%.

Preshower clusters are identified from the position of crystal
clusters in the EE. The energies in each ES plane are weighted, and
the total ES energy is given by:
\begin{equation}
\label{eq:esenergy}
E_{\mathrm{ES}}=G_{\mathrm{ES}} ( E_{\mathrm{ES}}^{\mathrm{clus1}}+\alpha_{\mathrm{ES}}\cdot E_{\mathrm{ES}}^{\mathrm{clus2}} )
\end{equation}
where $E_{\mathrm{ES}}^{\mathrm{clus1}}$ and
$E_{\mathrm{ES}}^{\mathrm{clus2}}$ are the energies in each
preshower plane, expressed in MIPs, and $G_{\mathrm{ES}}$ is a
coefficient in \GeVns{}/MIP. The coefficient $\alpha_{\mathrm{ES}}$ defines
the relative weight of the second ES plane with respect to the first.

Beam test results showed that the optimal energy resolution of ECAL is
achieved for $\alpha_{\mathrm{ES}}$ ranging between 0.6 and 0.8. The
coefficient $\alpha_{\mathrm{ES}}$ has been fixed to 0.7
~\cite{ES_testbeam}. The parameter $G_{\mathrm{ES}}$ was extracted
from a straight line fit to the EE cluster energy versus the
associated ES cluster energy using electrons from $\PW$-boson
decays~\cite{EGM-10-003}. The measured value of $G_{\mathrm{ES}}$ is
$0.023\pm 0.002\stat \pm 0.001\syst$\GeV{}/MIP.
The systematic uncertainty was calculated assuming an uncertainty of
4\% on the EE shower energy.

\subsection{Energy corrections, \texorpdfstring{$F_{\Pe,\Pgg}$}{F}}
\label{sec:energycorrections}

Superclusters are used to reconstruct the energies of photons and
electrons, and to form seeds for electron track reconstruction. A
correction function, $F_{\Pe,\Pgg}$, derived from MC simulation, is
applied to the supercluster energy to account for energy containment
effects, including both shower containment in the calorimeter, and
energy containment in the supercluster for particles that shower in
the material in front of ECAL.
The energy corrections have been tuned for electrons and photons
separately to account for the differences in the way they interact
with the material in front of the ECAL.

In this analysis, corrections for photons have been optimized using a
multivariate regression technique based on a Boosted Decision Tree
(BDT) implementation. The regression has been trained on prompt
photons (from $\Pgg$+jets MC samples) using the ratio of generator
level photon energy to the supercluster energy, including the energy
in the preshower for the endcaps, as the target variable. The input
variables are the $\eta$ and $\phi$ coordinates of the supercluster in
CMS, a collection of shower shape variables, and a set of local
cluster coordinates to measure the distance of the clusters from ECAL
boundaries. The local coordinates provide information on the amount of
energy which is likely to be lost in crystal and module gaps and
cracks, and drive the level of local containment corrections predicted
by the regression. The other variables provide information on the
likelihood and location of a photon conversion and the degree of
showering in the material. They are correlated with the global $\eta$
and $\phi$ position of the supercluster. These variables drive the
degree of global containment correction predicted by the regression.
The global and local containment corrections address different
effects. However, these corrections are allowed to be correlated in
the regression to account for the fact that a photon converted before
reaching ECAL is not incident at a single point on the calorimeter
face, and is therefore relatively less affected by local containment.
This approach leads to better energy resolution than factorized
parametric corrections of the different effects. The number of primary
vertices is also included as input to the BDT in order to correct for
the dependence of the shower energy on spurious energy deposits due to
pileup events.

The primary validation tool for the regression is to compare data and
MC simulation performance for electrons in $\cPZ$- and $\PW$-boson
decays. A BDT with identical training settings and input variables to
those described above has been trained on a MC sample of electrons
from $\cPZ$-boson decays. The consequent corrections are different from
the ones used for the electron reconstruction in CMS, where tracker
information is included in the energy measurement. However, they
enabled a direct comparison of the ECAL calibration and resolution in
data and MC simulation to be performed, as we discuss in
Sections~\ref{sec:scalecalib} and~\ref{sec:ereso}.

A cluster shape parameter, $R9$, is defined in order to distinguish
photons that convert upstream of ECAL from those entering ECAL
unconverted. It is defined as the ratio of the energy contained
within the 3$\times$3 array of crystals centred around the crystal
with maximum energy deposit to the total energy of the supercluster.
Showers from photons that interact with the tracker material will
spread out in the magnetic field reducing the value of $R9$.
A value of 0.94 has been chosen to distinguish between photons that
convert in the tracker material $(R9 < 0.94)$ and unconverted photons
$(R9 \ge 0.94)$. According to MC studies, about 70\% of the photons
with $R9 \ge 0.94$ in EB are unconverted~\cite{CMS_photons}.
For the purpose of studying the ECAL response, this variable is also
used in this paper to separate electrons in two categories with
supercluster topology similar to that of photons. On average,
electrons with $R9 \ge 0.94$ radiate less energy in the material
in front of ECAL, and are therefore more representative of the
unconverted photon sample than electrons with $R9 < 0.94$.

\begin{figure}[thp]
\begin{center}
\includegraphics[width=0.48\linewidth,height=8cm]{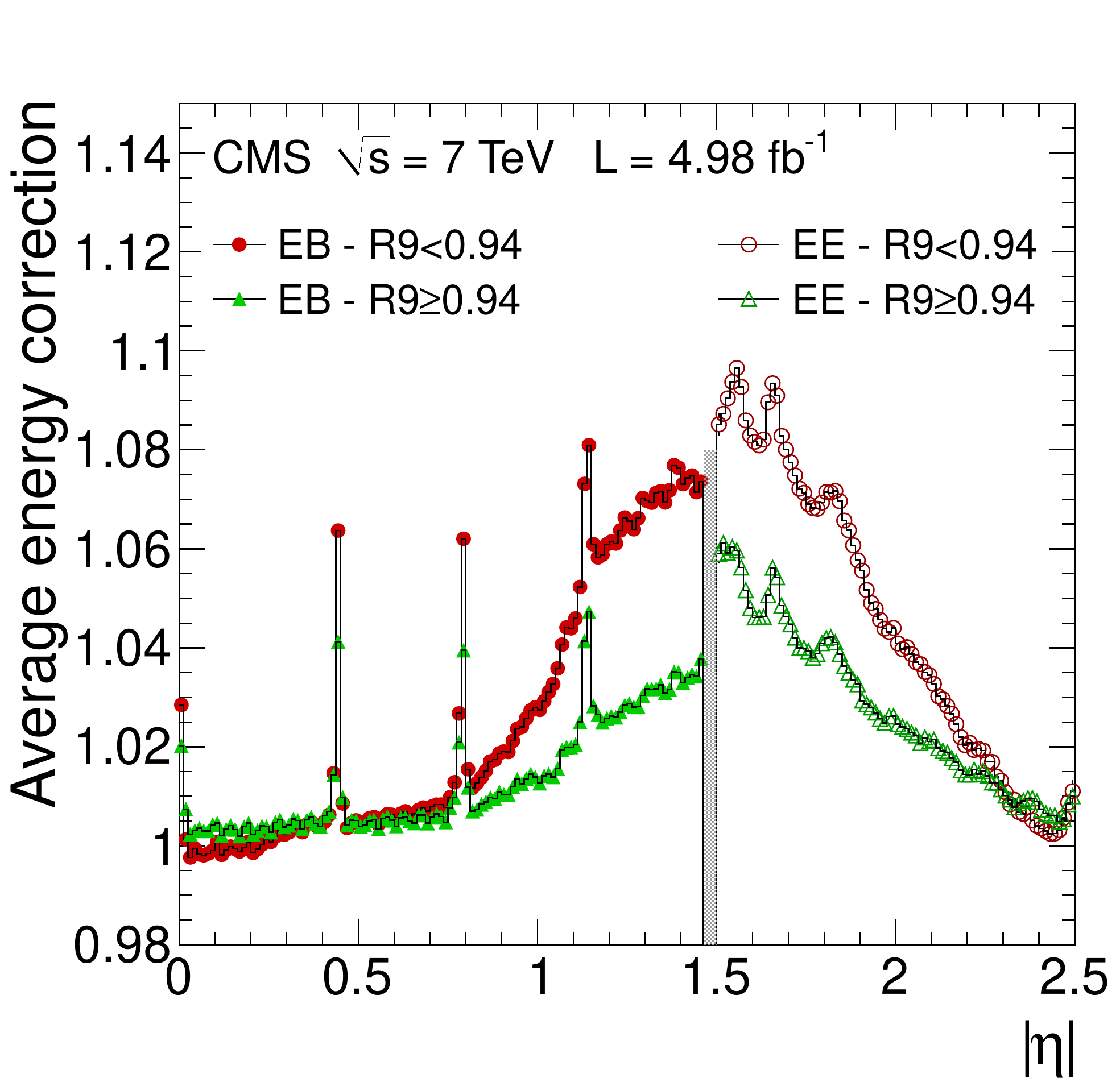}
\includegraphics[width=0.48\linewidth,height=7.6cm]{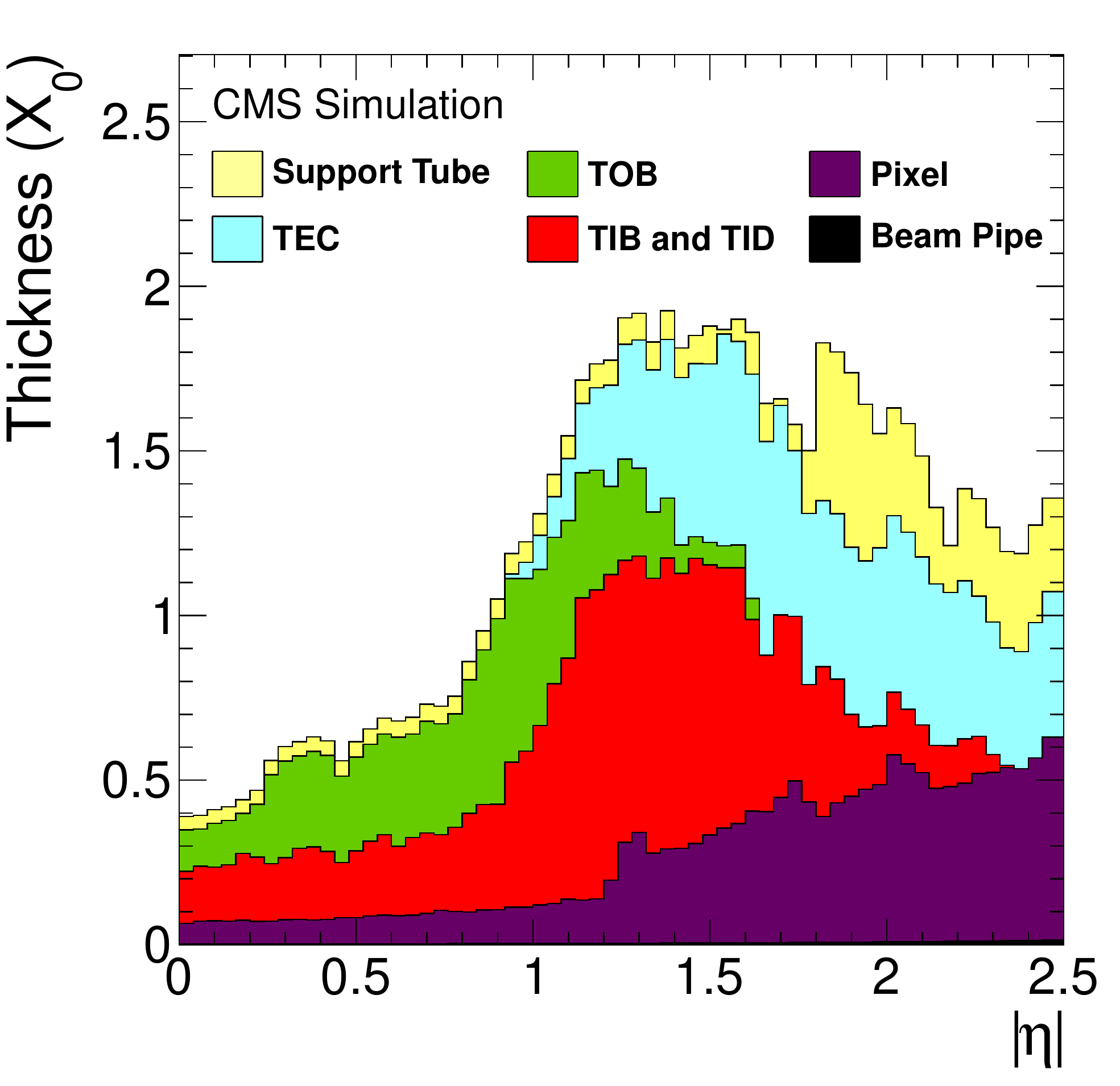}
\caption{Left: Average values of the $F_{\Pe}$ supercluster correction
  function plotted as a function of pseudorapidity for electrons
  from $\PW$-boson decays with $R9\ge0.94$ and $R9<0.94$,
  respectively. The steep  increase at $\abs{\eta} \approx 1$ is
  predominantly due to tracker material. Local structures correlate with
  the detector geometry (see text for details). Right: Material budget
  of the different components of the CMS tracker
  in front of the ECAL as a function of $\abs{\eta}$. The components are
  added to give the total tracker material budget. Notations in the
  legend correspond to TOB: tracker outer barrel, TIB: tracker inner
  barrel, TID: tracker inner discs, TEC: tracker endcaps.
  \label{fig:clustercorr}\label{fig:tracker_material}}
\end{center}
\end{figure}

Figure~\ref{fig:clustercorr} (left) shows the average value of the
$F_{\Pe}$ correction function as a function of supercluster
pseudorapidity evaluated from data using electrons from $\PW$-boson
decays. Events with both low and high values of $R9$ are shown. The
steep increase in the average value of the energy corrections between
$\abs{\eta} \approx 1$ and $\abs{\eta}\approx 2$ is dominated by the effects
of the tracker material. Figure~\ref{fig:clustercorr} (right) shows
the distribution of the tracker material in front of the ECAL as a
function of $\abs{\eta}$, which clearly exhibits a steep increase in the
material budget at the transition between the tracker barrel and
tracker endcap ($\abs{\eta} \approx 1$) and at $\abs{\eta}\approx 1.8$.
Correspondingly, the total material budget ranges from 0.4$\,X_0$ in the
central part of the barrel to about 2$\,X_0$ at $\abs{\eta} \approx 1.5$.
Local structures in Fig.~\ref{fig:clustercorr} (left) correlate with
inter-module boundaries in the barrel ($\abs{\eta}=0$, 0.45, 0.8, and
1.15). In the endcaps the structures at $\abs{\eta}\approx 1.55$, 1.65
and 1.8 correlate with the barrel-endcap transition, the preshower
edge, and the distribution of the tracker support tube material in
front of EE, respectively.

\begin{figure}[htbp]
\begin{center}
\includegraphics[width=0.49\linewidth]{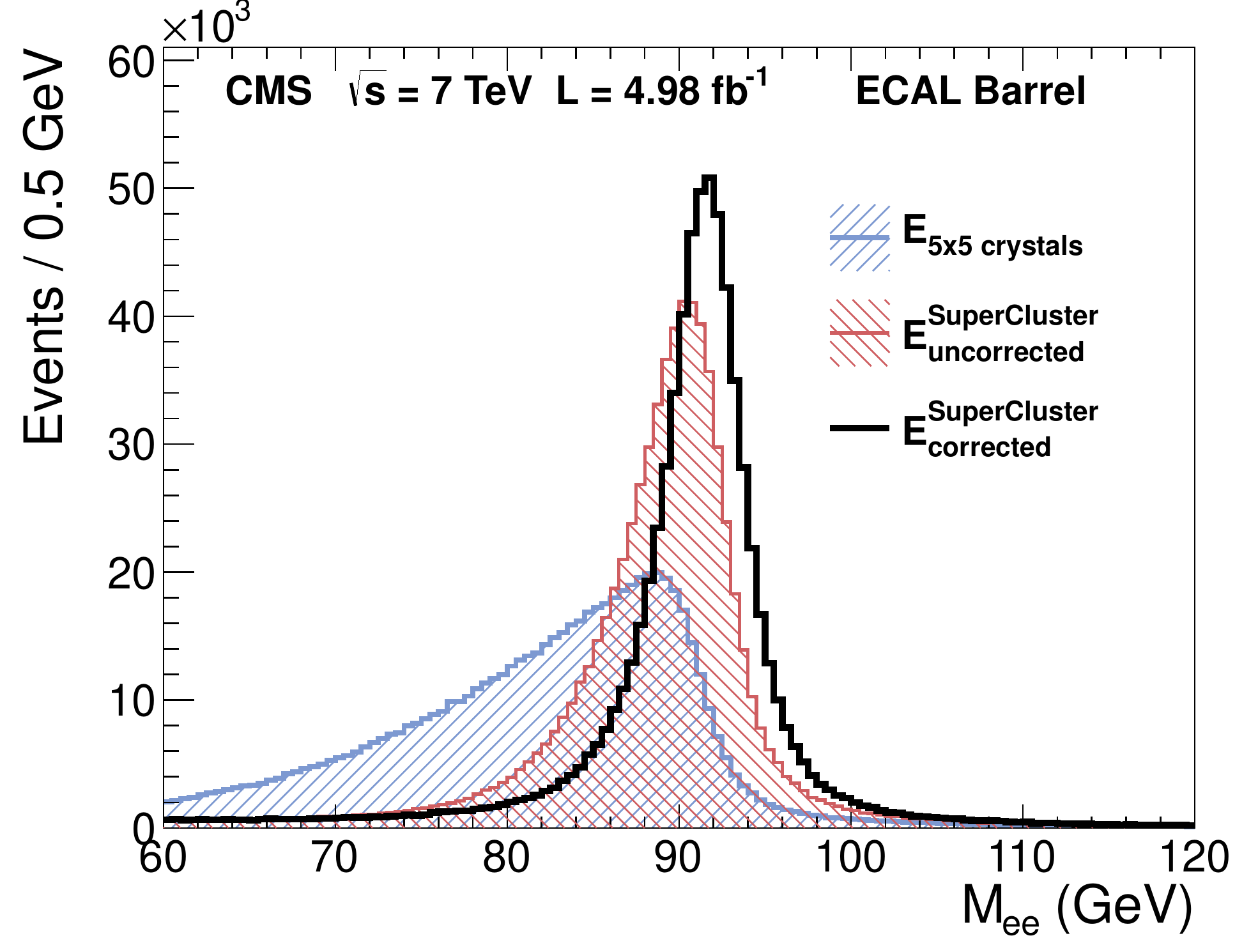}
\includegraphics[width=0.49\linewidth]{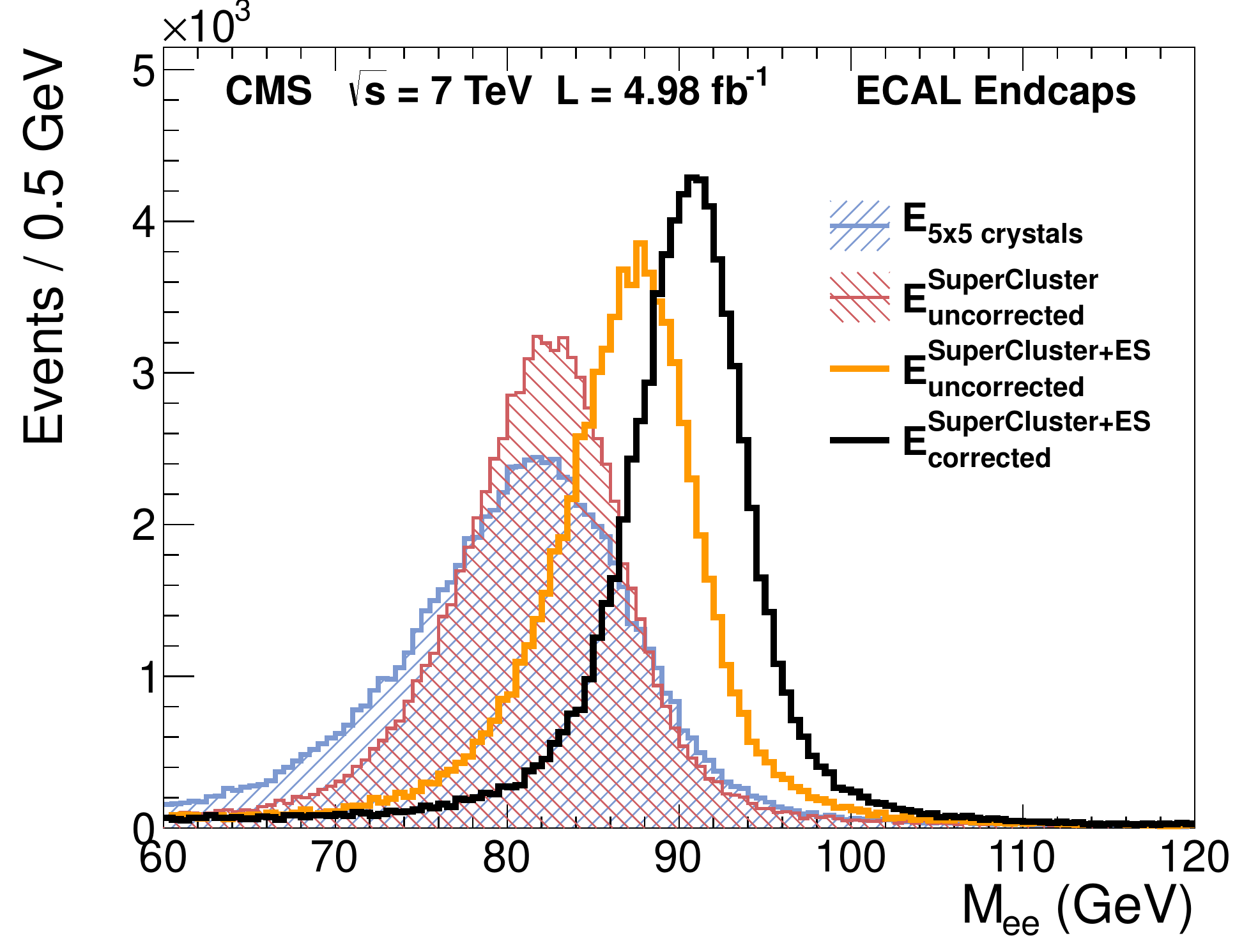}
\caption{Reconstructed dielectron invariant mass for electrons from $\cPZ
  \to \Pep\Pem$ events, applying a fixed-matrix clustering of 5x5 crystals,
  applying the supercluster reconstruction to recover radiated energy,
  and applying the supercluster energy corrections. For the EE the
  effect of adding the preshower detector energy is shown.
\label{fig:zee_clustercorr}}
\end{center}
\end{figure}

To illustrate the impact of the different steps in the energy
reconstruction, Fig.~\ref{fig:zee_clustercorr} shows the dielectron
invariant mass distribution for $\cPZ$-boson events, reconstructed
applying a fixed-matrix clustering of 5$\times$5 crystals with respect
to using the supercluster reconstruction to recover radiated energy,
and then applying the energy corrections. For the EE, the effect of
adding the preshower energy is also shown. The improvement in the
$\cPZ$-boson mass resolution is clearly demonstrated as the successive
steps are applied. This is particularly evident for the supercluster
reconstruction, which efficiently recovers the radiated energy and
reduces the low-energy tails of the distributions relative to the
5$\times$5 fixed-matrix clustering.

\subsection{Absolute energy calibration, \texorpdfstring{$G$}{G}}
\label{sec:scalecalib}

Nine EB supermodules and 500 EE crystals were exposed to high-energy
electron beams prior to being installed in CMS. From these data,
the absolute energy calibration for the ECAL was established by
equalizing the energy sum of a 5$\times$5 crystal matrix to the
electron beam energy. This calibration, which corresponds in CMS to
that relevant for unconverted photons, was adopted at the startup of
LHC operation in 2010.

In CMS, the absolute energy calibration ($G$) is computed in a
reference region of the ECAL where the effects of upstream material
and uncertainties in the energy corrections are minimal. The reference
region in the barrel is defined as the central 150 crystals in the
first module of each supermodule ($\abs{\eta}<0.35$), requiring a minimum
distance of 5 crystals from the border of each module in both $\eta$
and in $\phi$. This region is chosen because the material budget in
front of the first module is small, the geometry of these crystals is
very similar, and the centrality of the crystals in the module is
required so that there is no energy leakage due to the gaps between
modules or supermodules. In the EE, the reference region is defined as
the central region of each endcap ($1.7 < \abs{\eta} < 2.1$), to which
the crystals exposed to the beam test belong. The absolute energy
calibration in the MC simulation is computed using 50\GeV unconverted
photons. It is defined such that the energy reconstructed in a
5$\times$5 crystal matrix is equal to the true energy of the photon in
the reference region.
Decays of $\cPZ$-bosons into two electrons are used to set the
overall energy scale in EB and EE in data relative to the MC
simulation, and to validate the energy correction function $F_{\Pe}$
for electrons, using the $\cPZ$-boson mass constraint. Decays of
$\cPZ$-bosons into two muons where one muon radiates a photon, $\cPZ\to
\Pgm\Pgm\Pgg$, are used to cross-check the energy calibration of
photons.

\subsubsection{Energy scale calibration with $\cPZ \to \Pep\Pem$ events}
\label{sec:zeecalib}
The dielectron invariant mass in $\cPZ \to \Pep\Pem$ events is calculated
from the reconstructed supercluster energy in the ECAL, including the
energy corrections derived from MC simulation, and the opening angle
measured from tracks at the event vertex. The energy scale and
resolution are extracted from the dielectron invariant mass
distribution, for events with a reconstructed mass in the range
60--120\GeV. Electrons are selected if their transverse energy is
larger than 25\GeV as described in~\cite{EGM-10-004,Khachatryan:2010xn}.
With these selections, a background contamination of about 2\% is
estimated from MC simulation. The invariant mass distribution is
fitted with a Breit--Wigner line shape, convolved with a Crystal Ball
(CB) function~\cite{CB}:
\begin{equation}
\label{eq:bwcb}
\mathrm{CB}(x-\Delta m)=
\begin{cases}
\re^{-\frac{1}{2}(\frac{x-\Delta m}{\sigma_{\mathrm{CB}}})^2 }; &
\frac{x-\Delta m}{\sigma_{\mathrm{CB}}}>\alpha_{\mathrm{CB}}\\
\bigl(\frac{\gamma}{\alpha_{\mathrm{CB}}}\bigr)^\gamma
\cdot \re^{-\frac{\alpha_{\mathrm{CB}}^2}{2}}
\cdot \biggl( \frac{\gamma}{\alpha_{\mathrm{CB}}} -
\alpha_{\mathrm{CB}} -
\frac{x-\Delta m}{\sigma_{\mathrm{CB}}}\biggr)^{-\gamma}; &
\frac{x-\Delta m}{\sigma_{\mathrm{CB}}}<\alpha_{\mathrm{CB}} \\
\end{cases}
\end{equation}
where the parameter $\Delta m$ represents the displacement of the peak
with respect to the true $\cPZ$-boson mass, $\sigma_{\mathrm{CB}}$ is the
width of the Gaussian component of the CB function (a measure of the
energy resolution) and the parameters $\alpha_{\mathrm{CB}}$ and
$\gamma$ of the CB  tail function account for showering electrons
whose energy is not fully recovered by the clustering algorithms.

\begin{figure}[thbp]
\begin{center}
\begin{tabular}{ccc}
\hspace{-0.6cm} \includegraphics[width=0.34\linewidth]{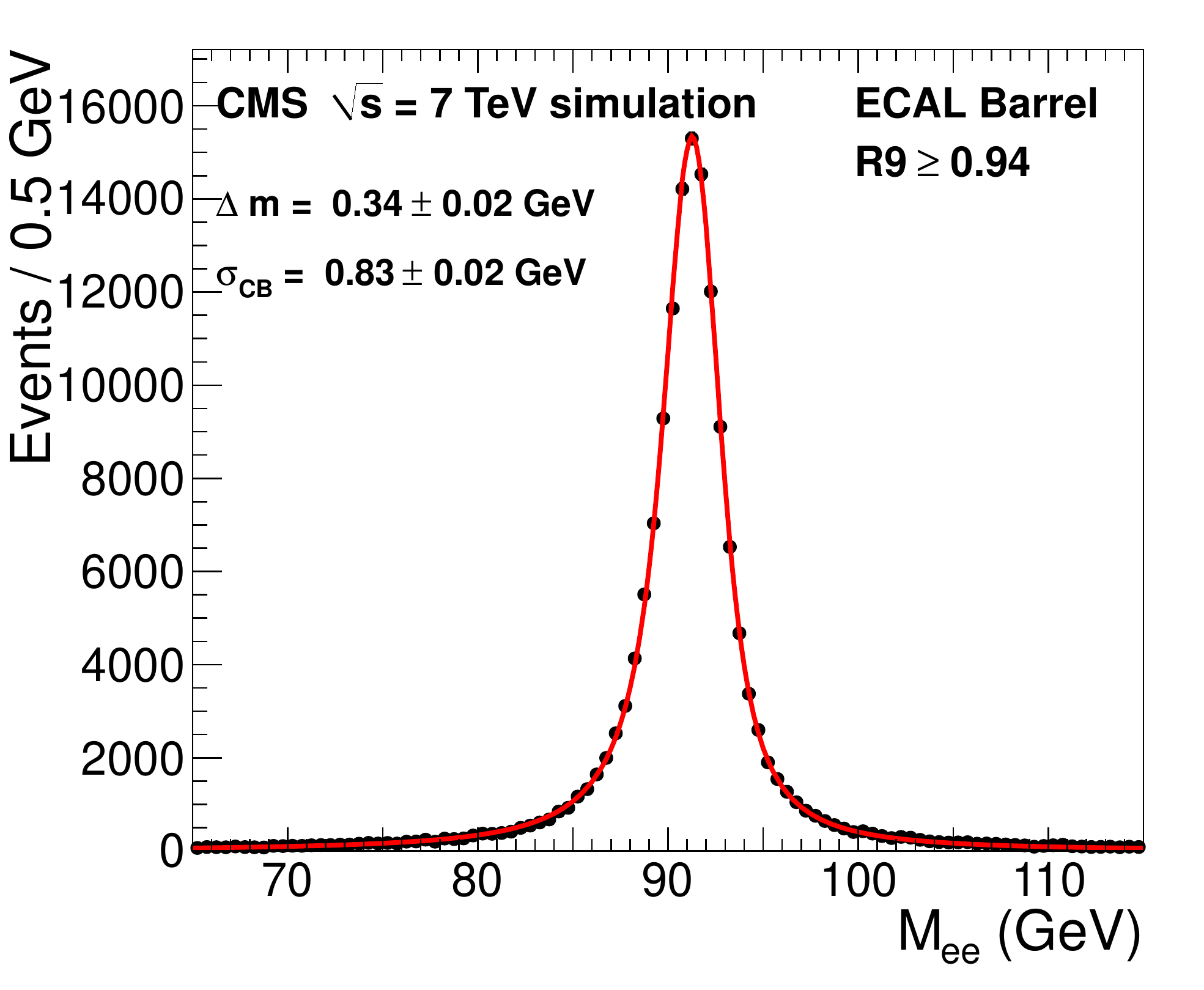} &
\hspace{-0.6cm} \includegraphics[width=0.34\linewidth]{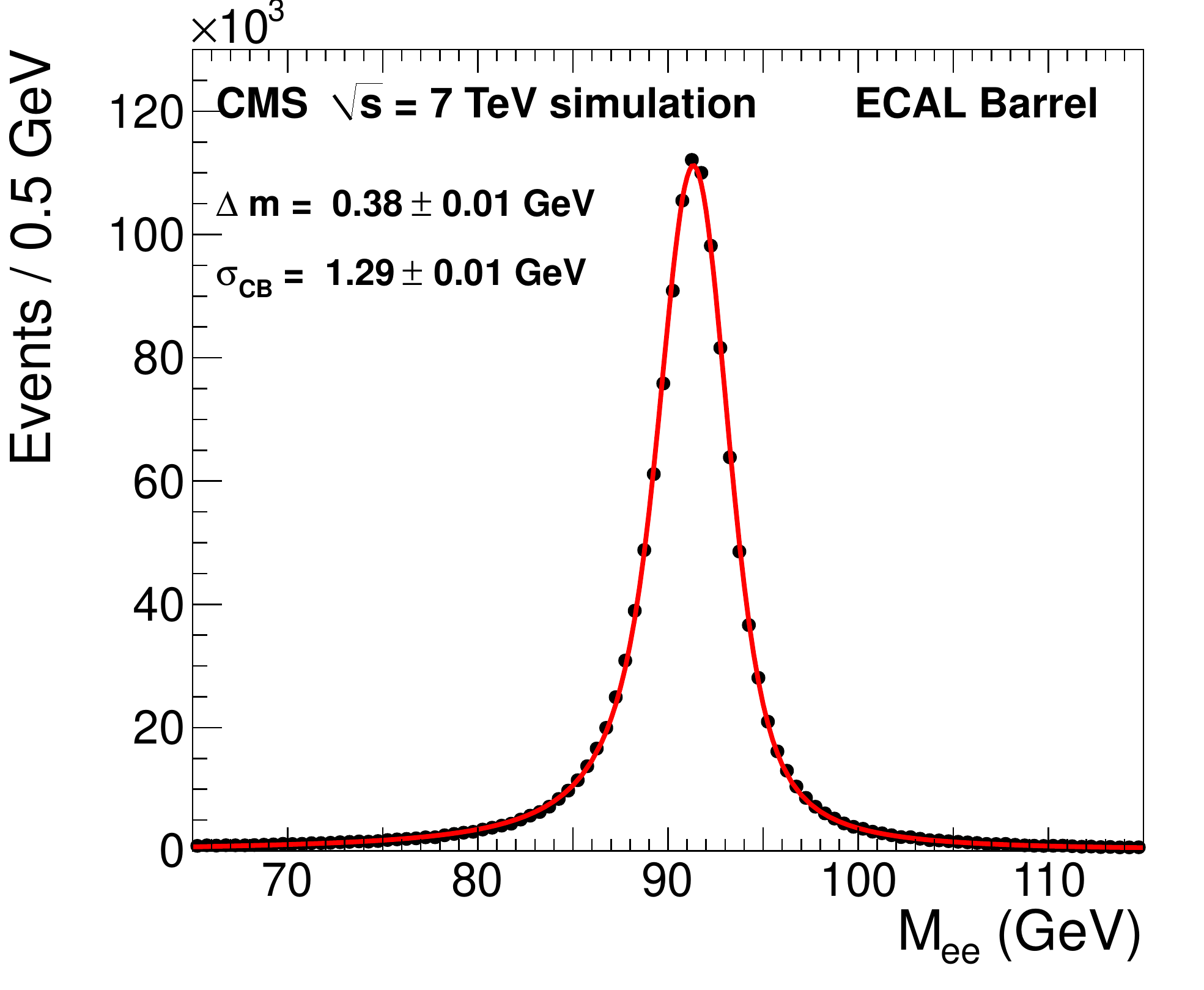} &
\hspace{-0.6cm} \includegraphics[width=0.34\linewidth]{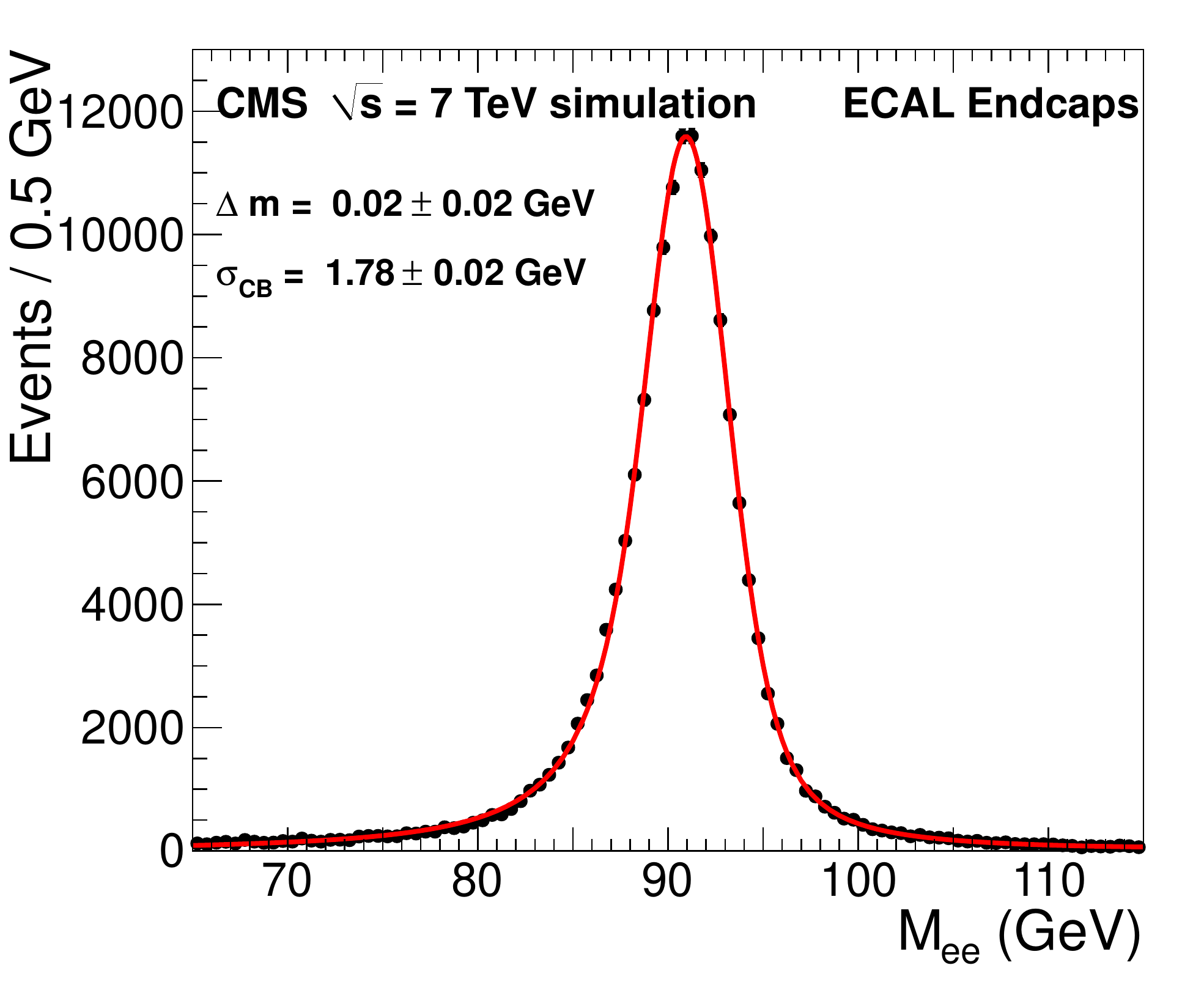} \\
\hspace{-0.6cm} \includegraphics[width=0.34\linewidth]{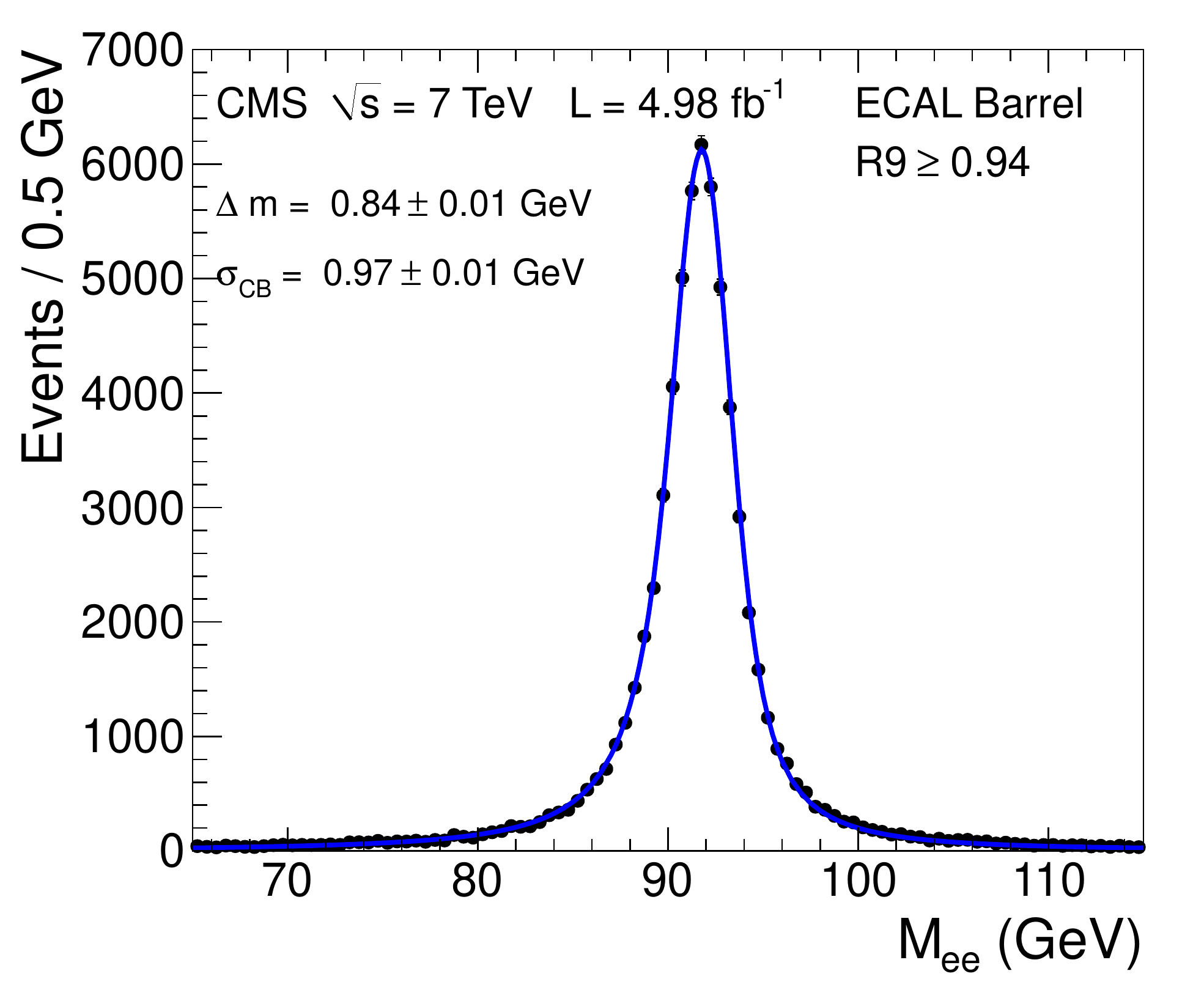}&
\hspace{-0.6cm} \includegraphics[width=0.34\linewidth]{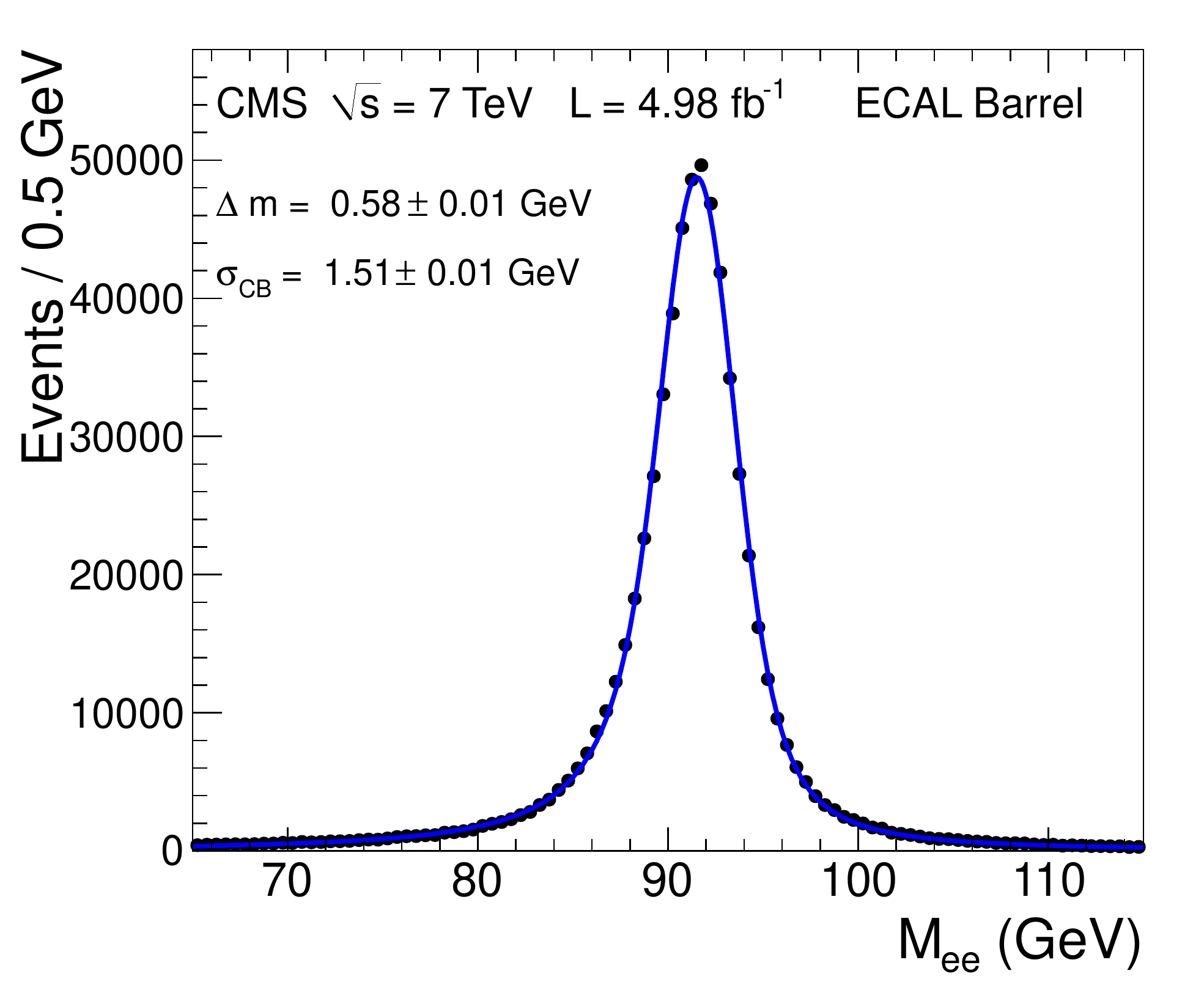}&
\hspace{-0.6cm} \includegraphics[width=0.34\linewidth]{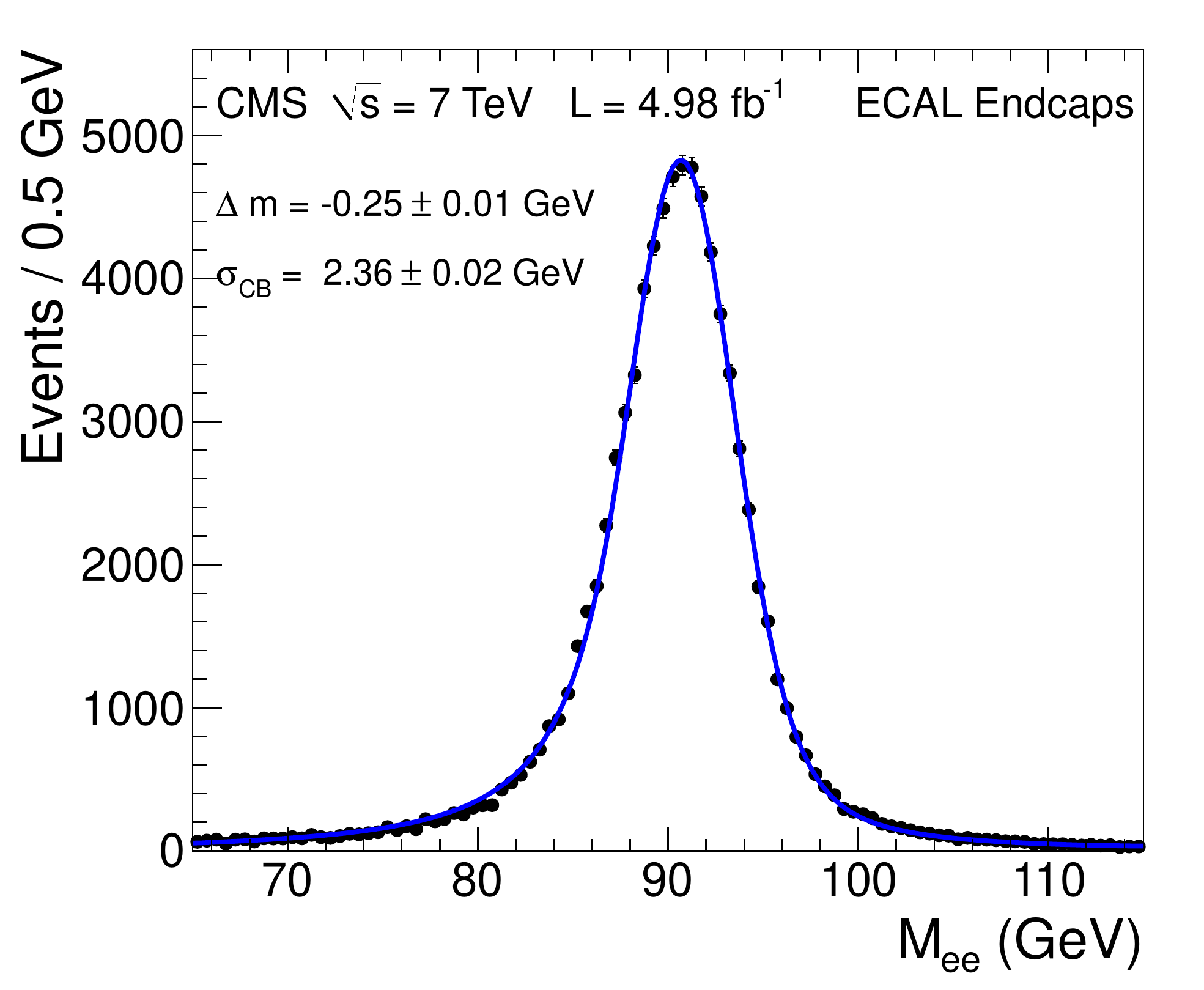}
\end{tabular}
\end{center}
 \caption{\label{fig:zee_scale}
The dielectron invariant mass distribution for $\cPZ$-boson decays with
both electrons in EB with $R9 \ge 0.94$ (left), both electrons in
the EB (centre) or both electrons in the EE (right). Distributions in
MC simulation (top row) and data (bottom row) are shown.
The parameters listed in each panel are $\Delta m$ --- the difference
between the CB mean and the true $\cPZ$-boson mass, and
$\sigma_{\mathrm{CB}}$ --- the width of the Gaussian term of the CB
function (see text for details).
}
\end{figure}

An unbinned maximum likelihood fit to the invariant mass distribution
is performed. The tail parameters $\alpha_{\mathrm{CB}}$ and $\gamma$
are constrained from MC simulation studies. The mass and width of the
Breit--Wigner function are fixed to $m_\cPZ = 91.188$\GeV and $\Gamma_\cPZ =
2.495$\GeV~\cite{PDG}. Figure~\ref{fig:zee_scale} shows the fitted
invariant mass distributions for data and simulation in EB and EE.
The ADC-to-GeV conversion factor $G$ of Eq.~(\ref{eq:one}) for data
is adjusted such that the fitted $\cPZ \to \Pep\Pem$ peak agrees with that
of the MC simulation separately for the barrel and endcap
calorimeters. In EB, $G$ is scaled by $(1 + (\Delta m_{\mathrm{MC}} -
\Delta m_{\mathrm{Data}}) / M_\cPZ)$, where $\Delta
m_{\mathrm{MC}}$ and $\Delta m_{\mathrm{Data}}$ are the results of
the fit on the MC simulation and data. In EE, the scaling is amplified
by the reciprocal of the mean fractional energy deposited in EE.

The systematic uncertainty associated to the absolute energy
calibration is estimated to be 0.4\% in EB and 0.8\% in EE for the
2011 data sample, and is dominated by the uncertain knowledge of the
energy correction function for the electrons ($F_{\Pe}$) in
the reference region. In order to determine the size of this
uncertainty, the energy
scale has been derived from the dielectron invariant mass
distributions reconstructed from the raw supercluster energy both in
data and MC events. Moreover, the analysis has been repeated using MC
samples generated with tracker material budget altered within its
uncertainty~\cite{TRACKER_NOTE, TRACKER_PAS}. The observed variation in
the results is taken as a systematic uncertainty. In the endcaps, the
uncertainty of the ES detector energy calibration also contributes to
the systematic uncertainty. The dependence of $G$ on a number of
additional effects has been also studied. They include the stability
of the result on changes in the event selection and in the
functional form used to describe the ECAL response. Each of these
effects causes an uncertainty on $G$ of about 0.1\% or less, for a
total uncertainty of 0.2\%.

\subsubsection{Verification of the energy calibration and corrections and linearity check}
The $\cPZ \to \Pgm\Pgm\Pgg$ decays, where the photon arises from muon
final-state radiation, are used to cross-check the photon energy
calibration. A data sample with about 98\% purity has been selected by
requiring a pair of identified muons of $\ET$ greater than 15\GeV, an
isolated photon of $\ET$ greater than 25\GeV, a separation $\Delta R =
\sqrt{\Delta\eta^2 + \Delta\phi^2}$ between the photon and the closest
muon lower than 0.8, an invariant mass of the $\Pgm\Pgm\Pgg$ system,
computed from the muon momenta and the photon energy measured by ECAL,
between 60\GeV and 120\GeV, and the sum of the $\Pgm\Pgm\Pgg$ and the
dimuon invariant masses lower than 180\GeV. The mean $\ET$ of the
photons in the events selected is approximately 32\GeV; the mean
energy is about 42\GeV in EB and 114\GeV in EE.

\begin{figure}[htbp]
\begin{center}
\hspace{-0.40cm}\includegraphics[width=0.345 \linewidth,height=4.8cm]{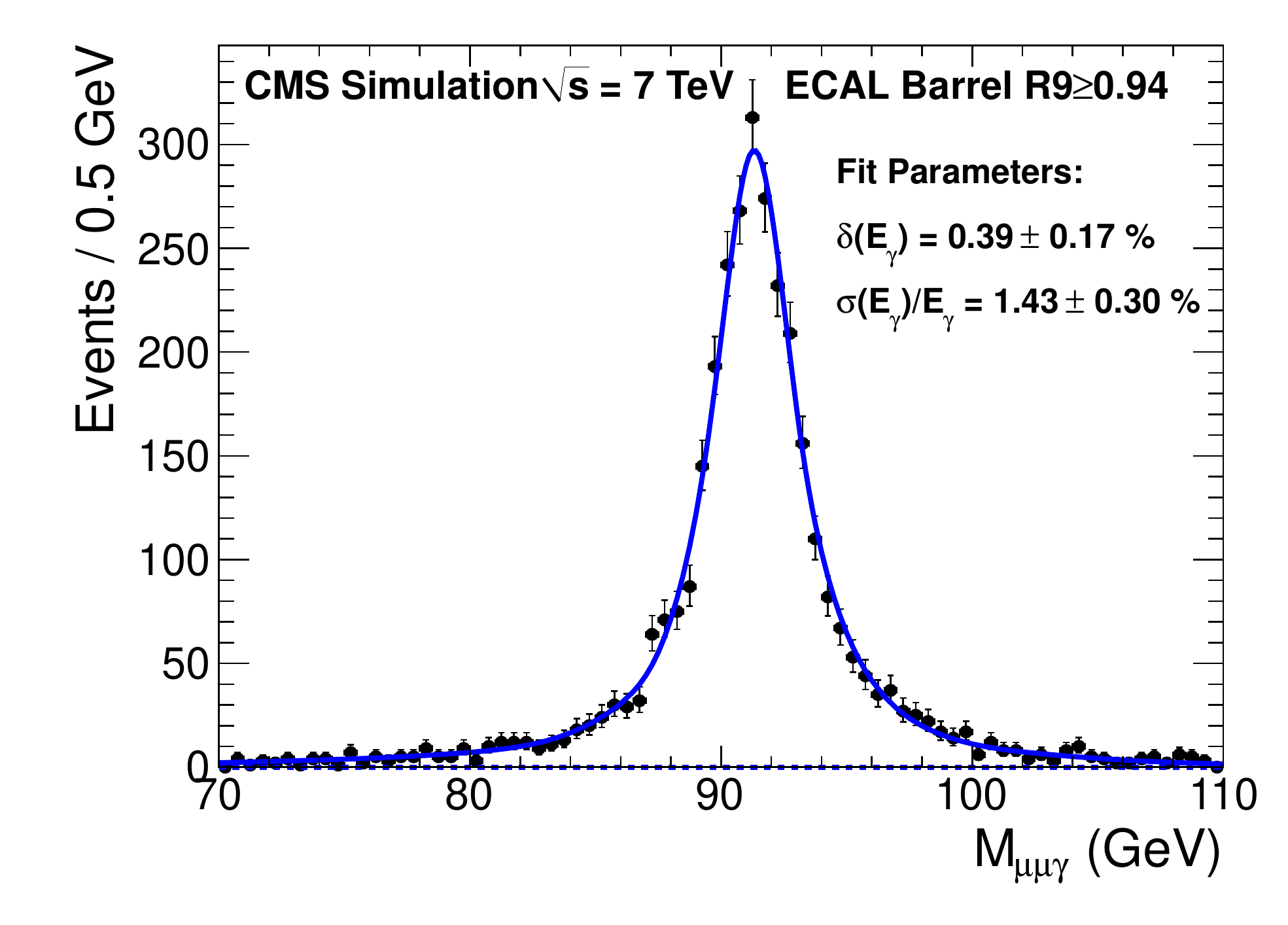}
\hspace{-0.25cm}\includegraphics[width=0.345 \linewidth,height=4.8cm]{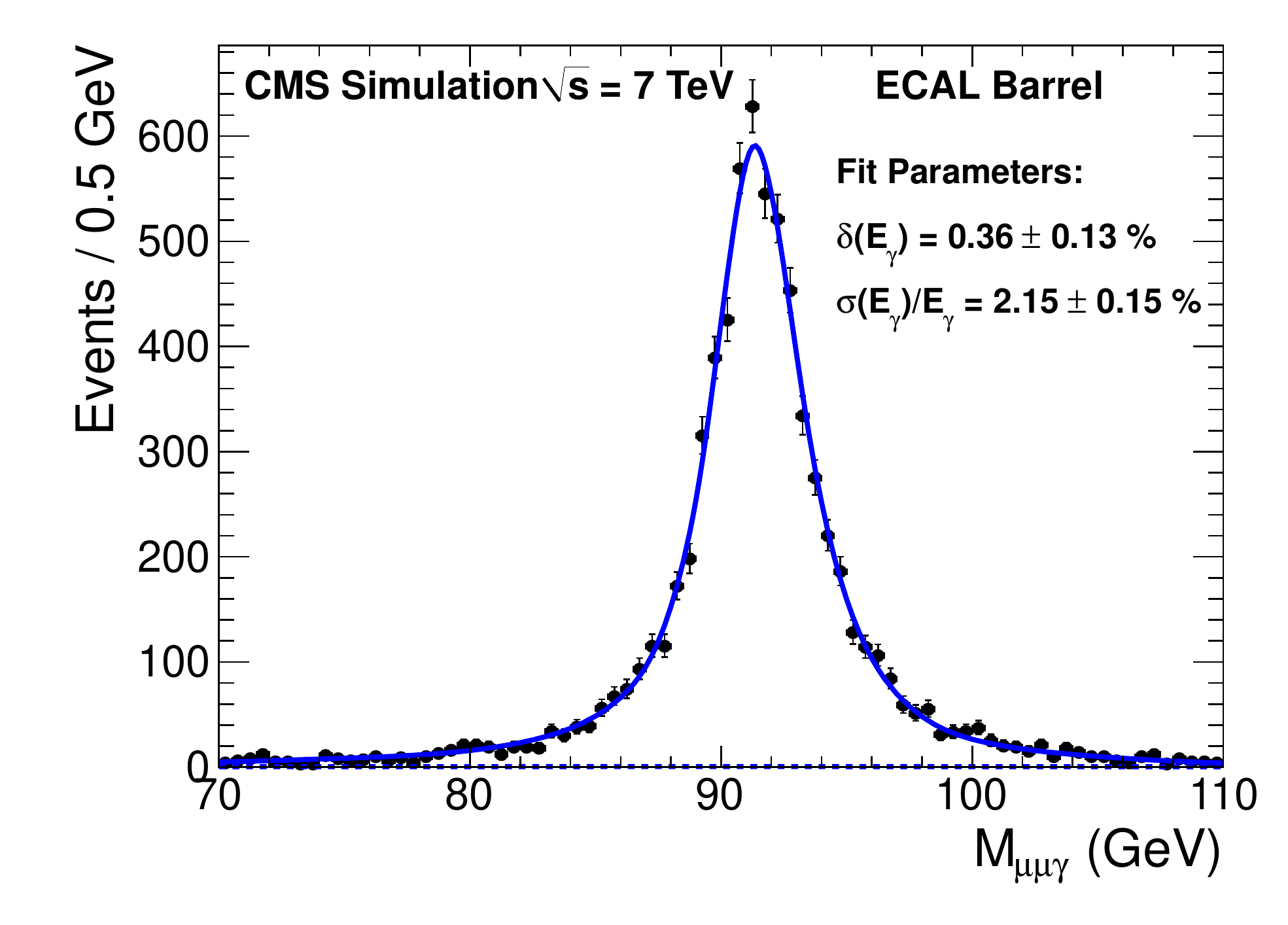}
\hspace{-0.10cm}\includegraphics[width=0.345 \linewidth,height=4.8cm]{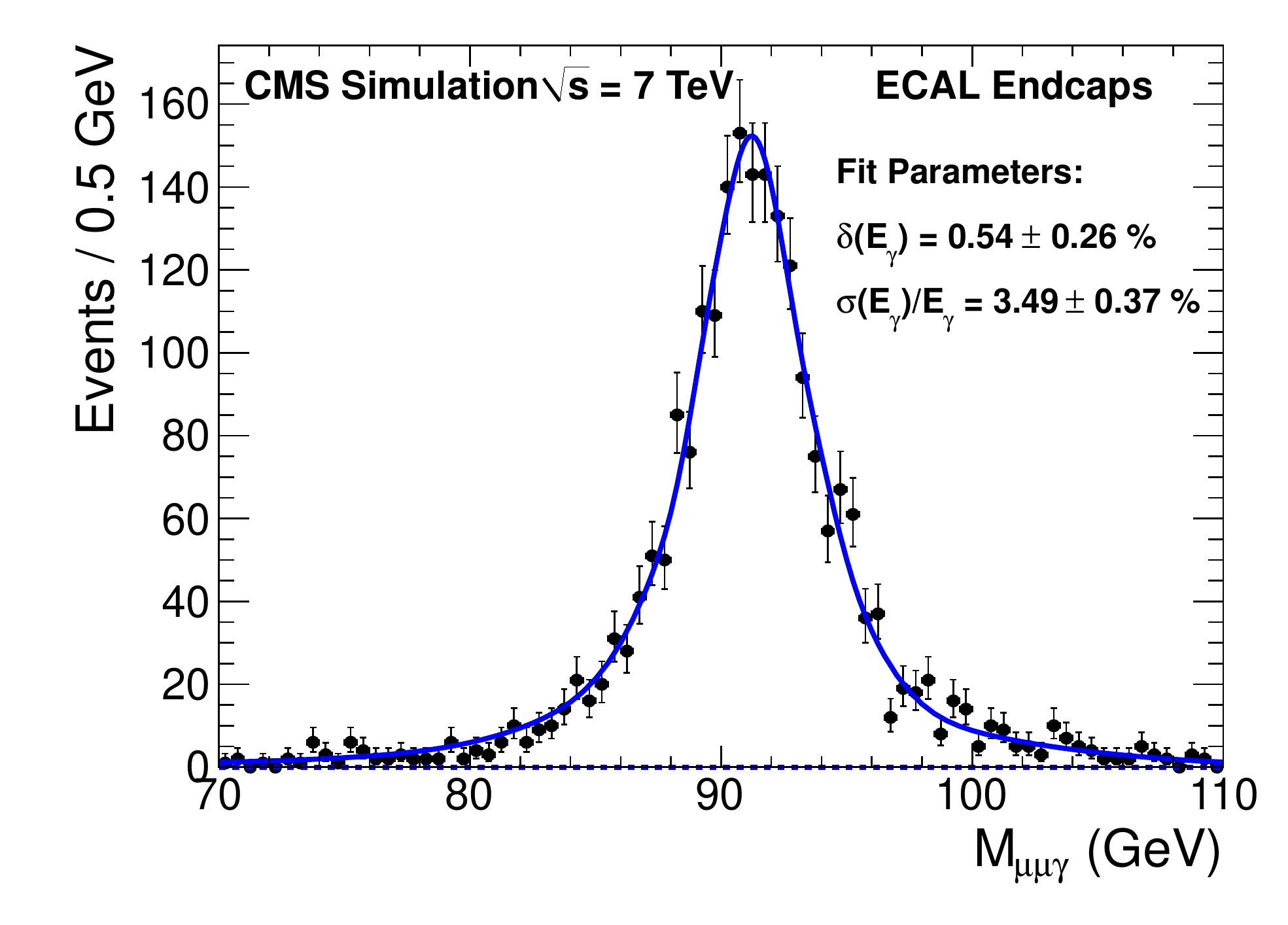}\\
\hspace{-0.40cm}\includegraphics[width=0.345 \linewidth,height=4.8cm]{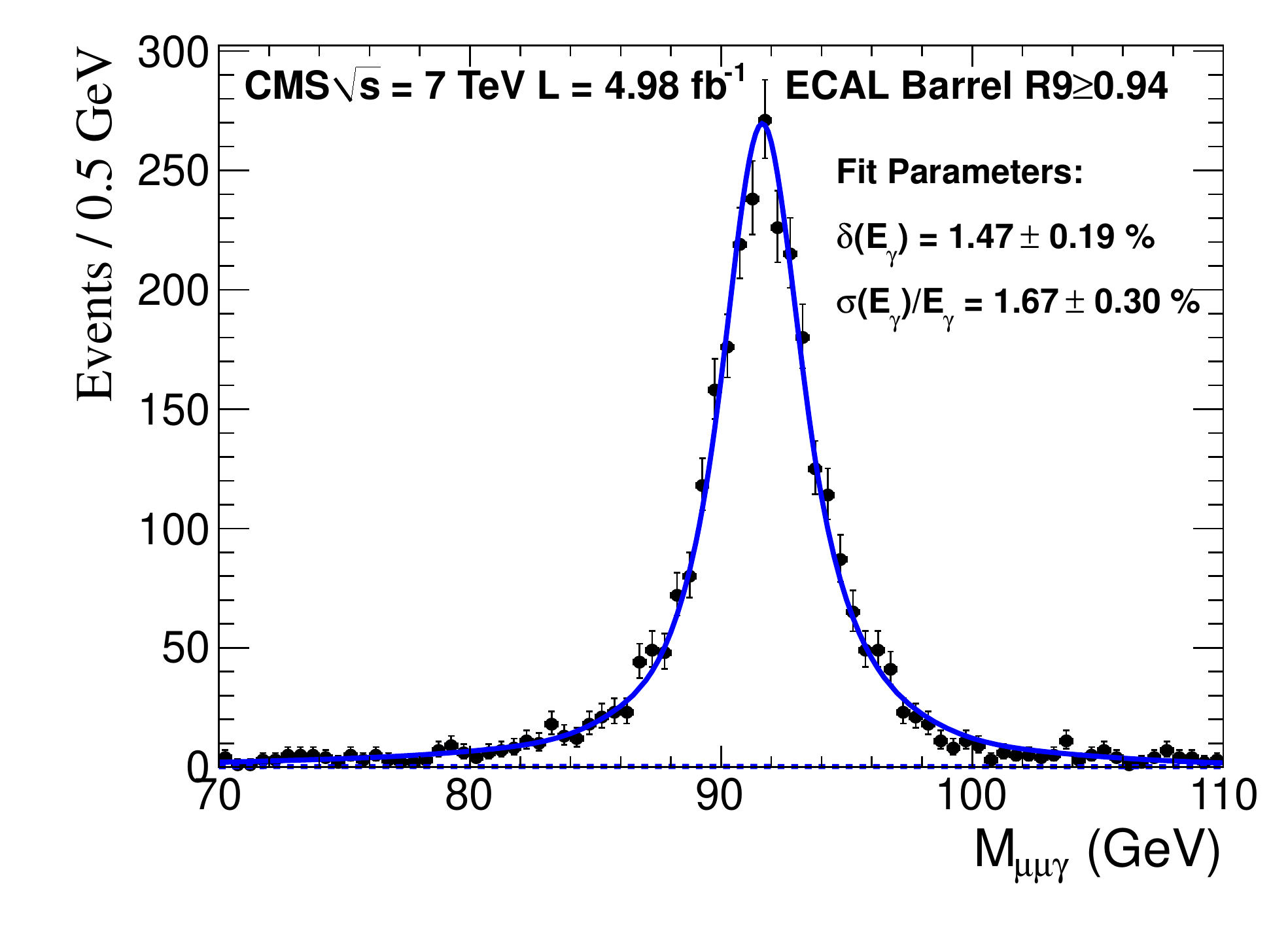}
\hspace{-0.25cm}\includegraphics[width=0.345 \linewidth,height=4.8cm]{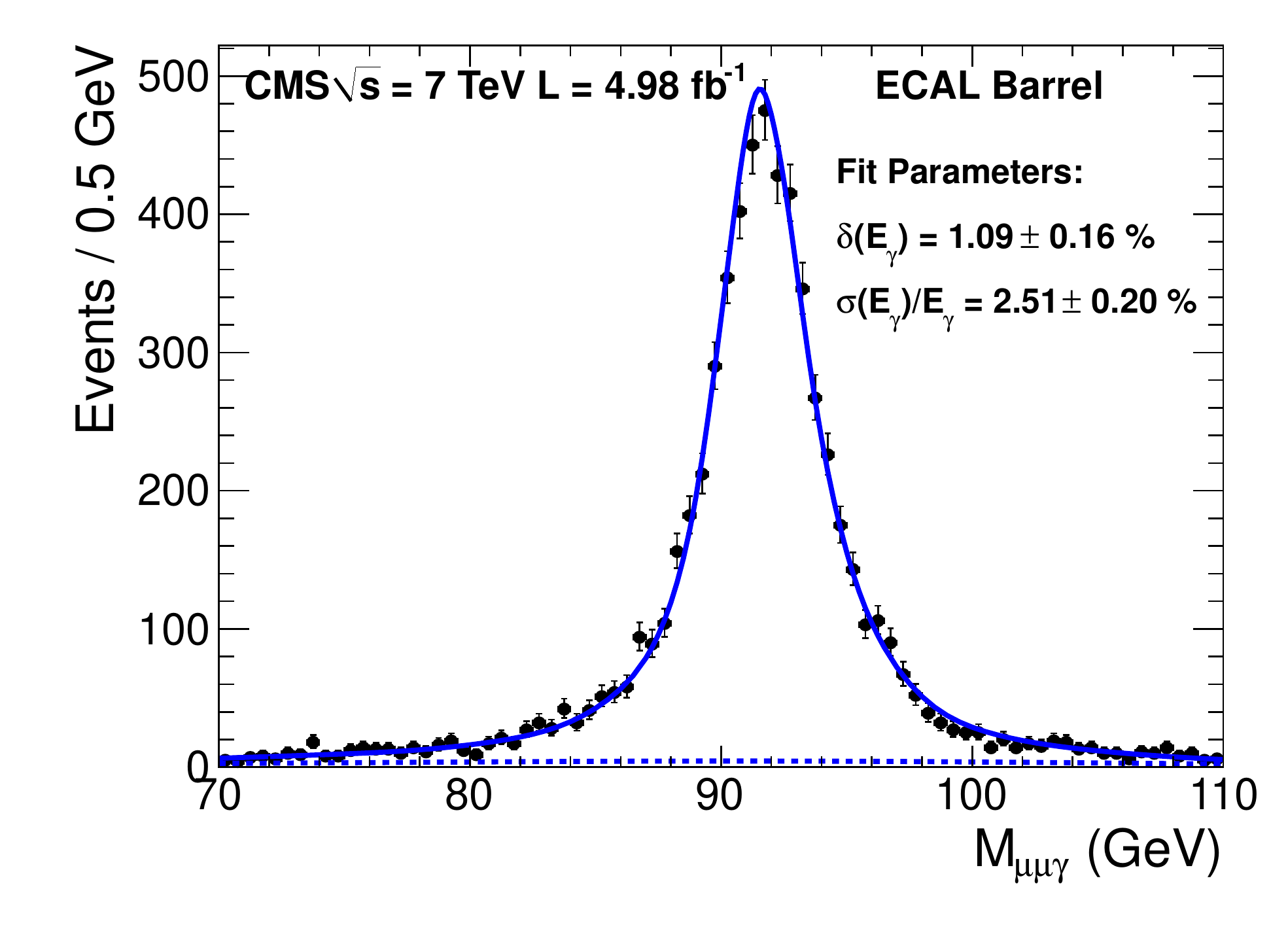}
\hspace{-0.10cm}\includegraphics[width=0.345\linewidth,height=4.8cm]{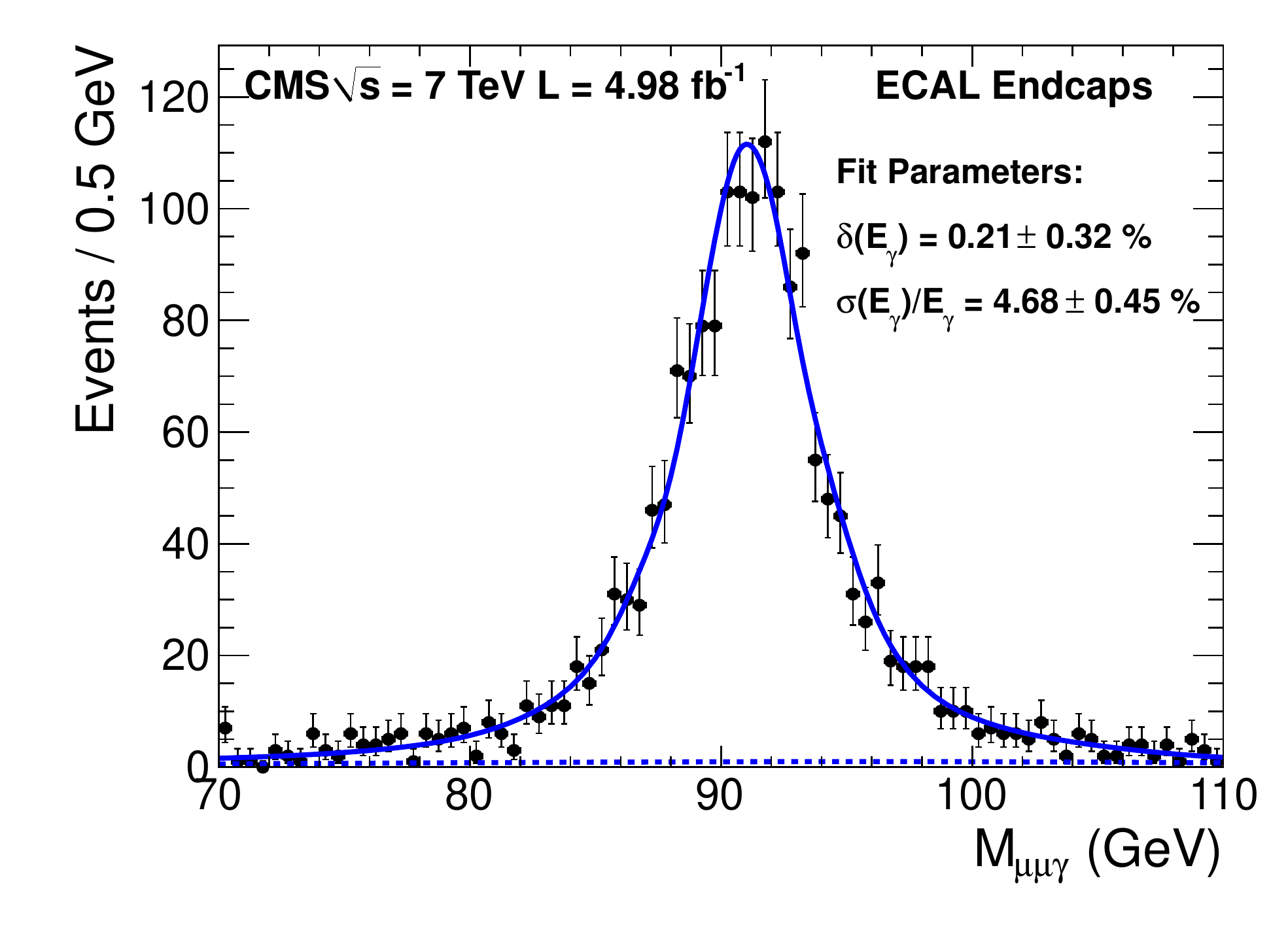}\\
\caption{
Invariant mass distribution of $\cPZ \to \Pgm \Pgm \Pgg$ events.
Plots show MC simulation (top row) and data (bottom row) for EB
photons with $R9 \ge 0.94$, EB inclusive and EE inclusive categories.
The relative mean deviation of the reconstructed photon energy from
that expected from the decay kinematics, $\delta$, and the mean
energy resolution of the selected events are listed. The continuous
lines show the fit results for the $\cPZ$-boson lineshape convolved
with a response function modelled from MC samples (see text for
details). \label{fig:zmumugamma} }
\end{center}
\end{figure}

Figure~\ref{fig:zmumugamma} shows the invariant mass distributions
reconstructed from two muons and the radiated photon.
In each plot, fitted values of the relative mean deviation of the
reconstructed photon energy from that expected from the kinematics of
$\cPZ\to\Pgm\Pgm\Pgg$ decays, $\delta$, and the mean energy resolution of
the selected events, $\sigma_E/E$, are listed.
The photon energy scale and resolution are extracted
simultaneously by unfolding the $\cPZ$-boson line shape from the detector
response function. The response function is modelled from MC samples
using a kernel density estimator~\cite{Rosenblatt1956,
  Parzen1962}. The scale and resolution dependence of the response
function is built by scaling the distribution of the differences of the
true and the reconstructed photon energy. The effective $\sigma$,
defined as the interval around the most probable value of the
normalized differences of the true and the reconstructed energy
containing 68\% of the events, is used to measure the resolution.
Alternatively, the photon energy scale is estimated from the mean of
the distribution of a per-event energy scale estimator defined as $s =
(m^2_{\Pgm\Pgm\Pgg} - m^2_{\Pgm\Pgm}) / (m^2_Z-m^2_{\Pgm\Pgm})-1$, where
the terms indicate the dimuon and the $\Pgm\Pgm\Pgg$ invariant
masses, and the nominal $\cPZ$-boson mass. The mean of the distribution
is extracted from a fit with a Breit--Wigner distribution convolved
with a Gaussian function.
A systematic uncertainty of 0.3\% on the photon energy scale is
ascribed to the analysis, due to the dependence of the result on the
fitting method. The effect of the muon momentum calibration
uncertainty and the contribution of various backgrounds in data has
been checked and found to be negligible.

Given the systematic uncertainty on the absolute energy scale factor,
$G$, extracted from the analysis of $\cPZ\to \Pep\Pem$ events, which is a common
term of the electron and photon calibration schema presented in
Eq.~(\ref{eq:one}), the relative mean deviations of the reconstructed
photon energy $\delta$ listed in the plots of Fig.~\ref{fig:zmumugamma}
show that the photon energy is consistently calibrated in data and MC
simulation within statistical and systematic uncertainties.

In order to assess the quality of the energy corrections in data, the
variation of $E/p$ with isolated electrons from $\PW$- and
$\cPZ$-boson decays and of the mass resolution in $\cPZ\to \Pep\Pep$
decays have been studied as a function of several
observables that impact on the energy reconstruction. This analysis
exploits the same methods discussed in Sec.~\ref{sec:wenu_laser}.
Before the application of energy corrections, the effect of pileup
generates a dependence of the shower energy on the number of collision
vertices of about 0.05\% per vertex in EB and 0.1\% per vertex in
EE. After corrections, no residual dependence of the energy
calibration and resolution on the number of collision vertices per
beam crossing is observed~\cite{DiemozCR}, showing the effectiveness
of the correction for pileup derived from MC simulation with the MVA
technique. A case of imperfect corrections has been identified in the
study of $E/p$ as a function of the impact point of the electron on
the crystal, showing that corrections based on the MC simulation do
not fully compensate for the energy leakage in the inter-crystal gaps,
yielding a residual response variation up to 1\% between showers
hitting the centre of a crystal and those close to a crystal
boundary. These effects are estimated to contribute to the current
energy resolution with an RMS of about 0.3\%-0.5\% and may indicate
that the shower width in MC simulation is not exactly matched to
data~\cite{TabarellideFatis:2012koa}.

The linearity of the energy response was checked by studying the
dependence of $E/p$ as a function of $\ET$ with isolated electrons
from $\cPZ$- and $\PW$-boson decays. Moreover, using boosted $\cPZ$-boson
events, the stability of the $\cPZ$-boson mass as a function of the
scalar sum of the transverse energies of the two electrons, \ie, $\HT
= \ET^1 + \ET^2$, was studied. In these analyses, the $E/p$
distribution in bins of $\ET$ and the dielectron invariant mass in
bins of $\HT$ from MC simulation were fitted to the corresponding
distributions in data. A scale factor was extracted from each fit,
whose difference from unity measures the residual non-linearity
of the energy response in data relative to the MC samples. This
non-linearity is found to vary from $-0.2$\% to $+0.2$\% for $\ET$
varying from 30\GeV to 110\GeV. The amount of data collected in 2011
did not permit the measurement to be extended to higher energies.

\section{Energy resolution}
\label{sec:ereso}

\subsection{Inclusive energy resolution from the \texorpdfstring{$\cPZ$}{Z}-boson line shape}
\label{sec:eresolutionz}

The energy resolution for electrons is measured using $\cPZ\to \Pep\Pem$
events. The electron energies are reconstructed from the
ECAL energy deposits with the calibrations and corrections described
in the previous sections. The dielectron invariant mass resolution
(which is dominated by the electron energy resolution) is related to
the single-electron energy resolution by an approximate scaling factor
of $\sqrt{2}$, verified using MC simulations. The intrinsic detector
resolution is estimated by the Gaussian width of the Crystal Ball
function, the $\sigma_{\mathrm{CB}}$ parameter in Eq.~(\ref{eq:bwcb}).

The dielectron invariant mass distributions for data and MC
samples are shown in Fig.~\ref{fig:zee_scale}.
The fitted  values of $\sigma_{\mathrm{CB}}$ are reported in
Table~\ref{tab:width_zpeakfits}.
The width of the Gaussian term of the Crystal Ball function is 1.51\GeV
when both electrons are in the barrel (0.97\GeV if both electrons have
$R9\ge0.94$), and 2.36\GeV when both electrons are in the
endcaps. These correspond to a relative mass resolution of 1.65\% in
the barrel and to 2.59\% in the endcaps for dielectrons from $\cPZ$-boson
decays.

\begin{table}[tp]
 \begin{center}
 \topcaption{ \label{tab:width_zpeakfits}
 Extracted values of the parameter $\sigma_{\mathrm{CB}}$ from fits to
 the $\cPZ \to\Pep\Pem$ invariant mass spectrum for simulation and
 data. The fit is performed with the line shape given in Eq.~(\ref{eq:bwcb}).}
  \begin{tabular}{|l|c|c|} \hline
   Event class     & $\sigma_{\mathrm{CB}}^{\mathrm{MC}}$ (\GeVns{}) & $\sigma_{\mathrm{CB}}^{\text{data}}$ (\GeVns{})  \\ \hline
   EB ($R_9>0.94$) & $ 0.83 \pm 0.02 $  & $ 0.97 \pm 0.01 $   \\
   EB              & $ 1.29 \pm 0.01 $  & $ 1.51 \pm 0.01 $   \\
   EE              & $ 1.78 \pm 0.02 $  & $ 2.36 \pm 0.02 $   \\ \hline
  \end{tabular}
 \end{center}
\end{table}

Similarly, the energy resolution for photons has been studied from the
line shape of $\cPZ\to \Pgm\Pgm\Pgg$ events, in an $\ET$ range slightly
lower, but comparable, to that of $\cPZ\to \Pep\Pem$ events. Results are
shown in Fig.~\ref{fig:zmumugamma}, for photons with $R9 \ge 0.94$ in EB,
and for the inclusive samples of photons in EB and EE separately. Because
of the $\abs{\eta}$ dependence of the material in front of the ECAL, shown
in Fig.~\ref{fig:tracker_material} right, the photon resolution for
$R9 \ge 0.94$ is dominated by photons with $\abs{\eta}<1$ while the
performance for $R9<0.94$ is dominated by photons with $\abs{\eta}>1$.
The measured mean energy resolution is 2.5\% in the barrel (1.7\% for
high $R9$) and 4.7\% in the endcaps. As with the electrons from
$\cPZ$-boson decays, the photon energy resolution in data is not
correctly described by the MC simulation.

For both the electrons from $\cPZ$-boson decays and the photons from
$\cPZ\to\Pgm\Pgm\Pgg$, the energy resolution in the data is not correctly
described by the MC simulation. The sources of this discrepancy are
thought to be common, and are discussed in
Section~\ref{sec:resolution}. These differences are accommodated in
CMS analyses by applying additional Gaussian smearing, in bins of
$\eta$ and $R9$, to the electron and photon energies in MC simulation,
as discussed in Sections~\ref{sec:unfolding} and~\ref{sec:hggresolution}.

\subsection{The energy resolution for electrons as a function of pseudorapidity}
\label{sec:unfolding}

A maximum likelihood fit is used to extract the ECAL energy resolution
as a function of the pseudorapidity of the final-state electrons, and
in two bins of $R9$. The fit is performed on $\cPZ \to \Pep\Pem$ decays,
with an invariant dielectron mass between 89\GeV and 100\GeV, and the
following likelihood function is maximized:
\begin{equation}
\mathcal{L} = \prod_i
{\mathrm{Voigt}}(M^i_{\Pe\Pe},\sigma^i_{M_{\Pe\Pe}};M_Z,\Gamma_Z),
\label{eq:resolutionlikelihood}
\end{equation}
where Voigt is a convolution of a Breit--Wigner distribution with a
Gaussian function, and the product is run over all the events. The
mass resolution $\sigma_{M_{\Pe\Pe}}$ can be written as:
\begin{equation}
 \sigma_{M_{\Pe\Pe}} = \frac{1}{2} \cdot M_{\Pe\Pe} \cdot
     \sqrt{\left[ \frac{\sigma_E}{E} (\eta_1,R9_1) \right]^2 +
     \left[ \frac{\sigma_E}{E} (\eta_2,R9_2) \right]^2}
\end{equation}
where the average values of $\sigma_E/E$  in several bins of $\eta$
and two bins of $R9$ for $\ET \approx 45$\GeV electrons from $\cPZ$-boson
decays are free parameters in the fit. The narrow mass window used in
the fit allows the resolution to be determined mostly from the high
energy side of the invariant mass distribution, where the Crystal-Ball
function used in Eq.~(\ref{eq:bwcb}) reduces to a Gaussian function.
The likelihood function adopted here is numerically simpler than that
in Eq.~(\ref{eq:bwcb}) and allows the number of parameters in the fit
to be made sufficiently large to extract a detailed map of the energy
resolution as a function of  $\abs{\eta}$.

\begin{figure}[htbp]
\begin{center}
\includegraphics[height=5.2cm]{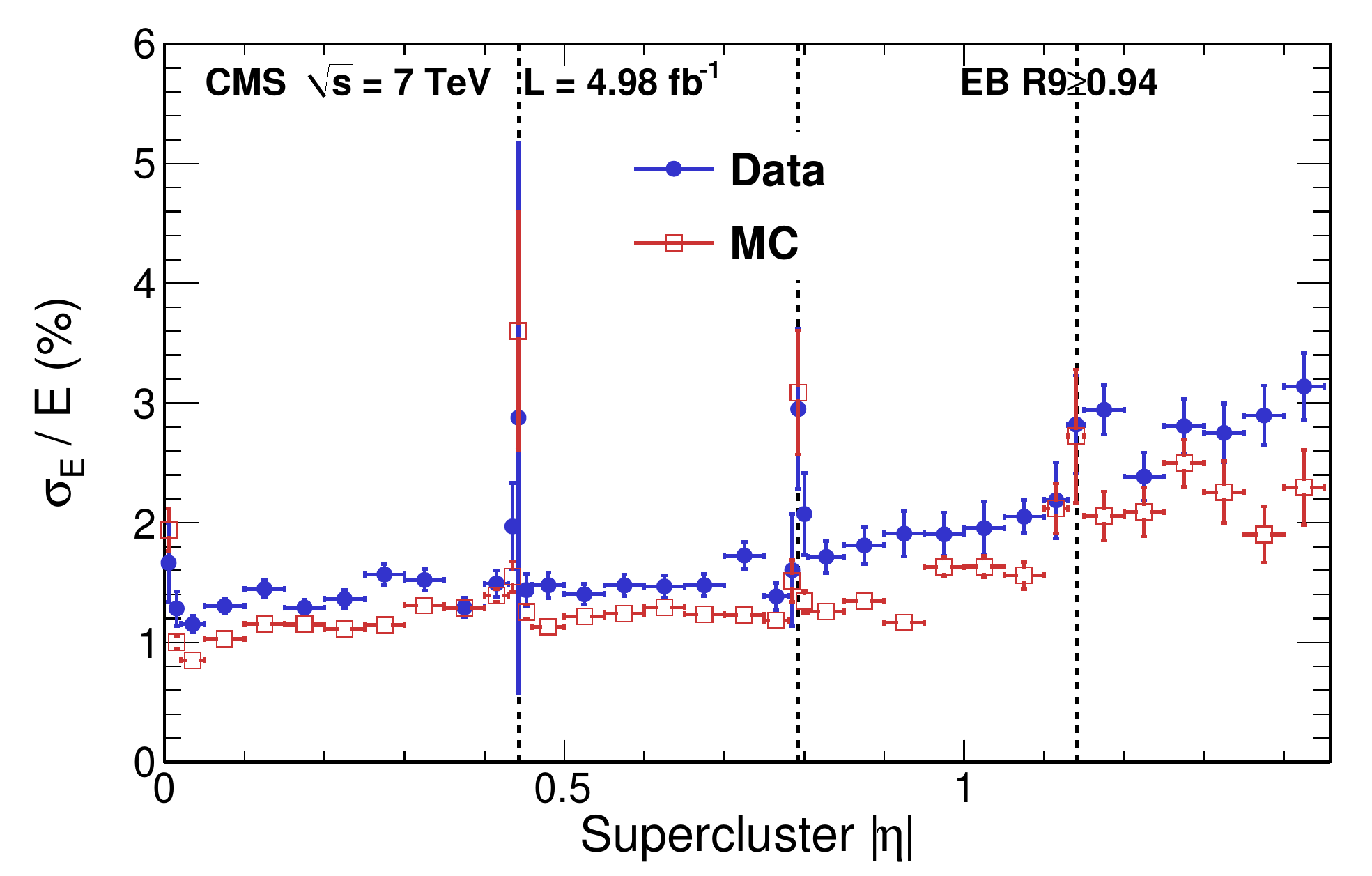}
\includegraphics[height=5.2cm]{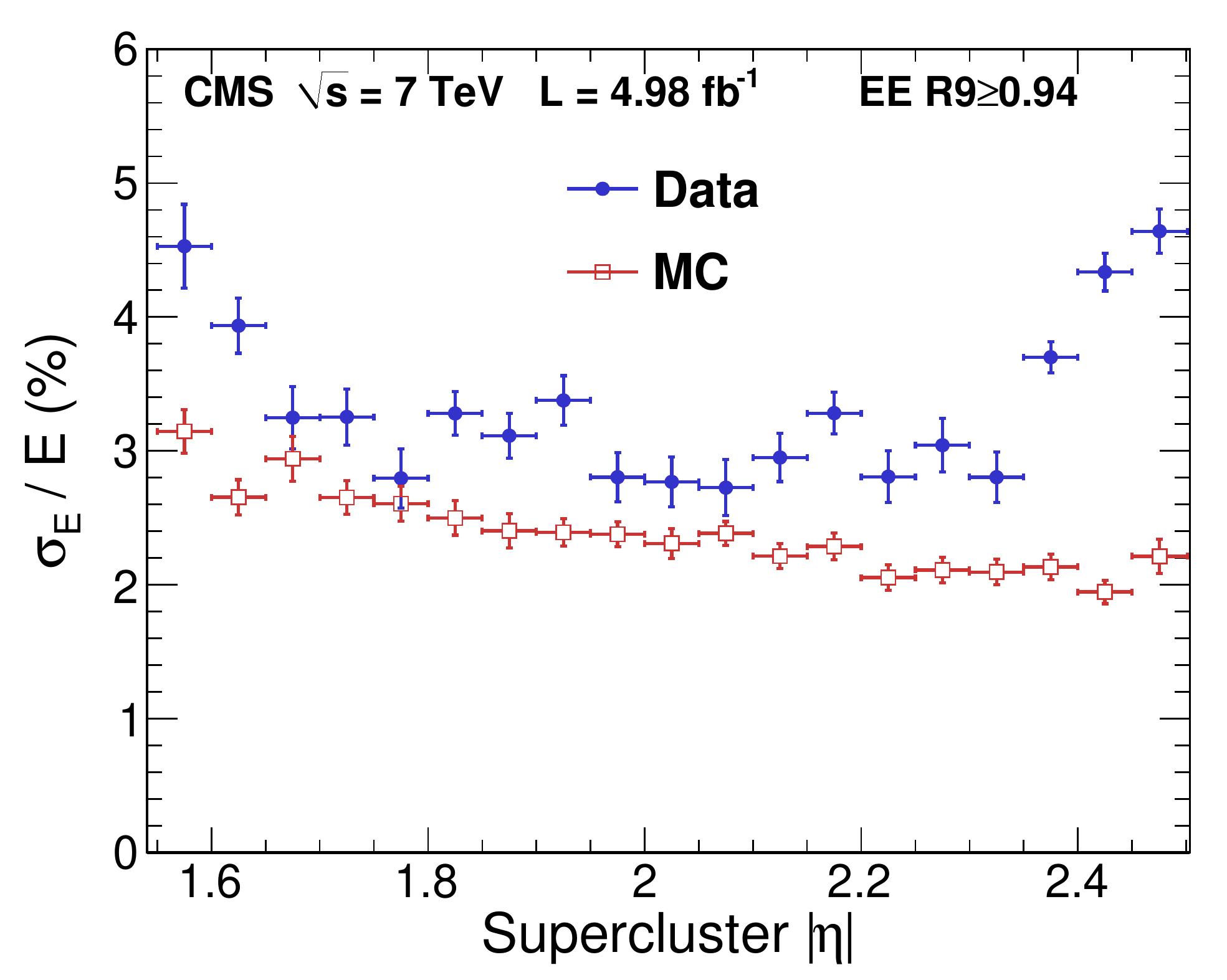}
\includegraphics[height=5.2cm]{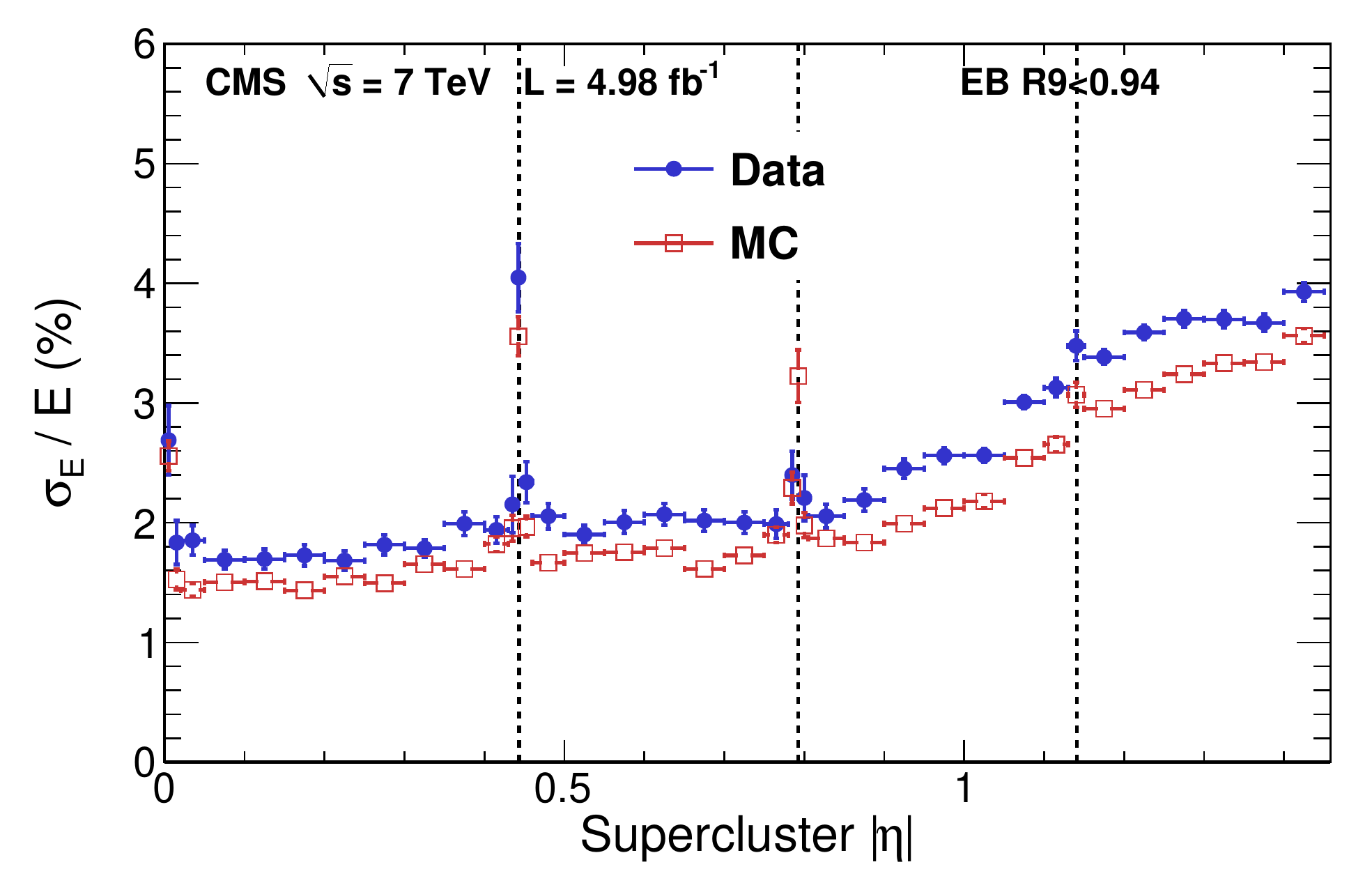}
\includegraphics[height=5.2cm]{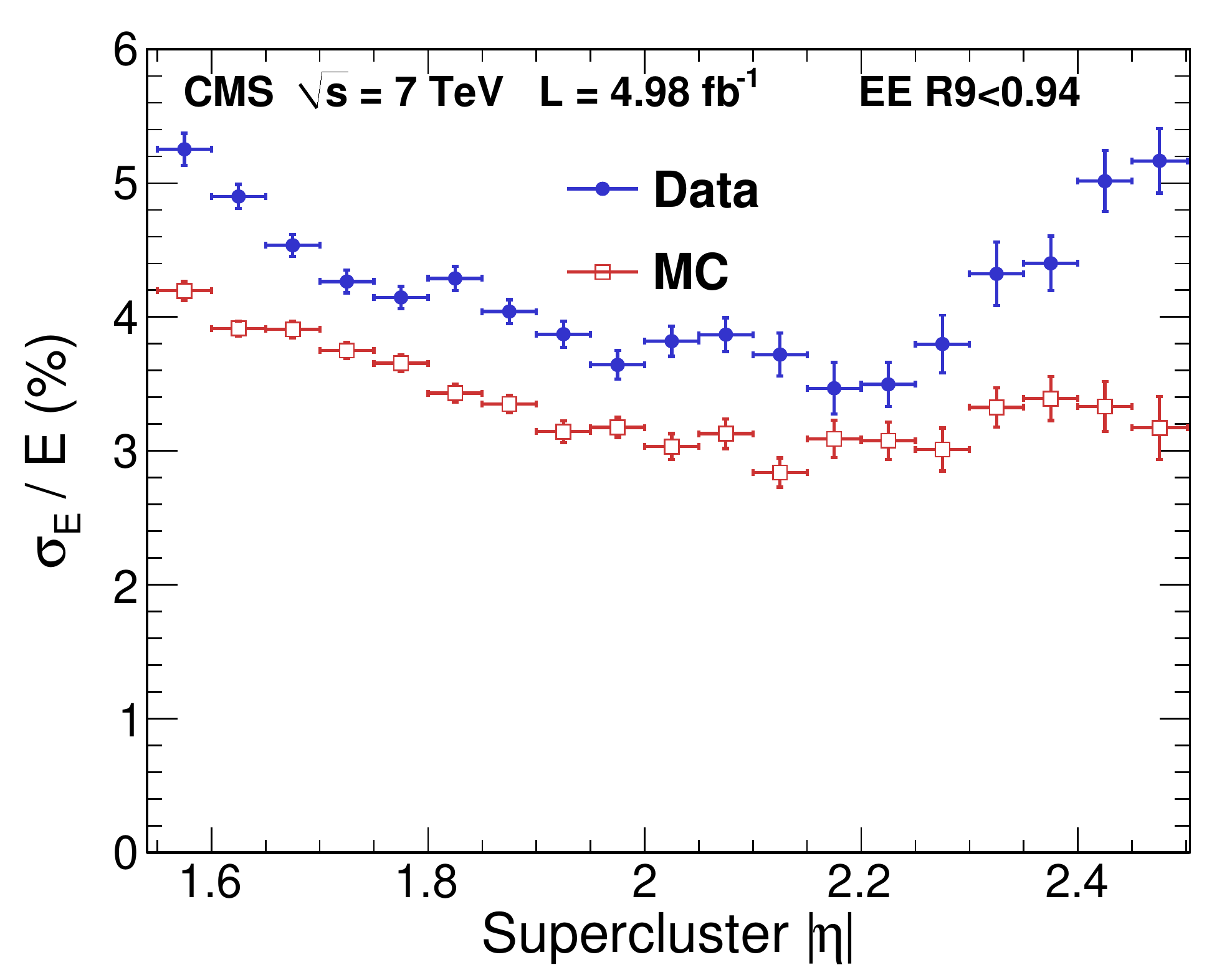}
\end{center}
\caption{\label{fig:zee_unfolding}
Relative electron energy resolution in data and MC events unfolded in
bins of pseudorapidity $\eta$ for the barrel and the endcaps, using
electrons from $\cPZ \to \Pep\Pem$ decays. The resolution is shown
separately for electrons with $R9 \ge 0.94$ and $R9 < 0.94$. The
resolution, $\sigma_E$, is extracted from a fit to $\cPZ \to \Pep\Pem$
events, using a Breit--Wigner distribution convolved with a Gaussian
function as the signal model. }
\end{figure}

Figure~\ref{fig:zee_unfolding} shows the energy resolution extracted
using this method for both data and MC simulation. The average
resolution $\sigma_E/E$ for electrons from $\cPZ$-boson decays is
plotted as a function of $\eta$ in the barrel and endcaps, and is
shown separately for electrons with $R9 \ge 0.94$ and $R9 < 0.94$. The
energy resolution obtained with this method is in agreement with the
fits to the $\cPZ$-boson invariant mass distribution in
Fig.~\ref{fig:zee_scale}, assuming a scaling of the mass resolution
by $\sqrt{2}$ to obtain the equivalent per-electron energy resolution.

The resolution in the barrel depends on the amount of material in
front of the ECAL (see Fig.~\ref{fig:clustercorr} right), and is
degraded in the vicinity of the ECAL module boundaries, as indicated
by vertical lines in the plots. The resolution in the endcaps shows an
$\eta$ dependence that is also correlated with the amount of material
in front of the ECAL, up to $\abs{\eta}\approx 2.0$. At larger pseudorapidity,
single-channel response variations, not fully modelled in simulation,
are also contributing to the difference between data and MC
simulation.

\begin{figure}[tbp]
\begin{center}
\includegraphics[width=0.49\linewidth]{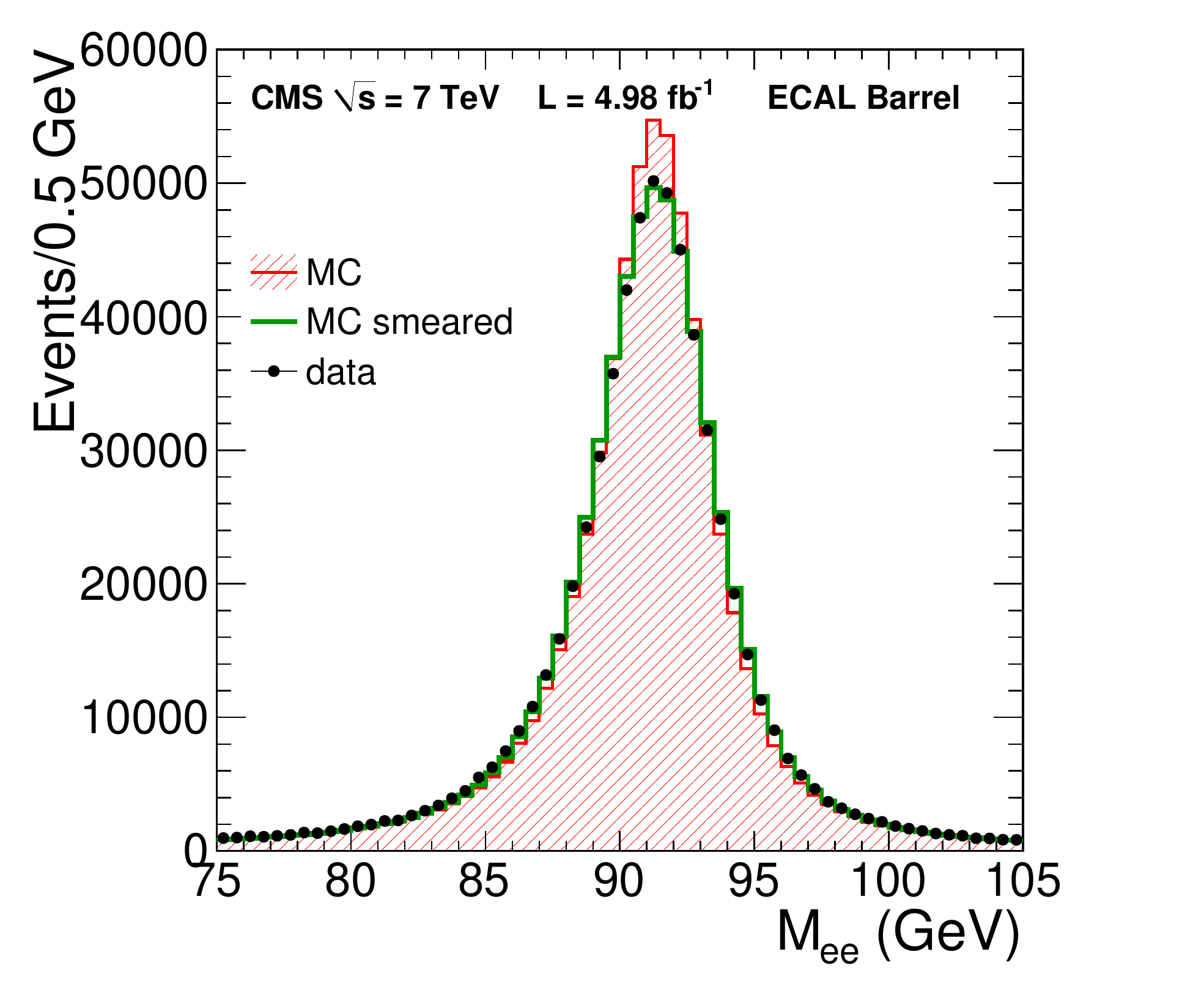}
\includegraphics[width=0.49\linewidth]{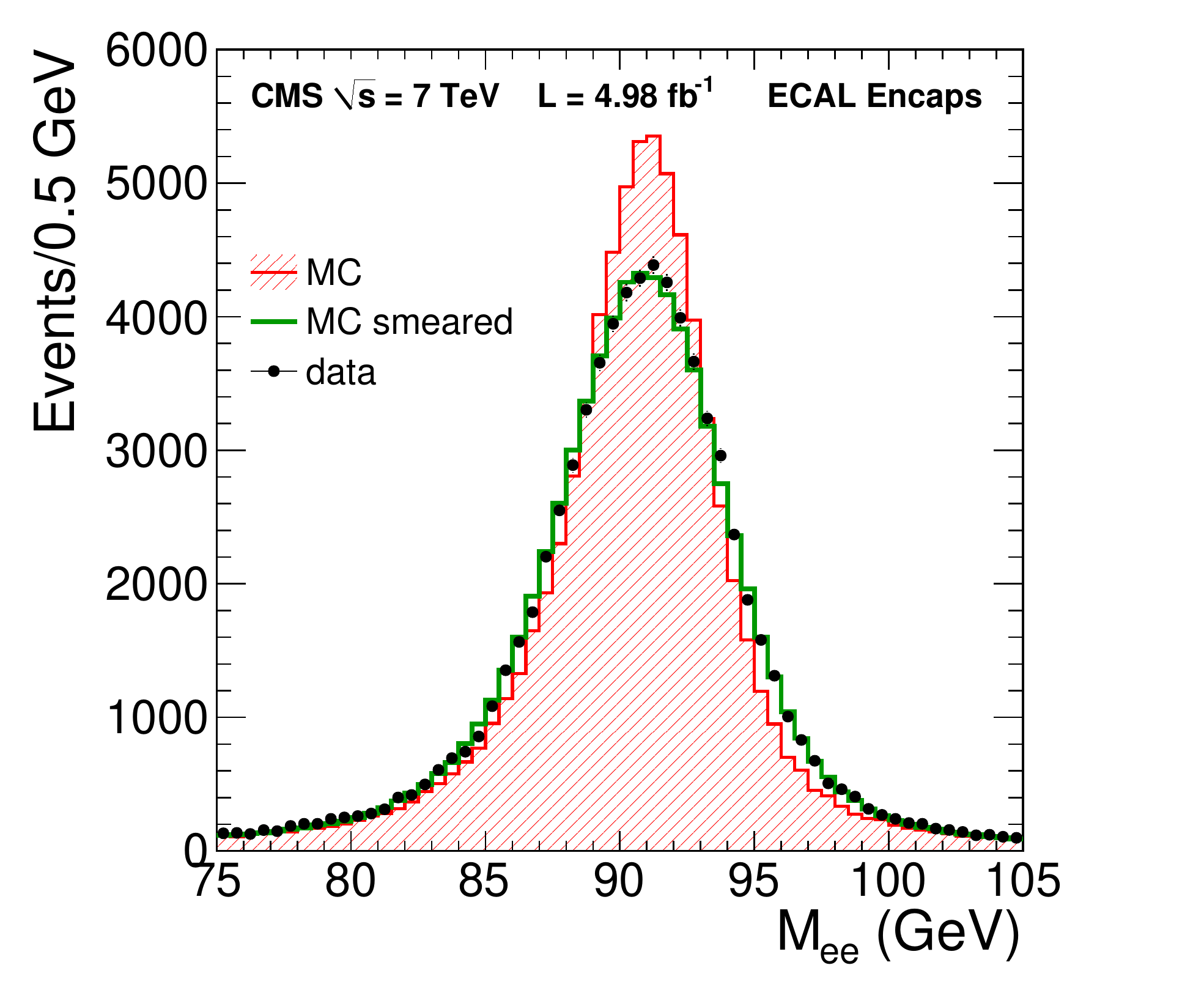}
\end{center}
\caption{\label{fig:Zee_smear}
Distribution of the dielectron invariant mass for the default MC
simulation (filled line), for the MC simulation with additional
Gaussian smearing (green line), and for the data (dots). The
distributions for events with both electrons in EB (left) and in EE
(right) are displayed.}
\end{figure}

To accommodate the mismatch in the energy resolution between data and
simulation, an additional smearing term is extracted, which is the
quadratic difference between the electron resolution in data and MC
simulation of Fig.~\ref{fig:zee_unfolding}. This term is added in
quadrature as a constant Gaussian smearing to the electron and photon
energy in the MC events, assuming the same degradation in resolution
between data and MC events for photons and electrons. The consistency
of this method was checked by comparing the mass resolution in  $\cPZ\to
\Pep\Pem$ and $\cPZ\to \Pgm\Pgm\Pgg$ events in data and in a MC sample
with this smearing term applied. Fig.~\ref{fig:Zee_smear} shows the
dielectron invariant mass for $\cPZ\to \Pep\Pem$ events, for the MC
samples with and without this smearing, compared to data. The
agreement between data and the smeared MC confirms that the smeared MC
sample correctly models the detector response. For $\cPZ\to\Pgm\Pgm\Pgg$
events, the resolution of the smeared MC sample is also consistent
with data. This supports the compatibility of the resolution
measurements for electrons and photons from $\cPZ$-boson decays.

\subsection{Energy resolution for photons from simulated \texorpdfstring{$\PH \to \Pgg
  \Pgg$}{H->gg} events}
\label{sec:hggresolution}

\begin{figure}[t]
\begin{center}
\includegraphics[height=5.2cm]{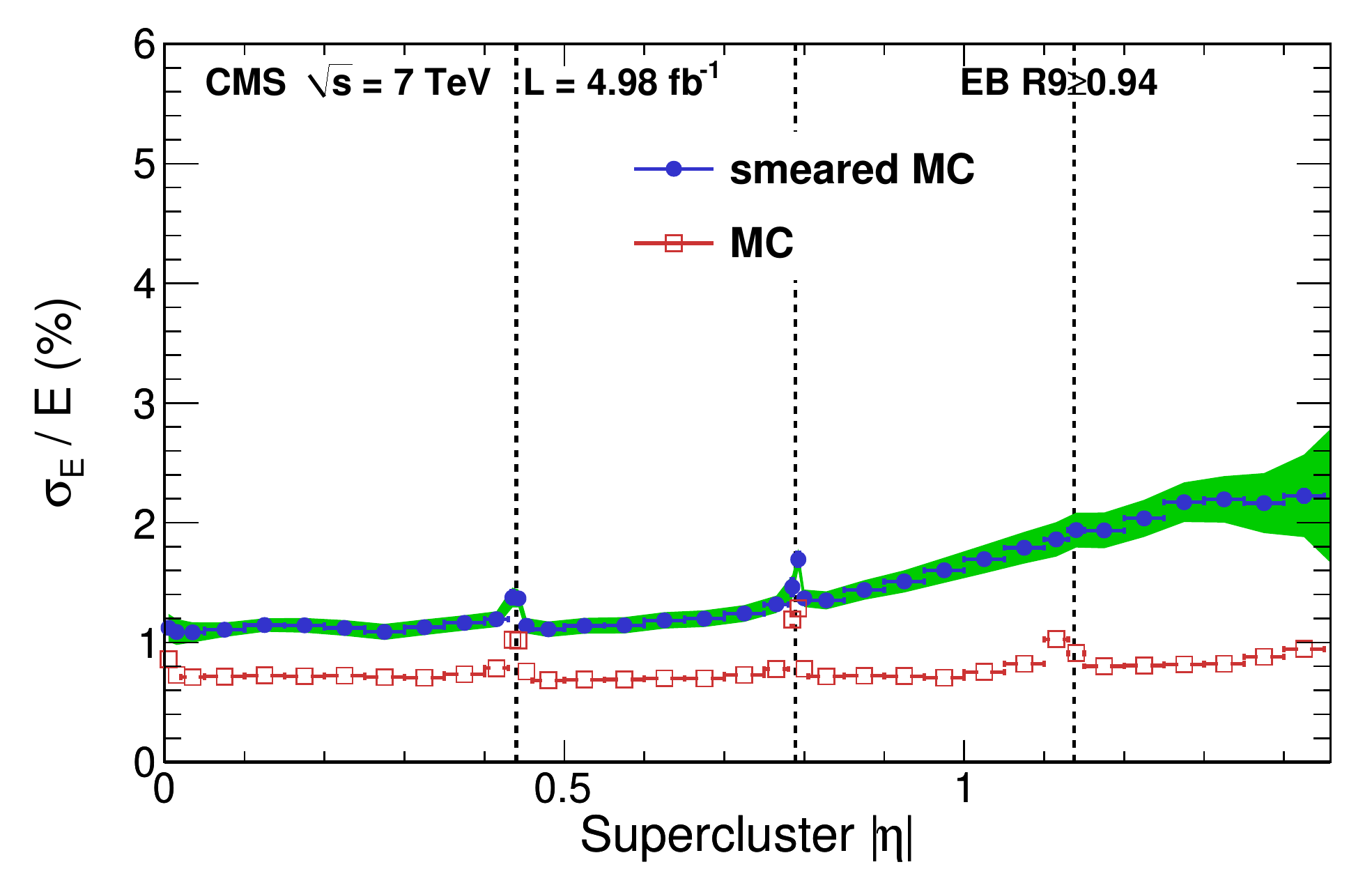}
\includegraphics[height=5.2cm]{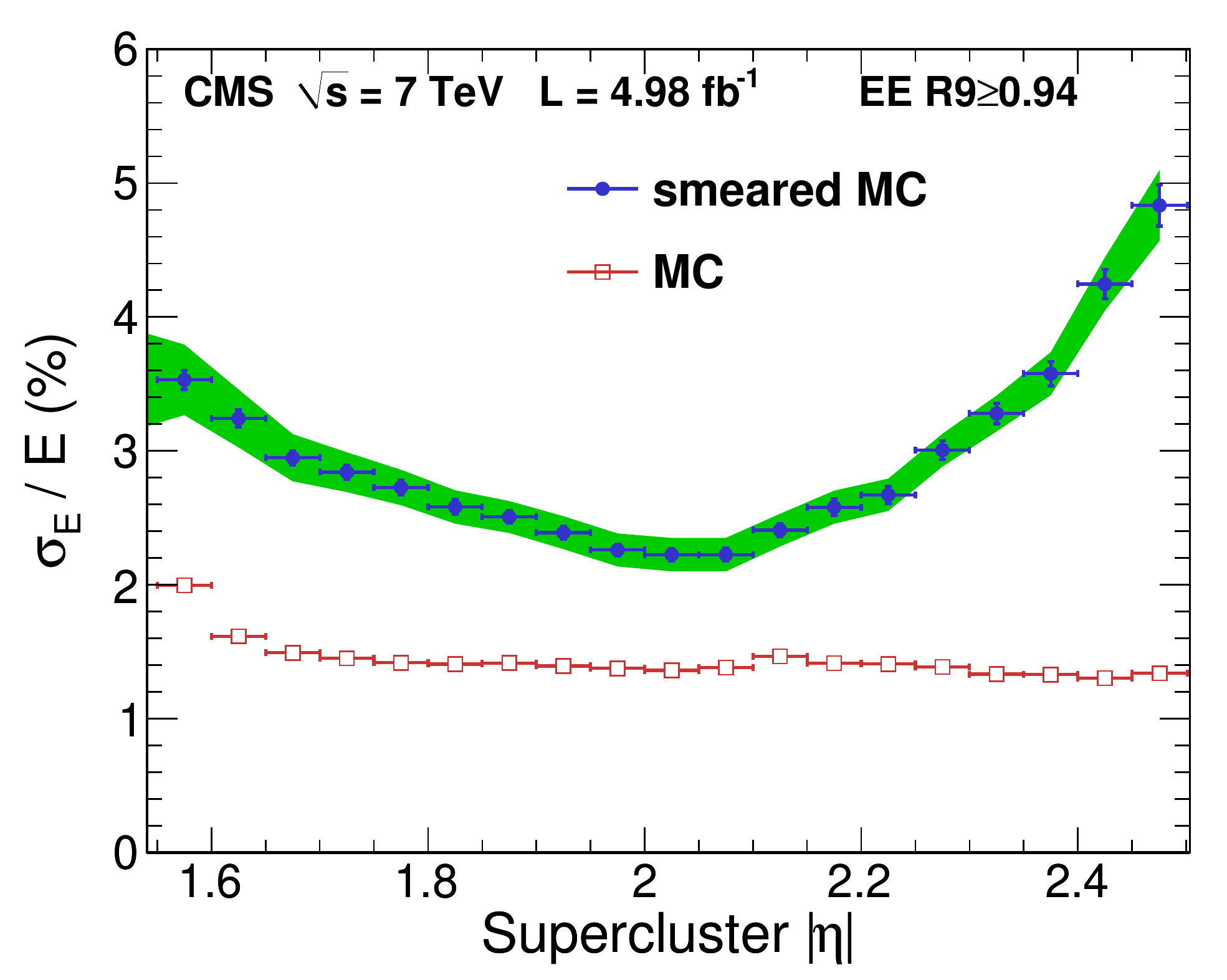}
\includegraphics[height=5.2cm]{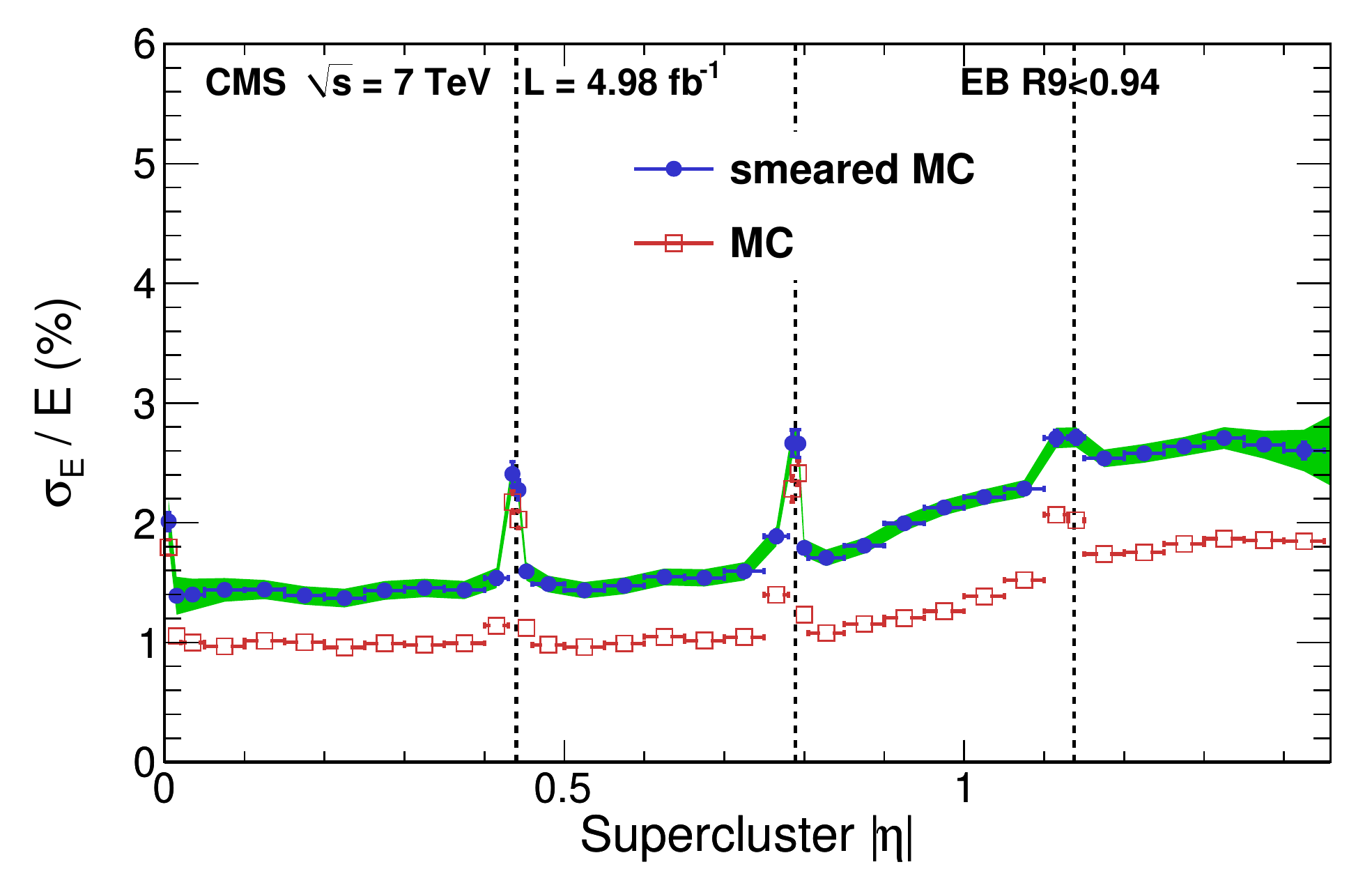}
\includegraphics[height=5.2cm]{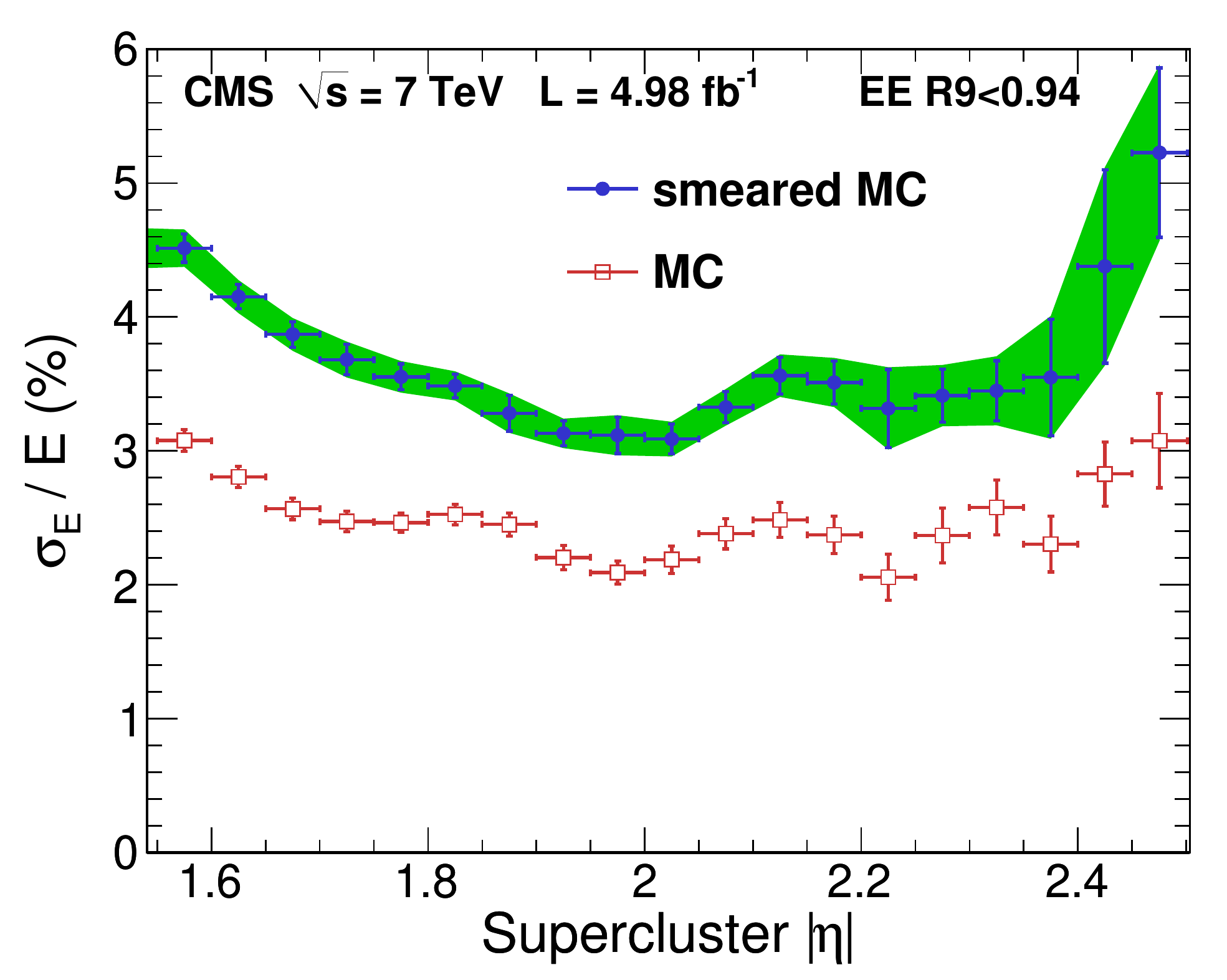}
\end{center}
\caption{\label{fig:photon_resolution}
Photon energy resolution in bins of pseudorapidity $\abs{\eta}$ for the
barrel (left column) and the endcaps (right column). The resolution is
shown separately for photons having  $R9 \ge 0.94$ (top row) and $R9 <
0.94$ (bottom row). The energy resolution is plotted for the simulated
$\PH\to\Pgg\Pgg$ events for the default MC simulation and for MC
simulation with the addition of Gaussian smearing. The green band
shows the uncertainty on the photon resolution calculated as the
quadratic sum of the uncertainty on the smearing term and the
statistical uncertainty in the photon resolution (shown by the
vertical error bars).}
\end{figure}

The energy resolution for photons of $\ET \approx 60$\GeV, predicted
by MC simulation of $\PH \to \Pgg \Pgg$ events for a 125\GeV Higgs
boson, is shown in Fig.~\ref{fig:photon_resolution} with and without
the smearing term discussed in the previous section. The photon
selection and cluster corrections are identical to those used in the
CMS $\PH \to \Pgg \Pgg$ analysis of 2011 data~\cite{HGG-PAPER}. The
resolution, $\sigma_E/E$, is extracted from a fit to the distribution
of the ratio of the reconstructed and the true photon energies.
The resolution as a function of $\abs{\eta}$ is plotted separately in EB
and EE for photons with $R9 \ge 0.94$ and $R9 < 0.94$, which are
samples enhanced in unconverted and converted photons, respectively.
The energy resolution for photons from the $\PH\to\Pgg\Pgg$ decay in the
default MC samples varies between 0.7\% for unconverted photons in the
central part of the barrel to 1.8\% towards the end of the barrel for
converted photons. In the endcap the resolution in MC samples varies
between 1.5\% and 3\%. In the MC sample with smearing added, the
resolution varies between 1.1\% and 2.6\% in the barrel and from 2.2\%
up to 5\% in the endcaps.

In this approach the difference in resolution between data and
simulation, observed using electrons of $\ET\approx 45$\GeV from
$\cPZ$-boson decays, has been ascribed to the constant term in the
resolution function. With this assumption, the results with smearing
shown in Fig.~\ref{fig:photon_resolution} should be regarded as an
upper limit to the resolution for photons from $\PH\to\Pgg\Pgg$
decays. The smearing required to correctly describe the data is higher
where the material budget is higher, suggesting that the effect of the
material is not properly simulated or that the material budget is not
completely realistic. This indicates that a component of the data to
MC simulation difference is related to the interaction of electrons
and photons with the material in front of ECAL, whose effect on the
resolution has also an $\ET$ dependence.

\subsection{Discussion on the energy resolution in data and simulation}
\label{sec:resolution}
As has been discussed above, the resolution predicted in MC simulation is
better than that in data. There is a continuing effort to improve the
detector modelling in the MC simulation and to improve the resolution
in data. Four specific areas have been identified for further study
and will be addressed when additional data become available:

\paragraph{Tracker material description:} The amount of material in front
of the ECAL included in the MC simulation has been verified by
comparing the number of conversions and nuclear interactions observed
in the data with those expected from the simulation. They agree to
better than 5\% on average, with local differences up to
10\%~\cite{TRACKER_NOTE,TRACKER_PAS}. However,
as shown in Fig.~\ref{fig:zee_unfolding}, the energy resolution is
better for electrons that do not interact significantly with the
tracker material, as characterized by $R9 \ge 0.94$. Also, in
Fig.~\ref{fig:zee_unfolding}, the difference in quadrature in the
measured resolution between data and MC simulation is larger, in the
barrel and in much of the endcaps, in regions which have more upstream
tracker material. These observations suggest that the deteriorating
effect of the material in front of the ECAL has a larger impact on
data than on MC events. Further improvements in the modelling of the
material geometry and its effect in MC simulation as well as
mitigating the impact of the material in the reconstruction algorithms
and corrections are being investigated.

\paragraph{Clustering improvements and energy corrections:}
A key element in the mitigation of material effects is a clustering
algorithm that efficiently collects all the energy deposits of the
electrons and photons, and the subsequent optimization of the energy
corrections. The clustering algorithms are very sensitive to the
precise modelling of the showering process and the geometric
description of the tracker material as well as of the ECAL. A better
tuning of the description of electromagnetic showers in the MC
simulation is being pursued, using CMS data as input to the
simulations. In addition, with increased data size, possible
improvements in the determination of cluster corrections may be
achieved by measuring them directly from data. In particular, the
local effect of module borders in the barrel and 5$\times$5 crystal
structures in the endcaps shows differences in the reconstructed
energy as a function of the impact point between data and MC samples
at the level of 1\%, which are not yet accounted for using data-driven
corrections.

\paragraph{Imperfect knowledge of the $\alpha$ parameter:} The limited
knowledge of the channel-to-channel spread in the  $\alpha$  parameter
in Eq.~(\ref{eq:alpha}) would degrade the resolution in data. Assuming
an uncertainty on $\alpha$ of 10\%, the typical response change at the
level of 3\%  in the barrel and 15\% on the average in the endcaps (up
to 25\% at $\abs{\eta}\approx 2.5$) would result in an additional contribution
to the resolution of 0.3\%, and 1.5\% (2.5\% at $\abs{\eta}\approx 2.5$),
respectively. A first-order correction was performed in the endcaps by
optimizing $\alpha$ using events in data, as described in
Section~\ref{sec:lasercorrection}. Further gains are anticipated by
measuring $\alpha$ at the per crystal or per $\phi$ ring level, using
the large sample of events ($\Pgpz/\Pgh, \PW\to \Pe \nu, \cPZ$ and minimum
bias data) collected in 2011 and 2012.

\paragraph{Imperfect knowledge of the intercalibration systematic uncertainty:}
The intercalibration constants are determined with several independent
methods, which exploit different events, and are then combined. In the
combination it is assumed that the methods are completely independent.
However some experimental effects, for example those related to the
detector geometry, may lead to common systematic uncertainties. The
$E/p$ intercalibration method is so far statistically limited, and it
also needs a larger event sample to study the systematic uncertainties
in detail with data. It is expected that additional data will help
clarify whether common sources of systematic uncertainties could lead
to an overestimate of the intercalibration precision.

There are also a number of additional small effects that are not
modelled in the MC simulation, which may affect the energy resolution
when comparing data to MC samples. The crystal transparency change is
not implemented in the MC simulation, resulting in an underestimate of
the stochastic and constant term in the energy resolution.

\section{Conclusions}
\label{sec:summary}
The energy calibration and resolution of the electromagnetic
calorimeter of the CMS detector have been determined using
proton-proton collision data from LHC operation in 2010 and 2011 at a
centre-of-mass energy of $\sqrt{s}=7$\TeV with integrated luminosities
of about 5\fbinv.

The stability of the cooling, high voltage and readout electronics has
fully met requirements, with an impact on the resolution of less than
0.2\%. The spatial alignment with the tracker and the synchronization
of the ECAL readout match expectations, and are at the level required
for $\Pe/\Pgg$ identification and signal amplitude reconstruction. The
laser monitoring system was successfully exploited to correct for
changes in channel response due to the radiation damage. The
stability of the corrected ECAL response was better than 0.2\%
in the barrel and 0.4\% in the endcaps throughout this
period. Physical processes such as $\pi^0$ and $\eta$ decays to
two photons, and $\PW$- and $\cPZ$-boson decays to electrons, have been
used to monitor the stability and to carry out channel-to-channel
intercalibrations. The $\Pgpz$ and $\Pgh$ data were collected on
dedicated trigger streams. The contribution to the constant term of
the energy resolution, due to the intercalibration uncertainty, is
between 0.3--0.5\% in EB and 1--1.5\% in EE, depending on $\eta$.

The energy resolution has been compared in detail to that obtained
with the full CMS Monte Carlo simulation. The resolution in simulation
is better than in data. The difference is greater in regions where
there is a significant amount of material in front of the
ECAL. Although the origin of the difference is not fully understood,
disagreement between data and MC simulation is removed by applying an
additional contribution to the constant term of the energy resolution
of electrons and photons in the MC events.
The resolution for $\ET\approx 45$\GeV electrons from $\cPZ$-boson decays
is better than 2\% in the central region of the ECAL barrel
($\abs{\eta}<0.8$), and is between 2\% and 5\% elsewhere. For electrons
with little bremsstrahlung, where 94\% or more of the clustered energy
is contained within a 3$\times$3 array of crystals, the energy
resolution improves to 1.5\% for $\abs{\eta}<0.8$.  The mass resolution
for $\cPZ \to \Pep\Pem$ decays, when both electrons are in the barrel, is
1.6\% and is 2.6\% for decays when both electrons are in the endcaps.
The resulting energy resolution for photons with $\ET\approx 60$\GeV
from 125\GeV Higgs boson decays varies across the barrel from 1.1\% to
2.6\% and from 2.2\% to 5\% in the endcaps.

The analysis of 2010-2011 proton-proton collisions has shown that,
even in the challenging LHC environment, with high radiation and high
event pileup, the CMS electromagnetic calorimeter has been
successfully operated and calibrated to excellent precision. The
achievements on the energy resolution played a vital part in enabling
CMS to observe a new boson with a mass of
125\GeV~\cite{Chatrchyan:2012ufa, Chatrchyan:2013lba}.

\section*{Acknowledgements}
\hyphenation{Bundes-ministerium Forschungs-gemeinschaft
  Forschungs-zentren} We congratulate our colleagues in the CERN
accelerator departments for the excellent performance of the LHC and
thank the technical and administrative staffs at CERN and at other CMS
institutes for their contributions to the success of the CMS
effort. In addition, we gratefully acknowledge the computing centres
and personnel of the Worldwide LHC Computing Grid for delivering so
effectively the computing infrastructure essential to our
analyses. Finally, we acknowledge the enduring support for the
construction and operation of the LHC and the CMS detector provided by
the following funding agencies: the Austrian Federal Ministry of
Science and Research and the Austrian Science Fund; the Belgian Fonds
de la Recherche Scientifique, and Fonds voor Wetenschappelijk
Onderzoek; the Brazilian Funding Agencies (CNPq, CAPES, FAPERJ, and
FAPESP); the Bulgarian Ministry of Education, Youth and Science; CERN;
the Chinese Academy of Sciences, Ministry of Science and Technology,
and National Natural Science Foundation of China; the Colombian
Funding Agency (COLCIENCIAS); the Croatian Ministry of Science,
Education and Sport; the Research Promotion Foundation, Cyprus; the
Ministry of Education and Research, Recurrent financing contract
SF0690030s09 and European Regional Development Fund, Estonia; the
Academy of Finland, Finnish Ministry of Education and Culture, and
Helsinki Institute of Physics; the Institut National de Physique
Nucl\'eaire et de Physique des Particules~/~CNRS, and Commissariat \`a
l'\'Energie Atomique et aux \'Energies Alternatives~/~CEA, France; the
Bundesministerium f\"ur Bildung und Forschung, Deutsche
Forschungsgemeinschaft, and Helmholtz-Gemeinschaft Deutscher
Forschungszentren, Germ\-any; the General Secretariat for Research and
Technology, Greece; the National Scientific Research Foundation, and
National Office for Research and Technology, Hungary; the Department
of Atomic Energy and the Department of Science and Technology, India;
the Institute for Studies in Theoretical Physics and Mathematics,
Iran; the Science Foundation, Ireland; the Istituto Nazionale di
Fisica Nucleare, Italy; the Korean Ministry of Education, Science and
Technology and the World Class University program of NRF, Republic of
Korea; the Lithuanian Academy of Sciences; the Mexican Funding
Agencies (CINVESTAV, CONACYT, SEP, and UASLP-FAI); the Ministry of
Science and Innovation, New Zealand; the Pakistan Atomic Energy
Commission; the Ministry of Science and Higher Education and the
National Science Centre, Poland; the Funda\c{c}\~ao para a Ci\^encia e
a Tecnologia, Portugal; JINR (Armenia, Belarus, Georgia, Ukraine,
Uzbekistan); the Ministry of Education and Science of the Russian
Federation, the Federal Agency of Atomic Energy of the Russian
Federation, Russian Academy of Sciences, and the Russian Foundation
for Basic Research; the Ministry of Science and Technological
Development of Serbia; the Secretar\'{\i}a de Estado de
Investigaci\'on, Desarrollo e Innovaci\'on and Programa
Consolider-Ingenio 2010, Spain; the Swiss Funding Agencies (ETH Board,
ETH Zurich, PSI, SNF, UniZH, Canton Zurich, and SER); the National
Science Council, Taipei; the Thailand Center of Excellence in Physics,
the Institute for the Promotion of Teaching Science and Technology of
Thailand and the National Science and Technology Development Agency of
Thailand; the Scientific and Technical Research Council of Turkey, and
Turkish Atomic Energy Authority; the Science and Technology Facilities
Council, UK; the US Department of Energy, and the US National Science
Foundation.

Individuals have received support from the Marie-Curie programme and
the European Research Council and EPLANET (European Union); the
Leventis Foundation; the A. P. Sloan Foundation; the Alexander von
Humboldt Foundation; the Belgian Federal Science Policy Office; the
Fonds pour la Formation \`a la Recherche dans l'Industrie et dans
l'Agriculture (FRIA-Belgium); the Agentschap voor Innovatie door
Wetenschap en Technologie (IWT-Belgium); the Ministry of Education,
Youth and Sports (MEYS) of Czech Republic; the Council of Science and
Industrial Research, India; the Compagnia di San Paolo (Torino); the
HOMING PLUS programme of Foundation for Polish Science, cofinanced by
EU, Regional Development Fund; and the Thalis and Aristeia programmes
cofinanced by EU-ESF and the Greek NSRF.

\bibliography{auto_generated}   

\cleardoublepage \appendix\section{The CMS Collaboration \label{app:collab}}\begin{sloppypar}\hyphenpenalty=5000\widowpenalty=500\clubpenalty=5000\textbf{Yerevan Physics Institute,  Yerevan,  Armenia}\\*[0pt]
S.~Chatrchyan, V.~Khachatryan, A.M.~Sirunyan, A.~Tumasyan
\vskip\cmsinstskip
\textbf{Institut f\"{u}r Hochenergiephysik der OeAW,  Wien,  Austria}\\*[0pt]
W.~Adam, T.~Bergauer, M.~Dragicevic, J.~Er\"{o}, C.~Fabjan\cmsAuthorMark{1}, M.~Friedl, R.~Fr\"{u}hwirth\cmsAuthorMark{1}, V.M.~Ghete, N.~H\"{o}rmann, J.~Hrubec, M.~Jeitler\cmsAuthorMark{1}, W.~Kiesenhofer, V.~Kn\"{u}nz, M.~Krammer\cmsAuthorMark{1}, I.~Kr\"{a}tschmer, D.~Liko, I.~Mikulec, D.~Rabady\cmsAuthorMark{2}, B.~Rahbaran, C.~Rohringer, H.~Rohringer, R.~Sch\"{o}fbeck, J.~Strauss, A.~Taurok, W.~Treberer-Treberspurg, W.~Waltenberger, C.-E.~Wulz\cmsAuthorMark{1}
% ...
\vskip\cmsinstskip
% ...
1:~~Also at Vienna University of Technology, Vienna, Austria\\
2:~~Also at CERN, European Organization for Nuclear Research, Geneva, Switzerland\\
% ...

\end{sloppypar}
\end{document}